\documentclass[acmtocl]{acmtrans2m}

\usepackage{stmaryrd}

\usepackage{amssymb}
\usepackage{amsmath}
\usepackage{latexsym}
\usepackage{paralist}
\usepackage{calrsfs}

\usepackage[curve,graph,matrix]{xy}

\newtheorem{theorem}{Theorem}[section]
\newtheorem{lemma}[theorem]{Lemma}
\newtheorem{proposition}[theorem]{Proposition}
\newtheorem{corollary}[theorem]{Corollary}

\newdef{definition}[theorem]{Definition}
\newdef{observation}[theorem]{Observation}
\newdef{remark}[theorem]{Remark}
\newdef{para}[theorem]{}
\newdef{casana}[theorem]{Case Analysis}

\usepackage{pst-node,pstricks-add}

\usepackage{color}

\font\lemfont=cmr10 scaled 500

\def\lembox#1{\mbox{\lemfont #1}}

\def\looongrightarrow{-\mkern-7.5mu-\mkern-7.5mu-\mkern-8.5mu\longrightarrow}
\def\loongrightarrow{-\mkern-7.5mu-\mkern-7.5mu-\mkern-8.5mu\rightarrow}
\def\looongleftarrow{\longleftarrow\mkern-8.5mu-\mkern-7.5mu-\mkern-7.5mu-}

\def\lemuparrow{%
  \begin{array}{c}
    \uparrow\\
  \end{array}
}

\newcommand\quadsteptol[2]{%
  \!
  \begin{array}{c}
    \phantom{#2}\\
    \phantom{\lemuparrow}\\
    \smash{\stackrel{\hbox{#1}}{\looongrightarrow}}\\
    \lemuparrow\\
    #2
  \end{array}
  \!
}

\newcommand\quadstepfroml[2]{%
  \!
  \begin{array}{c}
    \phantom{#2}\\
    \phantom{\lemuparrow}\\
    \smash{\stackrel{\hbox{#1}}{\looongleftarrow}}\\
    \lemuparrow\\
    #2
  \end{array}
  \!
}

\newcommand{\ie}{i.e.,\ }

\newcommand{\wolg}{without loss of generality}
\newcommand{\bwoc}{by way of contradiction}

\newcommand{\vlpar}{\mathchoice
      {\mathbin{\raise.3ex \hbox{$\scriptstyle      \lpar  $}}}
      {\mathbin{\raise.3ex \hbox{$\scriptstyle      \lpar  $}}}
      {\mathbin{\raise.12ex\hbox{$\scriptscriptstyle\lpar  $}}}
      {\mathbin{           \hbox{$\scriptscriptstyle\lpar  $}}}}

\newcommand{\vlten}{\mathchoice
      {\mathbin{\raise.3ex \hbox{$\scriptstyle      \ltens$}}}
      {\mathbin{\raise.3ex \hbox{$\scriptstyle      \ltens$}}}
      {\mathbin{\raise.12ex\hbox{$\scriptscriptstyle\ltens$}}}
      {\mathbin{           \hbox{$\scriptscriptstyle\ltens$}}}}

\let\seqsym\vartriangleleft

\newcommand{\vlseq}{\mathchoice
      {\mathbin{\raise.3ex \hbox{$\scriptstyle      \seqsym   $}}}
      {\mathbin{\raise.3ex \hbox{$\scriptstyle      \seqsym   $}}}
      {\mathbin{\raise.12ex\hbox{$\scriptscriptstyle\seqsym   $}}}
      {\mathbin{           \hbox{$\scriptscriptstyle\seqsym   $}}}}

\def\vldots{\cdots}

\newdimen\dercldim                                
\newdimen\derccdim                                
\newdimen\dercrdim                                
\newdimen\derldim                                 
\newdimen\dercdim                                 
\newdimen\derrdim                                 
\newdimen\derdim                                  
\newdimen\derdldim                                
\newdimen\derdrdim                                
\newbox\derboxone                                 
\newbox\derboxtwo                                 
\newbox\derboxthree                               
\newbox\derboxfour                                
\newdimen\derquad\derquad=\fontdimen6\textfont2
\newdimen\deropen\deropen=\fontdimen5\textfont2\divide\deropen by3
\def\leaf #1{\global\setbox\derboxone=\hbox{\strut$#1$}%
   \global\derldim=0pt                            
   \global\dercdim=\wd\derboxone                  
   \global\derrdim=0pt                            
   }%
\def\rootaux #1#2#3{\setbox\derboxtwo=\hbox{\unhbox\derboxone}%
   \setbox\derboxthree=\hbox 
      {$\smash{\lower\fontdimen22\textfont2\hbox{$#1$}}$}%
   \setbox\derboxfour=\hbox 
      {$\smash{\lower\fontdimen22\textfont2\hbox{$#2$}}$}%
   \leaf{#3}
   \derdim=\dercdim\advance\derdim by-\derccdim\divide\derdim by2 
   \global\derldim=\dercldim\global\advance\derldim by-\derdim
   \global\derrdim=\dercrdim\global\advance\derrdim by-\derdim
   \deropen=\fontdimen5\textfont2\divide\deropen by3
   \setbox\derboxone=\hbox{\vbox{\offinterlineskip
         \hbox{\ifdim\derldim<0pt\kern-\derldim\fi
               \box\derboxtwo
               \ifdim\derrdim<0pt\kern-\derrdim\fi}%
         \kern\deropen
         \hbox{\ifdim\dercldim>\derldim
                  \ifdim\derldim>0pt\kern\derldim\fi
                  \else\kern\dercldim\fi
               \hbox to0pt{\hss\copy\derboxthree}%
               \vbox{\ifdim\derccdim>\dercdim\hsize=\derccdim
                                        \else\hsize=\dercdim \fi
                    \hrule height.2pt depth.2pt width\hsize}%
               \hbox to0pt{\copy\derboxfour\hss}%
               \ifdim\dercrdim>\derrdim
                  \ifdim\derrdim>0pt\kern\derrdim\fi
                  \else\kern\dercrdim\fi}%
         \kern\deropen
         \hbox{\ifdim\derldim>0pt\kern\derldim\fi
               \box\derboxone
               \ifdim\derrdim>0pt\kern\derrdim\fi}}}%
   \ifdim\derldim<0pt\global\derldim=0pt\fi       
   \ifdim\derrdim<0pt\global\derrdim=0pt\fi       
   \derdldim=\wd\derboxthree\advance\derdldim by-\dercldim
   \derdrdim=\wd\derboxfour \advance\derdrdim by-\dercrdim
   \ifdim\derdim<0pt
      \ifdim\derdldim<0pt
         \derdldim=0pt                            
      \fi
      \ifdim\derdrdim<0pt
         \derdrdim=0pt                            
      \fi
   \else
      \ifdim\derldim>0pt
         \ifdim\derdldim>-\derdim
            \advance\derdldim by\derdim           
         \else			                          
            \derdldim=0pt                         
         \fi                                      
      \else
         \advance\derdldim by\dercldim            
      \fi
      \ifdim\derrdim>0pt
         \ifdim\derdrdim>-\derdim
            \advance\derdrdim by\derdim           
         \else			                          
            \derdrdim=0pt                         
         \fi                                      
      \else
         \advance\derdrdim by\dercrdim            
      \fi
   \fi
   \global\setbox\derboxone=\hbox
      {\kern\derdldim\unhbox\derboxone\kern\derdrdim}%
   \global\advance\derldim by\derdldim            
   \global\advance\derrdim by\derdrdim            
   }%
\def\rootr #1#2#3#4{{#4}%
   \dercldim=\derldim
   \derccdim=\dercdim
   \dercrdim=\derrdim
   \rootaux{#1}{#2}{#3}}%
\def\rrootr #1#2#3#4#5{\derquad=\fontdimen6\textfont2
   {#4}%
           \dercldim  =\derldim
   \setbox\derboxtwo=\hbox{\unhbox\derboxone\kern\derquad}%
           \derccdim  =\dercdim
   \advance\derccdim by\derrdim
   \advance\derccdim by\derquad
   {#5}%
   \setbox\derboxone=\hbox{\unhbox\derboxtwo\unhbox\derboxone}%
   \advance\derccdim by\derldim
   \advance\derccdim by\dercdim
           \dercrdim  =\derrdim
   \rootaux{#1}{#2}{#3}}%
\def\rrrootr #1#2#3#4#5#6{\derquad=\fontdimen6\textfont2
   {#4}%
           \dercldim  =\derldim
   \setbox\derboxtwo=\hbox{\unhbox\derboxone\kern\derquad}%
           \derccdim  =\dercdim
   \advance\derccdim by\derrdim
   \advance\derccdim by\derquad
   {#5}%
   \setbox\derboxtwo=\hbox{\unhbox\derboxtwo\unhbox\derboxone\kern\derquad}%
   \advance\derccdim by\derldim
   \advance\derccdim by\dercdim
   \advance\derccdim by\derrdim
   \advance\derccdim by\derquad
   {#6}%
   \setbox\derboxone=\hbox{\unhbox\derboxtwo\unhbox\derboxone}%
   \advance\derccdim by\derldim
   \advance\derccdim by\dercdim
           \dercrdim  =\derrdim
   \rootaux{#1}{#2}{#3}}%
\def\root       #1#2#3{\rootr  {#1\;}{\;}{#2}{#3}}%
\def\rroot    #1#2#3#4{\rrootr {#1\;}{\;}{#2}{#3}{#4}}%

\newbox\stembox
\def\stemaux #1#2#3{\setbox\derboxtwo=\hbox{\unhbox\derboxone}%
   \setbox\derboxthree=\hbox{$#1$}   
   \setbox\derboxfour =\hbox{$#2$}   
      {\global\setbox\derboxone=\hbox{$#3$}%
      \global\derldim=0pt                         
      \global\dercdim=\wd\derboxone               
      \global\derrdim=0pt                         
      }
   \derdim=\dercdim\advance\derdim by-\derccdim\divide\derdim by2 
   \global\derldim=\dercldim\global\advance\derldim by-\derdim
   \global\derrdim=\dercrdim\global\advance\derrdim by-\derdim
   \deropen=\fontdimen5\textfont2\divide\deropen by3
   \setbox\derboxone=\hbox{\vbox{\offinterlineskip
         \hbox{\ifdim\derldim<0pt\kern-\derldim\fi
               \box\derboxtwo
               \ifdim\derrdim<0pt\kern-\derrdim\fi}%
         \kern-\deropen\kern-\ht\strutbox\kern-\dp\strutbox
         \hbox{\ifdim\dercldim>\derldim
                  \ifdim\derldim>0pt\kern\derldim\fi
                  \else\kern\dercldim\fi
               \hbox to0pt{\hss\copy\derboxthree}%
               \vbox{\ifdim\derccdim>\dercdim\hsize=\derccdim
                                        \else\hsize=\dercdim \fi
                    \hbox{$\vcenter{\vbox{\offinterlineskip
                       \hbox{$\copy\stembox$}
                       \hbox{$\copy\stembox$}}}$}}%
               \hbox to0pt{\copy\derboxfour\hss}%
               \ifdim\dercrdim>\derrdim
                  \ifdim\derrdim>0pt\kern\derrdim\fi
                  \else\kern\dercrdim\fi}%
         \kern-\deropen
         \hbox{\ifdim\derldim>0pt\kern\derldim\fi
               \box\derboxone
               \ifdim\derrdim>0pt\kern\derrdim\fi}}}%
   \ifdim\derldim<0pt\global\derldim=0pt\fi       
   \ifdim\derrdim<0pt\global\derrdim=0pt\fi       
   \derdldim=\wd\derboxthree\advance\derdldim by-\dercldim
   \derdrdim=\wd\derboxfour \advance\derdrdim by-\dercrdim
   \ifdim\derdim<0pt
      \ifdim\derdldim<0pt
         \derdldim=0pt                            
      \fi
      \ifdim\derdrdim<0pt
         \derdrdim=0pt                            
      \fi
   \else
      \ifdim\derldim>0pt
         \ifdim\derdldim>-\derdim
            \advance\derdldim by\derdim           
         \else			                  
            \derdldim=0pt                         
         \fi                                      
      \else
         \advance\derdldim by\dercldim            
      \fi
      \ifdim\derrdim>0pt
         \ifdim\derdrdim>-\derdim
            \advance\derdrdim by\derdim           
         \else			                  
            \derdrdim=0pt                         
         \fi                                      
      \else
         \advance\derdrdim by\dercrdim            
      \fi
   \fi
   \global\setbox\derboxone=\hbox
      {\kern\derdldim\unhbox\derboxone\kern\derdrdim}%
   \global\advance\derldim by\derdldim            
   \global\advance\derrdim by\derdrdim            
   }%
\def\stemauxx #1#2#3{\setbox\derboxtwo=\hbox{\unhbox\derboxone}%
   \setbox\derboxthree=\hbox{$#1$}   
   \setbox\derboxfour =\hbox{$#2$}   
   \leaf{#3}
   \derdim=\dercdim\advance\derdim by-\derccdim\divide\derdim by2 
   \global\derldim=\dercldim\global\advance\derldim by-\derdim
   \global\derrdim=\dercrdim\global\advance\derrdim by-\derdim
   \deropen=\fontdimen5\textfont2\divide\deropen by3
   \setbox\derboxone=\hbox{\vbox{\offinterlineskip
         \hbox{\ifdim\derldim<0pt\kern-\derldim\fi
               \box\derboxtwo
               \ifdim\derrdim<0pt\kern-\derrdim\fi}%
         \kern\deropen
         \hbox{\ifdim\dercldim>\derldim
                  \ifdim\derldim>0pt\kern\derldim\fi
                  \else\kern\dercldim\fi
               \hbox to0pt{\hss\copy\derboxthree}%
               \vbox{\ifdim\derccdim>\dercdim\hsize=\derccdim
                                        \else\hsize=\dercdim \fi
                    \hbox{\hfil}}%
               \hbox to0pt{\copy\derboxfour\hss}%
               \ifdim\dercrdim>\derrdim
                  \ifdim\derrdim>0pt\kern\derrdim\fi
                  \else\kern\dercrdim\fi}%
         \kern\deropen
         \hbox{\ifdim\derldim>0pt\kern\derldim\fi
               \box\derboxone
               \ifdim\derrdim>0pt\kern\derrdim\fi}}}%
   \ifdim\derldim<0pt\global\derldim=0pt\fi       
   \ifdim\derrdim<0pt\global\derrdim=0pt\fi       
   \derdldim=\wd\derboxthree\advance\derdldim by-\dercldim
   \derdrdim=\wd\derboxfour \advance\derdrdim by-\dercrdim
   \ifdim\derdim<0pt
      \ifdim\derdldim<0pt
         \derdldim=0pt                            
      \fi
      \ifdim\derdrdim<0pt
         \derdrdim=0pt                            
      \fi
   \else
      \ifdim\derldim>0pt
         \ifdim\derdldim>-\derdim
            \advance\derdldim by\derdim           
         \else			                  
            \derdldim=0pt                         
         \fi                                      
      \else
         \advance\derdldim by\dercldim            
      \fi
      \ifdim\derrdim>0pt
         \ifdim\derdrdim>-\derdim
            \advance\derdrdim by\derdim           
         \else			                  
            \derdrdim=0pt                         
         \fi                                      
      \else
         \advance\derdrdim by\dercrdim            
      \fi
   \fi
   \global\setbox\derboxone=\hbox
      {\kern\derdldim\unhbox\derboxone\kern\derdrdim}%
   \global\advance\derldim by\derdldim            
   \global\advance\derrdim by\derdrdim            
   }%
\def\stemr #1#2#3#4{{#4}%
   \dercldim=\derldim
   \derccdim=\dercdim
   \dercrdim=\derrdim
   \stemaux{#1}{#2}{#3}}%
\def\stemrr #1#2#3#4{{#4}%
   \dercldim=\derldim
   \derccdim=\dercdim
   \dercrdim=\derrdim
   \stemauxx{#1}{#2}{#3}}%
\def\stem #1#2#3#4{\setbox\stembox=\hbox{$\|$}%
   \stemrr{  }{  }{#3              }  {
   \stemr {#1\;}{\;#2}{\kern\wd\stembox} {
   \stemrr{  }{  }{\kern\wd\stembox}{
   #4                             }}}}%
\def\stempr #1#2#3{\setbox\stembox=\hbox{$\|$}%
   \stemrr{  }{  }{#3              }  {
   \stemr {#1\;}{\;#2}{\kern\wd\stembox} {
   \stemrr{  }{  }{\kern\wd\stembox}{
      {\global\setbox\derboxone=\hbox{%
         \vbox to0pt{\vss\kern3pt\hbox{$-$}\vss\kern-2\deropen}}%
      \global\derldim=0pt                            
      \global\dercdim=\wd\derboxone                  
      \global\derrdim=0pt                            
      }                             }}}}%

\def\deraux {\derldim=0pt\dercdim=0pt\derrdim=0pt}%
\def\der       #1#2#3{\deraux\root  {#1}{#2}{#3}        \box\derboxone}%
\def\dder    #1#2#3#4{\deraux\rroot {#1}{#2}{#3}{#4}    \box\derboxone}%
\def\dernote       #1#2#3#4{\deraux\rootr  {#1\;}{\;#2}{#3}{#4}\box\derboxone}%
\def\ddernote    #1#2#3#4#5{\deraux\rrootr {#1\;}{\;#2}{#3}{#4}{#5}\box
                                                                   \derboxone}%
\def\inf       #1#2#3{\der  {#1}{#2}{\leaf{#3}}}%
\def\iinf    #1#2#3#4{\dder {#1}{#2}{\leaf{#3}}{\leaf{#4}}}%
\def\infnote       #1#2#3#4{\dernote  {#1}{#4}{#2}{\leaf{#3}}}%
\def\iinfnote    #1#2#3#4#5{\ddernote {#1}{#5}{#2}{\leaf{#3}}{\leaf{#4}}}%

\def\strpr  #1#2#3{\stempr{#1}{#2}{#3}\box\derboxone}%
\def\strder #1#2#3#4{\stem{#1}{#2}{#3}{#4}\box\derboxone}%

\newbox\derskelboxone
\newbox\derskelboxtwo
\newbox\derskelboxthree
\newbox\derskelboxfour
\newdimen\derskeldimenone
\newdimen\derskeldimentwo
\newdimen\derskeldimenthree
\newdimen\derskeldimenfour
\newdimen\derskeldimenfive
\newdimen\derskeldimensix
\newdimen\derskeldimenseven
\newdimen\derskeldimeneight
\def\derskel #1#2#3#4{%
   \setbox\derskelboxone=\hbox{$#1$\strut}%
   \derskeldimenone=\ht\derskelboxone
   \advance\derskeldimenone by\dp\derskelboxone
   \derskeldimentwo=\wd\derskelboxone
   \divide\derskeldimentwo by2
   \setbox\derskelboxone=\hbox to0pt{%
      \hss\raise\dp\derskelboxone\box\derskelboxone\hss}%
   \ht\derskelboxone=0pt
   \dp\derskelboxone=0pt
   \setbox\derskelboxtwo=\hbox{$#3$\strut}%
   \derskeldimenthree=\ht\derskelboxtwo
   \advance\derskeldimenthree by\dp\derskelboxtwo
   \derskeldimenfour=\wd\derskelboxtwo
   \divide\derskeldimenfour by2
   \setbox\derskelboxtwo=\hbox to0pt{%
      \hss\raise\dp\derskelboxtwo\box\derskelboxtwo\hss}%
   \ht\derskelboxtwo=0pt
   \dp\derskelboxtwo=0pt
   \ifdim\derskeldimenone>\derskeldimenthree
      \else\derskeldimenone=\derskeldimenthree\fi
   \setbox\derskelboxthree=\hbox{$#4$\strut}%
   \derskeldimenfive=\ht\derskelboxthree
   \advance\derskeldimenfive by\dp\derskelboxthree
   \derskeldimensix=\wd\derskelboxthree
   \divide\derskeldimensix by2
   \setbox\derskelboxthree=\hbox to0pt{%
      \hss\lower\ht\derskelboxthree\box\derskelboxthree\hss}%
   \ht\derskelboxthree=0pt
   \dp\derskelboxthree=0pt
   \setbox\derskelboxfour=\hbox{$#2$\strut}%
   \derskeldimenseven=\ht\derskelboxfour
   \advance\derskeldimenseven by\dp\derskelboxfour
   \derskeldimeneight=\wd\derskelboxfour
   \divide\derskeldimeneight by2
   \setbox\derskelboxfour=\hbox to0pt{%
      \hss\raise\dp\derskelboxfour\box\derskelboxfour\hss}%
   \ht\derskelboxfour=0pt
   \dp\derskelboxfour=0pt
   \ifdim\derskeldimenone>\derskeldimenseven
      \else\derskeldimenone=\derskeldimenseven\fi
   \derskeldimenthree=\derskeldimentwo
   \advance\derskeldimenthree by2\derskeldimeneight
   \advance\derskeldimenthree by\derskeldimenfour
   \advance\derskeldimenthree by2em
   \divide\derskeldimenthree by2
   \advance\derskeldimensix by-\derskeldimenthree
   \derskeldimenseven=\derskeldimensix
   \advance\derskeldimensix by-\derskeldimentwo
   \advance\derskeldimenseven by-\derskeldimenfour
   \ifdim\derskeldimensix>0pt
      \else\derskeldimensix=0pt\fi
   \ifdim\derskeldimenseven>0pt
      \else\derskeldimenseven=0pt\fi
   \vbox{\kern\derskeldimenone\hbox{\kern\derskeldimensix
         \kern\derskeldimentwo
         \xy
         <-\derskeldimenthree,\derskeldimenthree>="here"
            *{\box\derskelboxone}**\dir{-};
         "here"+<\derskeldimentwo,0pt>="here"**\dir{-};
         "here"+<1em,0pt>="here"**\dir{-};
         "here"+<\derskeldimeneight,0pt>="here"
            *{\box\derskelboxfour}**\dir{-};
         "here"+<\derskeldimeneight,0pt>="here"**\dir{-};
         "here"+<1em,0pt>="here"**\dir{-};
         "here"+<\derskeldimenfour,0pt>*{\box\derskelboxtwo}**\dir{-};
         0*{\box\derskelboxthree}**\dir{-};
         <-\derskeldimenthree,\derskeldimenthree>**\dir{-}
         \endxy
         \kern\derskeldimenfour\kern\derskeldimenseven}%
      \kern\derskeldimenfive}}%

\catcode`@=11
\def\upsmash{\relax 
  \ifmmode\def\next{\mathpalette\mathupsm@sh}\else\let\next\makeupsm@sh
  \fi\next}
\def\makeupsm@sh#1{\setbox\z@\hbox{#1}\finupsm@sh}
\def\mathupsm@sh#1#2{\setbox\z@\hbox{$\m@th#1{#2}$}\finupsm@sh}
\def\finupsm@sh{\ht\z@\z@ \box\z@}
\def\downsmash{\relax 
  \ifmmode\def\next{\mathpalette\mathdownsm@sh}\else\let\next\makedownsm@sh
  \fi\next}
\def\makedownsm@sh#1{\setbox\z@\hbox{#1}\findownsm@sh}
\def\mathdownsm@sh#1#2{\setbox\z@\hbox{$\m@th#1{#2}$}\findownsm@sh}
\def\findownsm@sh{\dp\z@\z@ \box\z@}
\catcode`@=12

\newbox\DerivOneBox
\newbox\DerivTwoBox
\newbox\DerivThreeBox
\newbox\DerivFourBox
\newdimen\DerivOneDimen
\newdimen\DerivTwoDimen
\newdimen\DerivThreeDimen
\newdimen\DerivFourDimen
\def\Derivationleaf #1#2#3#4#5{\global\setbox\derboxone=\hbox{\strut
                                    $\DerivationFactors{#1}{#2}{#3}{#4}{#5}11$}}%
\def\DerivationFactors #1#2#3#4#5#6#7{%
   \setbox\DerivOneBox=\hbox{$#1\strut$}%
      \DerivOneDimen=\wd\DerivOneBox\divide\DerivOneDimen by2
   \setbox\DerivThreeBox=\hbox{$#3\strut$}%
      \DerivThreeDimen=\wd\DerivThreeBox\divide\DerivThreeDimen by2
   \setbox\DerivTwoBox=\hbox{\box\DerivOneBox\hbox{$#2$}\box\DerivThreeBox}%
      \DerivTwoDimen=\wd\DerivTwoBox
   \setbox\DerivFourBox=\hbox{$#4\strut$}%
      \DerivFourDimen=\wd\DerivFourBox
   \ifdim\DerivFourDimen>\DerivTwoDimen
      \global\dercdim=\DerivFourDimen                
      \global\derldim=0pt                            
      \global\derrdim=0pt                            
      \advance\DerivFourDimen by-\DerivTwoDimen
      \divide \DerivFourDimen by2
      \advance\DerivTwoDimen  by-\DerivOneDimen
      \advance\DerivTwoDimen  by-\DerivThreeDimen
      \divide \DerivTwoDimen  by 2
   \else
      \global\dercdim=\DerivFourDimen                
      \DerivFourDimen=0pt
      \advance\DerivTwoDimen  by-\DerivOneDimen
      \advance\DerivTwoDimen  by-\DerivThreeDimen
      \global\derldim=\DerivTwoDimen
         \global\advance\derldim by-\dercdim
         \global\divide\derldim by2
         \global\advance\derldim by\DerivOneDimen    
      \global\derrdim=\DerivTwoDimen
         \global\advance\derrdim by-\dercdim
         \global\divide\derrdim by2
         \global\advance\derrdim by\DerivThreeDimen  
      \divide \DerivTwoDimen  by 2
   \fi
   \vbox{\offinterlineskip\hbox{\kern\DerivFourDimen\box\DerivTwoBox}%
         \hbox{\kern\DerivFourDimen\kern\DerivOneDimen
               \kern\DerivTwoDimen\kern-#6\DerivTwoDimen\hbox{$\xy
               0;<#6\DerivTwoDimen,0pt>:<0pt,#7\DerivTwoDimen>::
               (0,1);(2,1)**\crv{(1.25,1.1875)&(0.75,0.8125)};
               (1,0)**@{-};(0,1)**@{-};
               (1,0.625)*{\scriptstyle #5}
               \endxy$}}%
         \hbox{\kern\DerivFourDimen\kern\DerivOneDimen\kern\DerivTwoDimen
               \hbox to0pt{\hss\box\DerivFourBox\hss}%
               \kern\DerivFourDimen\kern\DerivOneDimen\kern\DerivTwoDimen}}}%

\newcommand{\vcinf}[3]
{\mathchoice
   {\vcenter{\inf{#1}{#2}{#3}}}
   {\hskip.9ex\vcenter{\inf{#1}{#2}{#3}}\hskip.9ex}
   {}{}
}

\newcommand{\vcdernote}[4]
{\vcenter{\dernote{#1}{#2}{#3}{#4}}}

\newcommand{\vcstrder}[4]
{\vcenter{\strder{#1}{#2}{#3}{#4}}}

\newcommand{\vcstrpr}[3]
{\vcenter{\strpr{#1}{#2}{#3}}}

\newcommand*{\simplederi}[4]{
  \vcstrder{#2}{#3}{#4}{\leaf{#1}}
  }

\def\sqn  #1{{\vdash #1}}%

\def\quadfs {\rlap{\rm\quad.}}%
\def\quadcm {\rlap{\rm\quad,}}%
\def\qquand {\qquad\mbox{and}\qquad}%
\def\nqquand {\qquad\mbox{\normalsize and}\qquad}%

\def\quadto{\quad\to\quad}%
\def\quadlto{\quad\leadsto\quad}%

\def\qquadto{\qquad\to\qquad}%
\def\qquadlto{\qquad\leadsto\qquad}%
\definecolor{rdxbackcolor}{gray}{0.88}
\def\rdx#1{\smash{\colorbox{rdxbackcolor}{\strut\smash{$#1$}}}}%
\def\clap#1{\hbox to 0pt{\hss#1\hss}}
\def\qlap#1{\hbox to 1em{\hss#1\hss}}
\def\qqlap#1{\hbox to 2em{\hss#1\hss}}
\def\qqqlap#1{\hbox to 3em{\hss#1\hss}}
\def\qqqqlap#1{\hbox to 4em{\hss#1\hss}}
\def\qqqqqlap#1{\hbox to 5em{\hss#1\hss}}
\def\qqqqqqlap#1{\hbox to 6em{\hss#1\hss}}
\def\qqqqqqqlap#1{\hbox to 7em{\hss#1\hss}}
\def\qqqqqqqqlap#1{\hbox to 8em{\hss#1\hss}}
\def\qqqqqqqqqlap#1{\hbox to 9em{\hss#1\hss}}

\def\qlapm#1{\qlap{$#1$}}

\def\rlapm#1{\hbox to 0pt{$#1$\hss}}
\def\llapm#1{\hbox to 0pt{\hss$#1$}}

\def\qqquad{\quad\qquad}
\def\qqqquad{\qquad\qquad}
\def\qqqqquad{\qqquad\qquad}
\def\qqqqqquad{\qqquad\qqquad}

\def\proofadjust{\vadjust{\nobreak\vskip-2.7ex\nobreak}}

\def\grammareq {\mathrel{\raise.4pt\hbox{::}{=}}}%

\let\turnstile=\vdash
\def\lone{1}
\def\lbot{\bot}
\def\ltens{\varotimes}

\def\lpar{\mathop\bindnasrepma}
\def\lneg{^\bot}

\def\limp{\multimap}

\def\loc{\mathord{!}}
\def\lwn{\mathord{?}}
\def\lbef{\mathop{\triangleleft}}

\def\Sl #1{\underline{#1\phantom{\hbox to0pt{,\hss}}}{}\lower1ex\hbox
  {$\scriptscriptstyle\Lsss$}}%
\def\ls #1{\underline{#1\phantom{\hbox to0pt{,\hss}}}{}\lower1ex\hbox
  {$\scriptscriptstyle\Ssss$}}%

\def\set#1{\{#1\}}
\def\mset#1{\mathopen{\mathord\{\mkern-5mu|}#1\mathclose{|\mkern-5mu\mathord\}}}

\def\tuple#1{\langle#1\rangle}
\def\Nat{\mathbb{N}}

\def\conldel {\{}%
\def\conrdel {\}}%
\def\lrgldel {\mathchoice{(}{(}{\langle}{\langle}}%
\def\lrgrdel {\mathchoice{)}{)}{\rangle}{\rangle}}%
\def\aprldel {\mathchoice
    {\mathopen {\setbox0=\hbox{$\displaystyle     \lrgldel$}\hbox to\wd0
                         {\hfil$\displaystyle     (       $\hfil}}}%
    {\mathopen {\setbox0=\hbox{$\textstyle        \lrgldel$}\hbox to\wd0
                         {\hfil$\textstyle        (        $\hfil}}}%
    {\mathopen {\setbox0=\hbox{$\scriptstyle      \lrgldel$}\hbox to\wd0
                         {\hfil$\scriptstyle      (        $\hfil}}}%
    {\mathopen {\setbox0=\hbox{$\scriptscriptstyle\lrgldel$}\hbox to\wd0
                         {\hfil$\scriptscriptstyle(        $\hfil}}}}%
\def\aprrdel {\mathchoice
    {\mathclose{\setbox0=\hbox{$\displaystyle     \lrgrdel$}\hbox to\wd0
                         {\hfil$\displaystyle     )       $\hfil}}}%
    {\mathclose{\setbox0=\hbox{$\textstyle        \lrgrdel$}\hbox to\wd0
                         {\hfil$\textstyle        )        $\hfil}}}%
    {\mathclose{\setbox0=\hbox{$\scriptstyle      \lrgrdel$}\hbox to\wd0
                         {\hfil$\scriptstyle      )        $\hfil}}}%
    {\mathclose{\setbox0=\hbox{$\scriptscriptstyle\lrgrdel$}\hbox to\wd0
                         {\hfil$\scriptscriptstyle)        $\hfil}}}}%
\def\seqldel {\mathchoice
    {\mathopen {\setbox0=\hbox{$\displaystyle     \lrgldel$}\hbox to\wd0
                         {\hfil$\displaystyle     \langle  $\hfil}}}%
    {\mathopen {\setbox0=\hbox{$\textstyle        \lrgldel$}\hbox to\wd0
                         {\hfil$\textstyle        \langle  $\hfil}}}%
    {\mathopen {\setbox0=\hbox{$\scriptstyle      \lrgldel$}\hbox to\wd0
                         {\hfil$\scriptstyle      \langle  $\hfil}}}%
    {\mathopen {\setbox0=\hbox{$\scriptscriptstyle\lrgldel$}\hbox to\wd0
                         {\hfil$\scriptscriptstyle\langle  $\hfil}}}}%
\def\seqrdel {\mathchoice
    {\mathclose{\setbox0=\hbox{$\displaystyle     \lrgrdel$}\hbox to\wd0
                         {\hfil$\displaystyle     \rangle  $\hfil}}}%
    {\mathclose{\setbox0=\hbox{$\textstyle        \lrgrdel$}\hbox to\wd0
                         {\hfil$\textstyle        \rangle  $\hfil}}}%
    {\mathclose{\setbox0=\hbox{$\scriptstyle      \lrgrdel$}\hbox to\wd0
                         {\hfil$\scriptstyle      \rangle  $\hfil}}}%
    {\mathclose{\setbox0=\hbox{$\scriptscriptstyle\lrgrdel$}\hbox to\wd0
                         {\hfil$\scriptscriptstyle\rangle  $\hfil}}}}%
\def\parldel {\mathchoice
    {\mathopen {\setbox0=\hbox{$\displaystyle     \lrgldel$}\hbox to\wd0
                         {\hfil$\displaystyle     [       $\hfil}}}%
    {\mathopen {\setbox0=\hbox{$\textstyle        \lrgldel$}\hbox to\wd0
                         {\hfil$\textstyle        [        $\hfil}}}%
    {\mathopen {\setbox0=\hbox{$\scriptstyle      \lrgldel$}\hbox to\wd0
                         {\hfil$\scriptstyle      [        $\hfil}}}%
    {\mathopen {\setbox0=\hbox{$\scriptscriptstyle\lrgldel$}\hbox to\wd0
                         {\hfil$\scriptscriptstyle[        $\hfil}}}}%
\def\parrdel {\mathchoice
    {\mathclose{\setbox0=\hbox{$\displaystyle     \lrgrdel$}\hbox to\wd0
                         {\hfil$\displaystyle     ]       $\hfil}}}%
    {\mathclose{\setbox0=\hbox{$\textstyle        \lrgrdel$}\hbox to\wd0
                         {\hfil$\textstyle        ]        $\hfil}}}%
    {\mathclose{\setbox0=\hbox{$\scriptstyle      \lrgrdel$}\hbox to\wd0
                         {\hfil$\scriptstyle      ]        $\hfil}}}%
    {\mathclose{\setbox0=\hbox{$\scriptscriptstyle\lrgrdel$}\hbox to\wd0
                         {\hfil$\scriptscriptstyle]        $\hfil}}}}%

\def\bc{,}

\def\aprs #1{{\def\bc{\vlten}\aprldel #1\aprrdel}}%
\def\pars #1{{\def\bc{\vlpar}\parldel #1\parrdel}}%
\def\seqs #1{{\def\bc{\vlseq}\seqldel #1\seqrdel}}%
\def\cons #1{\conldel #1\conrdel}%
\def\copt #1{#1}%

\def\set#1{\{#1\}}

\definecolor{rdxbackcolor}{gray}{0.75}
\def\Black#1{#1}%
\def\cnt#1{#1}%
\def\rdc#1{#1}%
\def\nordx#1{#1}%
\def\yrdx#1{\smash{\colorbox{rdxbackcolor}{\strut\smash{$#1$}}}}
\def\nrdx#1{{\hskip.5pt\strut#1\hskip.5pt}}
\let\rdx\nrdx%

\def\conhole      {\cons{\enspace}}%

\def\un{\mathord{\circ}}

\let\aprun\un

\let\neg=\bar

\def\MLL{\mathsf{MLL}}

\def\MELL{\mathsf{MELL}}

\def\LK{\mathsf{LK}}
\def\NEL{\mathsf{NEL}}
\def\SNEL{\mathsf{SNEL}}

\def\SELS{\mathsf{SELS}}
\def\BV{\mathsf{BV}}
\def\SBV{\mathsf{SBV}}

\def\SNELsc{\mathsf{SNELc}}

\def\SNELhc{\mathsf{SNELh}}

\def\sysS {\cS}%

\def\sqn  #1{{\turnstile #1}}%
\def\mix {\mathsf{mix}}%
\def\mixzero {\mathsf{mix0}}%
\def\cut{\mathsf{cut}}

\def\intrd{\mathsf{i}\mathord{\downarrow}}
\def\intru{\mathsf{i}\mathord{\uparrow}}
\def\atird{\mathsf{ai}\mathord{\downarrow}}
\def\atiru{\mathsf{ai}\mathord{\uparrow}}
\def\swir{\mathsf{s}}
\def\promrd{\mathsf{p}\mathord{\downarrow}}
\def\promru{\mathsf{p}\mathord{\uparrow}}
\def\seqrd{\mathsf{q}\mathord{\downarrow}}
\def\seqru{\mathsf{q}\mathord{\uparrow}}
\def\weakrd{\mathsf{w}\mathord{\downarrow}}
\def\weakru{\mathsf{w}\mathord{\uparrow}}
\def\absrd{\mathsf{b}\mathord{\downarrow}}
\def\absru{\mathsf{b}\mathord{\uparrow}}
\def\erd{\mathsf{e}\mathord{\downarrow}}
\def\eru{\mathsf{e}\mathord{\uparrow}}
\def\dmrd{\mathsf{g}\mathord{\downarrow}}
\def\dmru{\mathsf{g}\mathord{\uparrow}}
\def\unr{\un\mathord{\downarrow}}

\def\fg#1{G_{\loc\lwn}(#1)}
\def\bqfg{!-?-flow-graph} 
\def\bqfgs{!-?-flow-graphs} 
\def\afg{atomic flow-graph} 

\def\pee{$\mathsf{p}$}
\def\bee{$\mathsf{b}$}
\newcommand{\pe}[2][]{\mathsf{p}_{#1}(#2)}
\newcommand{\be}[2][]{\mathsf{b}_{#1}(#2)}
\newcommand{\flip}[2][]{\mathsf{fl}_{#1}(#2)}
\newcommand{\onion}[2][]{\mathord{\varocircle}_{#1}(#2)}
\newcommand{\onionbe}[2][]{\mathsf{b}\mathord{\varocircle}_{#1}(#2)}
\newcommand{\status}[2][]{\mathsf{st}_{#1}(#2)}
\newcommand{\fork}[2][]{\mathsf{fk}_{#1}(#2)}

\newcommand{\rank}[2][]{\mathsf{rk}_{#1}(#2)}
\newcommand{\rankd}[2][]{\mathsf{rk}^{\mathord{\downarrow}}_{#1}(#2)}
\newcommand{\ranku}[2][]{\mathsf{rk}^{\mathord{\uparrow}}_{#1}(#2)}
\newcommand{\erank}[1]{\mathsf{rk}_{\erd}(#1)}

\def\Deltat{\tilde{\Delta}}

\def\cS{{\mathcal S}}

\def\bloc{\loc^\bullet}
\def\blwn{\lwn^\bullet}
\let\bdmark\blacktriangledown
\def\ubloc{\loc^\blacktriangle}
\def\dblwn{\lwn^\bdmark}

\newbox\smilebox
\newbox\fullsmilebox
\newcommand{\dreierpack}[3]{%
  \setbox\smilebox\hbox{$\mathord{#2}$}%
  \copy\smilebox
  \kern-\wd\smilebox%
  \raise1.4ex\hbox to\wd\smilebox{\hss$\scriptscriptstyle{#1\,}$\hss}%
  \kern-\wd\smilebox%
  \lower.5ex\hbox to\wd\smilebox{\hss$\scriptscriptstyle{#3\,}$\hss}%
}%

\newcommand{\edgesym}[1][]{\dreierpack{}{\curvearrowright}{\Delta}}
\newcommand{\pathsym}[1][]{\dreierpack{+}{\curvearrowright}{\Delta}}
\newcommand{\spathsym}[1][]{\dreierpack{*}{\curvearrowright}{\Delta}}
\newcommand{\edge}[3][]{#2\mathbin{\edgesym[#1]}#3}
\renewcommand{\path}[3][]{#2\mathbin{\pathsym[#1]}#3}

\def\normalvecheight{.6}
\def\loopvecheight{.7}
\def\normalarrpos{.48}
\def\normaluparrpos{.7}
\def\normaldownarrpos{.53}
\def\loopangleA{60}
\def\loopangleB{120}
\def\vertangleA{55}
\def\vertangleB{125}

\newcommand{\lineanglesheight}[5]{%
  \nccurve[ncurv=#5,angleA=#3,angleB=#4]{#1}{#2}}

\newcommand{\uline}[2]{%
  \lineanglesheight{#1}{#2}{90}{-90}{\normalvecheight}}

\newcommand{\vecanglesposheight}[6]{%
  \nccurve[ncurv=#6,ArrowInside=->,ArrowInsidePos=#5,angleA=#3,angleB=#4]
  {#1}{#2}}
\newcommand{\duvecanglesheight}[5]{%
  \nccurve[ncurv=#5,angleA=#3,angleB=#4]{#1}{#2}
  \lput{:0}{\psline{->}(.05,.005)(.1,.005)}
  }
\newcommand{\udvecanglesheight}[5]{%
  \nccurve[ncurv=#5,angleA=#3,angleB=#4]{#1}{#2}
  \lput{:0}{\psline{->}(.05,-.005)(.1,-.005)}
  }

\newcommand{\vecanglespos}[5]{%
  \vecanglesposheight{#1}{#2}{#3}{#4}{#5}{\normalvecheight}}

\newcommand{\vecangles}[4]{%
  \vecanglesposheight{#1}{#2}{#3}{#4}{\normalarrpos}{\normalvecheight}}

\newcommand{\uvec}[2]{%
  \vecanglesposheight{#1}{#2}{90}{-90}{\normaluparrpos}{\normalvecheight}}

\newcommand{\urvec}[2]{%
  \vecanglesposheight{#1}{#2}{\vertangleA}{-\vertangleB}
		     {\normaluparrpos}{\normalvecheight}}

\newcommand{\ulvec}[2]{%
  \vecanglesposheight{#1}{#2}{\vertangleB}{-\vertangleA}
		     {\normaluparrpos}{\normalvecheight}}

\newcommand{\dvec}[2]{%
  \vecanglesposheight{#1}{#2}{-90}{90}{\normaldownarrpos}{\normalvecheight}}

\newcommand{\dlvec}[2]{%
  \vecanglesposheight{#1}{#2}{-\vertangleB}{\vertangleA}
		     {\normalarrpos}{\normalvecheight}}

\newcommand{\druvec}[2]{%
  \duvecanglesheight{#1}{#2}{-\loopangleA}{-\loopangleB}{\loopvecheight}}
\newcommand{\dluvec}[2]{%
  \duvecanglesheight{#1}{#2}{-\loopangleB}{-\loopangleA}{\loopvecheight}}

\newcommand{\urdvec}[2]{%
  \udvecanglesheight{#1}{#2}{\loopangleA}{\loopangleB}{\loopvecheight}}
\newcommand{\uldvec}[2]{%
  \udvecanglesheight{#1}{#2}{\loopangleB}{\loopangleA}{\loopvecheight}}

\newgray{darkgrayline}{.7}
\newgray{lightgrayline}{.95}
\newrgbcolor{darkredline}{1 .3 .3}
\newrgbcolor{lightredline}{1 .8 .8}

\def\slimlinewidth{.08ex}
\def\thicklinewidth{.22ex}
\def\boldlinewidth{.5ex}

\psset{%
  arrowscale=1.4,
  arrowlength=1.2,
  nodesep=0ex,
  linewidth=\slimlinewidth,
  }
 
\newbox\nodebox
\newdimen\nodewd

\newcommand{\pathnode}[2]{%
  \setbox\nodebox=\hbox{$#2$}%
  \nodewd=\wd\nodebox%
  \raise.8ex\hbox to\nodewd{\hss\rnode{#1}{ }\hss}%
}

\newcommand{\textnode}[2]{#2}

\def\pathtype{p}
\def\pathderimode{p}

\newcommand{\singlenode}[3]{%
  \if\pathderimode\pathtype
    \pathnode{#2#3}{#1} 
  \else 
    \textnode{#2#3}{#1} 
  \fi
}

\newcommand{\noc}[1]{\singlenode{\loc}{oc}{#1}}
\newcommand{\nwn}[1]{\singlenode{\lwn}{wn}{#1}}
\newcommand{\nP}[1]{\singlenode{\hbox{$P$}}{P}{#1}}
\newcommand{\nR}[1]{\singlenode{R}{R}{#1}}
\newcommand{\nZ}[1]{\singlenode{Z}{Z}{#1}}
\newcommand{\nW}[1]{\singlenode{W}{W}{#1}}
\newcommand{\nU}[1]{\singlenode{U}{U}{#1}}
\newcommand{\nT}[1]{\singlenode{\hbox{$T$}}{T}{#1}}
\newcommand{\nS}[1]{\singlenode{S}{S}{#1}}
\newcommand{\nSp}[1]{\singlenode{S}{S}{#1}'}
\newcommand{\nph}[1]{\singlenode{\phantom{W}}{ph}{#1}}

\newcommand{\pathderivation}[2]{
  \hbox{\begin{psmatrix}
      \let\rdx\yrdx
      \def\pathderimode{p}%
      \rlap{#1}%
      #2%
      \let\rdx\nrdx
      \def\pathderimode{t}%
      #1%
    \end{psmatrix}%
  }
}

\newcommand{\vctpathderivation}[2]{\vcenter{\hbox{\hskip-1em\pathderivation{#1}{#2}}}}

\newcount\linecount
\newcount\linecountb

\newcommand{\longuvec}[3]{%
  \linecount=\number#3%
  \loop\ifnum\linecount>\number#2%
    \linecountb=\linecount%
    \advance\linecount by -1%
    \uvec{#1\number\linecount}{#1\number\linecountb}%
  \repeat%
}
\newcommand{\longurvec}[3]{%
  \linecount=\number#3%
  \loop\ifnum\linecount>\number#2%
    \linecountb=\linecount%
    \advance\linecount by -1%
    \urvec{#1\number\linecount}{#1\number\linecountb}%
  \repeat%
}

\newcommand{\longdvec}[3]{%
  \linecount=\number#3%
  \loop\ifnum\linecount<\number#2%
    \linecountb=\linecount%
    \advance\linecount by 1%
    \dvec{#1\number\linecount}{#1\number\linecountb}%
  \repeat%
}

\newcommand{\longdlvec}[3]{%
  \linecount=\number#3%
  \loop\ifnum\linecount<\number#2%
    \linecountb=\linecount%
    \advance\linecount by 1%
    \dlvec{#1\number\linecount}{#1\number\linecountb}%
  \repeat%
}

\newcommand{\longline}[3]{%
  \linecount=\number#3%
  \loop\ifnum\linecount>\number#2%
    \linecountb=\linecount%
    \advance\linecount by -1%
    \uline{#1\number\linecount}{#1\number\linecountb}%
  \repeat%
}

\def\slimvec{\psset{linecolor=darkredline, 
linewidth=\slimlinewidth}}
\def\thickvec{\psset{linecolor=darkredline, 
linewidth=\thicklinewidth}}
\def\boldvec{\psset{linecolor=lightredline,  
linewidth=\boldlinewidth}}

\def\figurebox{}
\def\smalltitle{}

\def\mysmall{\fontsize{8pt}{10pt}\selectfont}
\def\myssmall{\fontsize{9pt}{11pt}\selectfont}
\def\mytiny{\fontsize{7pt}{9pt}\selectfont}

\def\mysmall{\fontsize{9pt}{11pt}\selectfont}
\def\myvsmall{\fontsize{8pt}{10pt}\selectfont}

\let\dfn\emph

\title{A System of Interaction and Structure IV:\\ 
The Exponentials and Decomposition}

\author{Lutz Stra\ss burger\\ 
  INRIA Saclay--\^Ile-de-France and \'Ecole Polytechnique, France
  \and
  Alessio Guglielmi\\
    University of Bath, UK and INRIA Nancy--Grand Est, France
}

\begin{abstract}
We study a system, called NEL, which is the mixed
commutative/non-commutative linear logic BV augmented with linear
logic's exponentials. Equivalently, NEL is MELL augmented with the
non-commutative self-dual connective seq. In this paper, we show a
basic compositionality property of NEL, which we call
\emph{decomposition}. This result leads to a cut-elimination theorem,
which is proved in the next paper of this series. To control the
induction measure for the theorem, we rely on a novel technique that
extracts from NEL proofs the structure of exponentials, into what we
call !-?-Flow-Graphs.
\end{abstract}

\category{F.4.1}{Mathematical Logic and Formal Languages}%
         {Mathematical Logic}[proof theory]

\terms{Deep Inference, Calculus of Structures, Linear Logic, Noncommutativity}


\begin{document}



\maketitle

\sloppy



\section{Introduction}\label{sec:intro}

This is the fourth in a series of papers dedicated to the proof 
theory of a self-dual non-commutative operator, called \emph{seq}, in 
the context of linear logic.

The first paper \emph{``A System of Interaction and Structure''}
\cite{gug:SIS} introduced seq in the context of multiplicative linear
logic. The resulting logic is called $\BV$. The proof system for $\BV$
is presented in the formalism called the \emph{calculus of
structures}, which is the simplest formalism in the methodology of
\emph{deep inference}. In fact, deep inference was born precisely for
giving $\BV$ a normalization theory.

In the second paper \emph{``A System of Interaction and Structure~II:
  The Need for Deep Inference''} \cite{tiu:SIS-II}, Alwen Tiu shows
that deep inference is necessary to obtain analyticity for $\BV$. In
other words, traditional Gentzen proof theory is not sufficient to
deal with seq. 

The third paper, currently being elaborated, explores the connection 
between $\BV$ and pomset logic \cite{retore:97}. 

This fourth paper, and the fifth paper \emph{``A System of Interaction
  and Structure~V: The Exponentials and
  Splitting''}~\cite{SIS-splitting} 
are devoted to the proof theory of system $\BV$ when it
is enriched with linear logic's exponentials. We call $\NEL$
(non-commutative exponential linear logic) the resulting system. We
can also consider $\NEL$ as $\MELL$ (multiplicative exponential linear
logic \cite{girard:87}) plus seq. $\NEL$, which was first presented in
\cite{gug:str:02}, is conservative over $\BV$ and over $\MELL$
augmented by the mix and nullary mix rules
\cite{fleury:retore:94,retore:phd,abramsky:jagadeesan:94}.  Note that,
like $\BV$, $\NEL$ cannot be analytically expressed outside deep
inference. System $\NEL$ can be immediately understood by anybody
acquainted with the sequent calculus, and is aimed at the same range
of applications as $\MELL$, but it offers, of course, explicit
sequential composition.

$\NEL$ is especially interesting because it is 
Turing-complete \cite{str:WoLLIC03}. The complexity of $\MELL$ is currently 
unknown, but $\MELL$ is widely conjectured to be decidable. If that 
was the case, then the line towards Turing-completeness would clearly 
be crossed by seq, which, in fact, has been interpreted already as an 
effective mechanism to structure a Turing machine tape. This is 
something that $\MELL$, which is fully commutative, apparently cannot 
do.

This paper is devoted to the \emph{decomposition theorem}. Together
with the \emph{splitting theorem} in~\cite{SIS-splitting} it
immediately yields cut-elimination, which will be claimed
in~\cite{SIS-splitting}.

Decomposition (which was first pioneered in \cite{gug:str:01,str:MELL}
for $\BV$ and $\MELL$) is as follows: we can transform every $\NEL$
derivation into an equivalent one, composed of eleven derivations
carried into eleven disjoint subsystems of $\NEL$. This means that we
can study small subsystems of $\NEL$ in isolation and then compose
them together with considerable more freedom than in the sequent
calculus, where, for example, contraction can not be isolated in a
derivation.  Decomposition is made available in the calculus of
structures by exploiting the top-down symmetry of derivations that is
typical of deep inference. Such a result is unthinkable in formalisms
lacking locality, like Gentzen systems.

The technique by which we prove the result is an evolution and
simplification of a technique that was first developed
in~\cite{str:MELL} for $\MELL$, but that would not work unmodified in
the presence of seq. In fact, seq makes matters more complicated, due
to similar phenomena to those unveiled by Tiu~\cite{tiu:SIS-II}, and
that make seq intractable for Gentzen methods.

Some of the main results of this paper have already been presented,
without proof, in \cite{gug:str:02}.

\section{The System}\label{sec:system}

We define the language for system $\NEL$ and its variants, as an
extension of the language for $\BV$, defined in~\cite{gug:SIS}.
Intuitively, $\pars{S_1\bc \vldots\bc S_h}$ corresponds to a sequent
$\sqn{S_1,\dots,S_h}$ in linear logic, whose formulae are essentially
connected by pars, subject to commutativity (and associativity).  The
structure $\aprs{S_1\bc \vldots\bc S_h}$ corresponds to the
associative and commutative tensor connection of $S_1$, \dots, $S_h$.
The structure $\seqs{S_1\bc \vldots\bc S_h}$ is associative and {\it
  non-commutative\/}: this corresponds to the new logical connective,
called {\it seq\/}, that we add to those of $\MELL$.\footnote{Please
  note that we slightly change the syntax with respect
  to~\cite{gug:SIS,tiu:SIS-II}: In these papers commas where used in
  the places of the connectives $\lpar$, $\ltens$, and
  $\lbef$. Although there is some redundancy in having the connectives
  and the three different types of brackets, we think, it is easier
  to parse for the reader.}

\begin{definition}\label{def:structures} 
  There are countably many \dfn{positive} and \dfn{negative atoms}.
  They, positive or negative, are denoted by $a$, $b$, \dots.
  \dfn{Structures} are denoted by $S$, $P$, $Q$, $R$, $T$, $U$, $V$,
  $W$, $X$ and $Z$.  The structures of the \dfn{language $\NEL$} are
  generated by
$$
S \grammareq 
             a                                        \mid 
             \un                                      \mid 
             \pars{\,\underbrace{S\bc \vldots\bc S}_{{}>0}\,} \mid 
             \aprs{\,\underbrace{S\bc \vldots\bc S}_{{}>0}\,} \mid 
             \seqs{\,\underbrace{S\bc \vldots\bc S}_{{}>0}\,} \mid 
             \lwn S                                   \mid
             \loc S                                   \mid
             \neg S                                       \quadcm
$$ 
where $\un$, the \dfn{unit}, is not an atom
and ${\bar S}$ is the
\dfn{negation} of the structure $S$.  Structures with a hole that does not
appear in the scope of a negation are denoted by ${S\cons{\enspace}}$.  The
structure $R$ is a \dfn{substructure} of $S\cons R$, and $S\cons\enspace$ is
its \dfn{context}.  We simplify the indication of context in cases where
structural parentheses fill the hole exactly: for example, $S\pars{R\bc T}$
stands for $S\cons{\pars{R\bc T}}$.
\end{definition}

Structures come with equational theories establishing some basic,
decidable algebraic laws by which structures are indistinguishable.
These are analogous to the laws of associativity, commutativity,
idempotency, and so on, usually imposed on sequents.  The difference
is that we merge the notions of formula and sequent, and we extend the
equations to formulae.  The structures of the language $\NEL$ are
equivalent modulo the relation $=$, defined in
Figure~\ref{fig:nel_SC}. There, $\vec R$, $\vec T$ and $\vec U$ stand
for finite, non-empty sequences of structures (elements of the
sequences are separated by $\vlpar$, $\vlseq$, or $\vlten$, as
appropriate in the context).\footnote{For complexity issues related to
  the use of equations see~\cite{bru:gug:PC-DI}.}

\begin{figure}[t]
  \begin{center}
    \figurebox{ 
      \parbox{17em}{
        \smalltitle{Associativity}%
        \begin{eqnarray*}
          \pars{\vec R\bc \pars{\vec T}\bc \vec U} &=& \pars{\vec R\bc \vec T\bc \vec U}\\
          \aprs{\vec R\bc \aprs{\vec T}\bc \vec U} &=& \aprs{\vec R\bc \vec T\bc \vec U}\\
          \seqs{\vec R\bc \seqs{\vec T}\bc \vec U} &=& \seqs{\vec R\bc \vec T\bc \vec U} 
        \end{eqnarray*}
        \smalltitle{Commutativity}%
        \begin{eqnarray*}
          \pars{\vec R\bc \vec T} &=& \pars{\vec T\bc \vec R}\\
          \aprs{\vec R\bc \vec T} &=& \aprs{\vec T\bc \vec R}
        \end{eqnarray*}
        \smalltitle{Unit}%
        \begin{eqnarray*}
          \pars{\un\bc \vec R} &=& \pars{\vec R}\\
          \aprs{\un\bc \vec R} &=& \aprs{\vec R}\\
          \seqs{\un\bc \vec R} &=& \seqs{\vec R}\\
          \seqs{\vec R\bc \un} &=& \seqs{\vec R}
        \end{eqnarray*}
        } \hfill
      \parbox{17em}{
        \smalltitle{Singleton}%
        $$\pars{R} = \aprs{R} = \seqs{R} = R $$
        \smalltitle{Negation}%
        \begin{eqnarray*}
          \overline{\un}                        &=& \un             \\
          \overline{\strut\pars{R_1\bc \vldots\bc R_h}} &=& 
          \aprs{\neg R_1\bc \vldots\bc \neg R_h}\\
          \overline{\strut\aprs{R_1\bc \vldots\bc R_h}} &=& 
          \pars{\neg R_1\bc \vldots\bc \neg R_h}\\
          \overline{\strut\seqs{R_1\bc \vldots\bc R_h}} &=& 
          \seqs{\neg R_1\bc \vldots\bc \neg R_h}\\
          \overline{\lwn R}                     &=& \loc\neg R          \\
          \overline{\loc R}                     &=& \lwn\neg R          \\
          \neg{\neg R}                          &=& R
        \end{eqnarray*}
        \smalltitle{Contextual Closure}%

        \hspace{1ex}

        \hfil if $R=T$ then $S\cons{R}=S\cons{T}$ \hfil
        
        }
      }
    \caption{Basic equations for the syntactic equivalence $=$}
    \label{fig:nel_SC}
  \end{center}
\end{figure}

\begin{definition}\label{def:rule} 
    An \dfn{{\rm(}inference\/{\rm)} rule} is any scheme
    $$
    \myssmall{\vcenter{\infnote{\rho}{R}{T}{}}}
    \quadcm$$ 
    where $\rho$ is the
    \dfn{name} of the rule, $T$ is its \dfn{premise} and $R$ is its
    \dfn{conclusion}; $R$ or $T$, but not both, may be missing. A
    \dfn{{\rm(}proof\/{\rm)} system}, denoted by $\sysS$, is a set of
    rules.  A \dfn{derivation} in a system $\sysS$ is a finite chain
    of instances of rules of $\sysS$, and is denoted by $\Delta$; a
    derivation can consist of just one structure.  The topmost
    structure in a derivation is called its \dfn{premise}; the
    bottommost structure is called \dfn{conclusion}.  A derivation
    $\Delta$ whose premise is $T$, conclusion is $R$, and whose rules
    are in $\sysS$ is denoted by\proofadjust
    $$\mysmall\simplederi{T}{\sysS}{\Delta}{R}\quadfs$$
\end{definition}

The typical inference rules are of the kind
$$
\myssmall
\vcenter{\infnote{\rho}{S\cons R}{S\cons T}{}}\quadfs$$
This rule scheme
$\rho$ specifies that if a structure matches $R$, in a context
$S\cons\enspace$, it can be rewritten as specified by $T$, in the same context
$S\cons\enspace$ (or vice versa if one reasons top-down).  A rule corresponds
to implementing in the deductive system {\it any axiom\/} $T\Rightarrow R$, where
$\Rightarrow$ stands for the implication we model in the system, in our case
linear implication.  The case where the context is empty corresponds to the
sequent calculus. For example, the linear logic sequent calculus rule
$$
\myssmall
\vcenter{\iinf{\ltens}
{\sqn{A\ltens B,\Phi,\Psi}}{\sqn{A,\Phi}}{\sqn{B,\Psi}}}$$ 
could be simulated easily in
the calculus of structures by the rule
$$
\myssmall
\vcenter{\infnote{\ltens'}{\aprs{\Gamma\bc \pars{\aprs{A\bc B}\bc \Phi\bc \Psi}}}%
{\aprs{\Gamma\bc \pars{A\bc \Phi}\bc \pars{B\bc \Psi}}}{}}\quadcm$$ 
where $\Phi$ and $\Psi$ stand
for multisets of formulae or their corresponding par structures.  The structure
$\Gamma$ stands for the tensor structure of the other hypotheses in the
derivation tree.  More precisely, any sequent calculus derivation
\vadjust{\vskip-2ex}\par{\mysmall$$
\setbox0=\hbox{\mysmall\iinf{\llap{$\ltens$}}%
                    {\sqn{A\ltens B,\Phi,\Psi}}%
                    {\sqn{A,\Phi}}%
                    {\sqn{B,\Psi}}}%
\DerivationFactors{\sqn{\Gamma_1}\enspace\cdots\enspace\sqn{\Gamma_{i-1}}}%
                  {\quad\box0\quad}%
                  {\sqn{\Gamma_{i+1}}\enspace\cdots\enspace\sqn{\Gamma_h}}%
                  {\sqn{\Sigma}}%
                  {\Delta}{1}{.4}%
$$}
containing the $\ltens$ rule can be simulated by
$$\mysmall
\vcstrder{}{\Delta'}{\Sigma'}
{\root{\ltens'}
{\aprs{\Gamma_1'\bc \dots\bc \Gamma_{i-1}'\bc 
       \pars{\aprs{ A'\bc B'}\bc \Phi'\bc \Psi'}\bc 
       \Gamma_{i+1}'\bc \dots\bc \Gamma_h'}}  
{\leaf{\aprs{\Gamma_1'\bc \dots\bc \Gamma_{i-1}'\bc 
             \pars{A'\bc \Phi'}\bc \pars{B'      \bc \Psi'}  \bc 
             \Gamma_{i+1}'\bc \dots\bc \Gamma_h'}}}
}
\quadcm
$$ in the calculus of structures, where $\Gamma_j'$, $A'$, $B'$, $\Phi'$,
$\Psi'$, $\Delta'$ and $\Sigma'$ are obtained from their counterparts in the
sequent calculus by the obvious translation.  This means that by this method
every system in the one-sided sequent calculus can be ported trivially to the
calculus of structures.

Of course, in the calculus of structures, rules could be used as axioms of a
generic Hilbert system, where there is no special, structural relation between
$T$ and $R$: then all the good proof theoretical properties of sequent systems
would be lost.  We will be careful to design rules in a way that is
conservative enough to allow us to prove cut elimination, and such that they
possess the subformula property.

In our systems, rules come in pairs,
$$
{\myssmall\vcenter{\infnote{\rho{\downarrow}}{S\cons R}{S\cons T}{}}}
\mbox{ (down version)}
\qquand
{\myssmall\vcenter{\infnote{\rho{\uparrow}}{S\cons{\bar T}}{S\cons{\bar R}}{}}}
\mbox{ (up version)}
\quadfs
$$ Sometimes rules are self-dual, i.e., the up and down versions are
identical, in which case we omit the arrows.  This duality derives
from the duality between $T\Rightarrow R$ and $\bar R\Rightarrow\bar
T$, where $\Rightarrow$ is the implication and $\bar{(\cdot)}$ the
negation of the logic. In the case of $\NEL$ these are linear
implication and linear negation.  We will be able to get rid of the up
rules without affecting provability---after all, $T\Rightarrow R$ and
$\bar R\Rightarrow\bar T$ are equivalent statements in many logics.
Remarkably, the cut rule reduces into several up rules, and this makes
for a modular decomposition of the cut elimination argument because we
can eliminate up rules one independently from the other.

Let us now define system $\NEL$ by starting from an up-down symmetric
variation, that we call $\SNEL$.  It is made by two sub-systems that we will
call conventionally {\it interaction\/} and {\it structure}.  The interaction
fragment deals with negation, i.e., duality.  It corresponds to identity and
cut in the sequent calculus.  In our calculus these rules become mutually
top-down symmetric and both can be reduced to their atomic counterparts.

The structure fragment corresponds to logical and structural rules in the
sequent calculus; it defines the logical connectives.  Differently from the
sequent calculus, the connectives need not be defined in isolation, rather
complex contexts can be taken into consideration.  In the following system we
consider {\it pairs\/} of logical connectives, one inside the other.

\begin{figure}[t]
  \begin{center}
      \figurebox{
      \begin{tabular}{c@{}c@{\qquad}c@{}c}
          &$\vcinf{\atird}{S\copt{\pars{a\bc \neg a}}}{S\cons{\un}}$&
          $\vcinf{\atiru}{S\cons{\un}}{S\copt{\aprs{a\bc \neg a}}}$&\\ 
          &&&\\
          &\multicolumn{2}{c}{
            $\vcinf{\swir}{S\copt{\pars{\aprs{R\bc T}\bc U}}}
            {S\copt{\aprs{\pars{R\bc U}\bc T}}}$}&\\ 
          &&&\\
          &$\vcinf{\seqrd}{S\pars{\seqs{R\bc T}\bc \seqs{U\bc V}}}
          {S\seqs{\pars{R\bc U}\bc \pars{T\bc V}}}$&
          $\vcinf{\seqru}{S\seqs{\aprs{R\bc T}\bc \aprs{U\bc V}}}
          {S\aprs{\seqs{R\bc U}\bc \seqs{T\bc V}}}$&\\
          &&&\\
          &$\vcinf{\promrd}{S\copt{\pars{\loc R\bc  \lwn T }}}
          {S\cons{\loc\pars{R\bc T }}}$&
          $\vcinf{\promru}{S\cons{\lwn\aprs{R\bc T }}}
          {S\copt{\aprs{\lwn R\bc  \loc T }}}$&\\
          &&&\\
          &$\vcinf{\erd}{S\cons{\loc\un}}{S\cons{\un}}$&
          $\vcinf{\eru}{S\cons{\un}}{S\cons{\lwn\un}}$&\\ 
          &&&\\
          &$\vcinf{\weakrd}{S\cons{\lwn R}}{S\cons{\un}}$&
          $\vcinf{\weakru}{S\cons{\un}}{S\cons{\loc R}}$&\\
          &&&\\
          &$\vcinf{\absrd}{S\cons{\lwn R}}{S\copt{\pars{\lwn R\bc R }}}$&
          $\vcinf{\absru}{S\copt{\aprs{\loc R\bc R }}}{S\cons{\loc R}}$&\\
          &&&\\
          &$\vcinf{\dmrd}{S\cons{\lwn R}}{S\cons{\lwn\lwn R}}$&
          $\vcinf{\dmru}{S\cons{\loc\loc R}}{S\cons{\loc R}}$&\\
       \end{tabular} }  
    \caption{System $\SNEL$}
    \label{fig:SNEL}
  \end{center}
\end{figure}

\begin{definition}\label{def:NEL} 
    In Figure \ref{fig:SNEL}, \dfn{system\/ $\SNEL$} (\dfn{symmetric
      non-commutative exponential linear logic}) is shown.  The rules
    $\atird$, $\atiru$, $\swir$, $\seqrd$, $\seqru$, $\promrd$,
    $\promru$, $\erd$, $\erd$, $\weakrd$, $\weakru$, $\absrd$,
    $\absru$, $\dmrd$, and $\dmru$ are called respectively \dfn{atomic
      interaction}, \dfn{atomic cut}, \dfn{switch}, \dfn{seq},
    \dfn{coseq}, \dfn{promotion}, \dfn{copromotion}, \dfn{empty},
    \dfn{coempty}, \dfn{weakening}, \dfn{coweakening},
    \dfn{absorption}, \dfn{coabsorption}, \dfn{digging}, and
    \dfn{codigging}.  The \dfn{down fragment} of $\SNEL$ is
    $\set{\atird,\swir,\seqrd,\promrd,\erd,\weakrd,\absrd,\dmrd}$, the
    \dfn{up fragment} is
    $\set{\atiru,\swir,\seqru,\promru,\eru,\weakru,\absru,\dmru}$.
\end{definition}

There is a straightforward two-way correspondence between structures not
involving seq and formulae of $\MELL$: for example 
$$
\loc\pars{\aprs{\lwn a\bc b}\bc \neg c\bc \loc\neg d}
\quad\mbox{corresponds to}\quad
\loc((\lwn a\ltens b)\lpar c\lneg \lpar\loc d\lneg)
\quadcm
$$ and vice versa.  Units are mapped into $\un$ (since
$\lone\equiv\bot$, when $\mix$ and $\mixzero$ are added to $\MELL$).
System $\SNEL$ is just the merging of systems $\SBV$ (which is the
symmetric version of $\BV$) and $\SELS$ (which is the symmetric
presentation of $\MELL$ in the calculus of structures) shown in
\cite{gug:SIS,gug:str:01,str:MELL,dissvonlutz}; there one can find
details on the correspondence between our systems and linear
logic.\footnote{Note that there is a change of our system with respect
  to the system $\SELS$ in \cite{str:MELL} and the version of $\SNEL$
  presented in~\cite{gug:str:02}:
  Here we have added the rules $\erd$, $\eru$, $\dmrd$, and $\dmru$,
  whereas previously we used the equations $\lwn\lwn R=\lwn R$ and
  $\loc\loc R=\loc R$, as well as $\loc\un=\un=\lwn\un$ in
  \cite{gug:str:02} and $\loc\lone=\lone$ and $\lwn\lbot=\bot$ in
  \cite{str:MELL} (see also \cite{dissvonlutz}). From the viewpoint of
  provability, there is no difference between the two approaches, but
  certain properties of the system, in particular decomposition, can
  be demonstrated in a cleaner way. Also from the viewpoint of
  denotational semantics, our system is now more easily
  accessible. For example in coherence spaces \cite{girard:87} we do
  not have an isomorphism between $\loc R$ and $\loc\loc R$.}  The
rules $\swir$, $\seqrd$ and $\seqru$ are the same as in pomset logic
viewed as a calculus of cographs \cite{retore:99}.

All equations are typical of a sequent calculus presentation, save
those for units and contextual closure.  Contextual closure just
corresponds to equivalence being a congruence, it is a necessary
ingredient of the calculus of structures.  All other equations can be
removed and replaced by rules (see, e.g., \cite{str:SD05}), as in the
sequent calculus.  This might prove necessary for certain
applications.  For our purposes, this setting makes for a much more
compact presentation, at a more effective abstraction level.

Negation is involutive and can be pushed to atoms; it is convenient always to
imagine it directly over atoms.  Please note that negation does not swap
arguments of seq, as happens in the systems of Yetter~\cite{yetter:90} and
Abrusci-Ruet \cite{abrusci:ruet:00}.  The unit $\un$ is self-dual and common
to par, times and seq.  One may think of it as a convenient way of expressing
the empty sequence.  Rules become very flexible in the presence of the unit.
For example, the following notable derivation is valid:
$$\mysmall
\vcdernote{\seqrd }{}{\nrdx{\pars{a\bc b}}} {
\root   {\seqru}             {\rdc{\seqs{a\bc b}}}{
\leaf                        {\cnt{\aprs{a\bc b}}}}}
\qquad\equiv\qquad
\vcdernote{=}{}{\nrdx{\pars{a\bc b}}} {
  \root{\seqrd }{\nrdx{\pars{\seqs{a\bc \un}\bc \seqs{\un\bc b}}}}{
    \root{=}{\cnt{\seqs{\pars{a\bc \un}\bc \pars{\un\bc b}}}}{
      \root{=}{\rdc{\seqs{a\bc b}}}{
	\root{\seqru}{\nrdx{\seqs{\aprs{a\bc \un}\bc \aprs{\un\bc b}}}}{
	  \root{=}{\cnt{\aprs{\seqs{a\bc \un}\bc \seqs{\un\bc b}}}}{
	    \leaf{\cnt{\aprs{a\bc b}}}
}}}}}}
\quadfs
$$
The right-hand side above is just a complicated way of writing the left-hand
side. Using the ``fake inference rule $=$'' sometimes eases the reading of a
derivation. 

\begin{remark}\label{rem:equation}
  We will also use the following variants of the rules $\swir$,
  $\promrd$, and $\promru$, which allows us to save some space by
  omitting the $=$ rule:
  \begin{equation*}
    \myssmall
    \vcinf{\swir}{
      S\copt{\pars{U\bc \aprs{R\bc T}}}}{
      S\copt{\aprs{\pars{U\bc R}\bc T}}}
    \qqquad
    \vcinf{\swir}{
      S\copt{\pars{\aprs{R\bc T}\bc U}}}{
      S\copt{\aprs{R\bc \pars{T\bc U}}}}
    \qqquad
    \vcinf{\swir}{
      S\copt{\pars{U\bc \aprs{R\bc T}}}}{
      S\copt{\aprs{R\bc \pars{U\bc T}}}}
  \end{equation*}
  \begin{equation*}
    \myssmall
    \vcinf{\promrd}{
      S\copt{\pars{ \lwn T\bc  \loc R}}}{
      S\cons{\loc\pars{T\bc R }}}
    \qqqqquad
    \vcinf{\promru}{
      S\cons{\lwn\aprs{T\bc R }}}{
      S\copt{\aprs{\loc T\bc  \lwn R }}}
  \end{equation*}
\end{remark}

Each inference rule in Figure~\ref{fig:SNEL} corresponds to a linear
implication that is sound in $\MELL$ plus $\mix$ and $\mixzero$.  For example,
promotion corresponds to the implication $\loc(R\lpar T)\limp(\loc R\lpar\lwn
T)$.  Notice that interaction and cut are atomic in $\SNEL$; we can define
their general versions as follows.

\begin{definition}\label{def:interaction} 
    The following rules are called \dfn{interaction} and \dfn{cut}:
$$\myssmall
\vcenter{\inf{\intrd}
    {S\copt{\pars{R\bc \neg R}}}
    {S\cons   {\un}}}
\qquad\hbox{and}\qquad
\vcenter{\inf{\intru}
    {S\cons   {\un}}
    {S\copt{\aprs{R\bc \neg R}}}}
\quadcm
$$
where $R$ and $\neg R$ are called \dfn{principal structures}.
\end{definition}

The sequent calculus cut rule\proofadjust
$$\myssmall
\vcenter{\iinfnote{\cut}
                  {\sqn{       \Phi,\Psi}}
                  {\sqn{A     ,\Phi     }}
                  {\sqn{A\lneg,     \Psi}}{}}
\qquad
\mbox{\normalsize is realized as}
\qquad
\vcdernote{\intru}{}{     \pars{                                              \Phi         \bc \Psi } }  {
\root   {\swir }         {\rdx{\pars{\rdc{\aprs{      A\bc             \neg A}}\bc \Phi\Black{{}\bc \Psi}}}} {
\root   {\swir }         {\rdx{\pars{\rdc{\aprs{\pars{A\bc \Phi}      \bc \neg A}}              \bc \Psi }}}{
\leaf                    {\cnt{           \aprs{\pars{A\bc \Phi}\bc \pars{\neg A                \bc \Psi}}}}}}}
\quadcm
$$ 
where $\Phi$ and $\Psi$ stand for multisets of formulae or their
corresponding par structures.  Notice how the tree shape of
derivations in the sequent calculus is realized by making use of
tensor structures: in the derivation above, the premise corresponds to
the two branches of the cut rule.  For this reason, in the calculus of
structures rules are allowed to access structures deeply nested into
contexts.

The cut rule in the calculus of structures can mimic the classical cut
rule in the sequent calculus in its realization of transitivity, but
it is much more general.  We believe a good way of understanding it is
thinking of the rule as being about lemmas {\it in context}.  The
sequent calculus cut rule generates a lemma which is valid in the most
general context; the new cut rule does the same, but the lemma only
affects the limited portion of structure that can interact with it.

\begin{definition}\label{def:derivable} 
  A rule $\rho$ is \dfn{derivable} in the system $\sysS$ if
  $\rho\notin\sysS$ and for every instance 
  ${\myssmall\vcenter{\infnote{\rho}{R}{T}{}}}$
  there exists a derivation 
  ${\mysmall\simplederi{T}{\sysS}{\Delta}{R}}$.
  The systems $\sysS$ and $\sysS'$ are \dfn{strongly equivalent} if 
  for every derivation
  ${\mysmall\smash{\simplederi{T}{\sysS}{\Delta}{R}}}$
  there exists a derivation
  ${\mysmall\smash{\simplederi{T}{\sysS'}{\Delta'}{R}}}$
  and vice versa.
\end{definition}

We easily get the next two propositions, which say: 1) The interaction
and cut rules can be reduced into their atomic forms---note that in
the sequent calculus it is possible to reduce interaction to atomic
form, but not cut.  2) The cut rule is as powerful as the whole up
fragment of the system, and vice versa.

\begin{proposition}\label{prop:atomic_int}
  The rule\/ $\intrd$ is derivable in\/
  $\set{\atird,\swir,\seqrd,\promrd,\erd}$, and, dually, the rule\/
  $\intru$ is derivable in the system\/
  $\set{\atiru,\swir,\seqru,\promru,\eru}$.
\end{proposition}

\begin{proof}
Induction on principal structures.  We show the inductive cases for $\intru$:
$$\mysmall
\vcdernote{=}{}{S\cons{\rdx\un}}   {
\root{\intru,\intru}{S\pars{\un\bc \un}} {
\root{\swir}{S\rdx{\pars{\rdc{\aprs{P\bc \neg P}}\bc \aprs{Q\bc \neg Q}} } } {
\root{\swir}{S\cnt{\aprs{  Q\bc \rdc{\pars{\aprs{P\bc \neg P}\bc \neg Q }}}}}{
\leaf{S\cnt{\aprs{P\bc \Black{Q\bc {}}\pars{   \neg P  \bc         \neg Q } }}}
}}}}
\qqquad
\vcdernote{=}{}{S\cons{\rdx\un}}   {
\root{\intru\bc \intru}{S\seqs{\un\bc \un}} {
\root{\seqru}{S \rdx{\seqs{\rdc{\aprs{P\bc \neg P}}\bc \aprs{Q\bc \neg Q}}} }{
\leaf{S\cnt{\aprs{\seqs{P\bc Q}\bc \seqs{\neg P\bc \neg Q }}}}
}}}
\qqquad
\vcdernote{\eru}{}{S\cons{\rdx\un}}   {
\root{\intru}{S\cons{\lwn\un}} {
\root{\promru}{S\cons{\rdx\lwn\rdc{\aprs{P\bc \neg P}}}}{
\leaf{S\cnt{\aprs{\lwn P\bc \loc\neg P}}}
}}}
\quad.
$$
The cases for $\intrd$ are dual.
\end{proof}

Note that in the proof above we tacitly used (for the sake of saving
paper) another helpful notation: writing
$\intru,\intru$ just means that two instances of $\intru$ applied one after
the other, where the order does not matter.

\begin{proposition}\label{prop:updown}
  Each rule\/ $\rho{\uparrow}$ in\/ $\SNEL$ is derivable in\/
  $\{\intrd,\intru,\swir,\rho{\downarrow}\}$, and, dually, each rule\/
  $\rho{\downarrow}$ in\/ $\SNEL$ is derivable in the system\/
  $\{\intrd,\intru,\swir,\rho{\uparrow}\}$.
\end{proposition}

\begin{proof}
Each instance 
\proofadjust
$$
\myssmall
\vcinf{\rho{\uparrow}}{S\cons R}{S\cons T}
\qqquad
\mbox{\normalsize can be replaced by}
\qqquad
\vcdernote{\intru          }{}{S     \cons R                              }   {
\root{\rho{\downarrow}}{S     \pars{R\bc \cnt{\aprs{T\bc \rdc{\neg T}}}} }  {
\root{\swir           }{S\rdx{\pars{R\bc      \aprs{T\bc \rdc{\neg R }}}}} {
\root{\intrd           }{S\cnt{\aprs{T\bc \rdc{\pars{R\bc      \neg R }}}}}{
\leaf                  {S     \cons T                              }}}}}
$$
and dually.
\end{proof}

In the calculus of structures, we call \dfn{core} the set of rules that is
used to reduce interaction and cut to atomic form. We use the term \emph{hard
core} to denote the set of rules in the core other than atomic interaction/cut
and empty/coempty. Rules that are not in the core are called \emph{non-core}.

\begin{definition}\label{def:core} 
The \dfn{core} of $\SNEL$ is
$\set{\atird,\atiru,\swir,\seqrd,\seqru,\promrd,\promru,\erd,\eru}$, denoted
by $\SNELsc$. The \dfn{hard core}, denoted by $\SNELhc$, is
$\set{\swir,\seqrd,\seqru,\promrd,\promru}$, and the \dfn{non-core} is
$\set{\weakrd,\weakru,\absrd,\absru,\dmrd,\dmru}$.
\end{definition}
 
System $\SNEL$ is up-down symmetric, and the properties we saw are also
symmetric.  Provability is an asymmetric notion: we want to observe the
possible conclusions that we can obtain from a unit premise.  We now break the
up-down symmetry by adding an inference rule with no premise, and we join
this logical axiom to the down fragment of $\SNEL$.

\begin{figure}[t]
  \begin{center}
      \figurebox{
    \begin{tabular}{c@{}c@{\qquad}c@{\qquad}c@{}c}
        &
        $\vcinf{\unr}{\enspace\un\enspace} {}$&
        $\vcinf{\atird} {S\copt{\pars{a\bc \neg a}}}{S\cons{\un}}$&
        $\vcinf{\erd} {S\cons{\loc\un}}{S\cons{\un}}$&
	\\ &&&\\ &
        $\vcinf{\swir}
          {S\copt{\pars{\aprs{R\bc T}\bc U}}} 
          {S\copt{\aprs{\pars{R\bc U}\bc T}}}$&
        $\vcinf{\seqrd}
          {S\pars{\seqs{R\bc T}\bc \seqs{U\bc V}}}
          {S\seqs{\pars{R\bc U}\bc \pars{T\bc V}}}$&    
        $\vcinf{\promrd}
          {S\copt{\pars{\loc R\bc  \lwn T}}}
          {S\cons{\loc\pars{R\bc T }}}$&
	\\ &&&\\ &
        $\vcinf{\weakrd}
          {S\cons{\lwn R}} 
          {S\cons{\un}}$&
        $\vcinf{\absrd}
          {S\cons{\lwn R}} 
          {S\copt{\pars{\lwn R\bc R }}}$&
        $\vcinf{\dmrd}
          {S\cons{\lwn R}} 
          {S\cons{\lwn\lwn R}}$&
          \\
    \end{tabular} }
    \caption{System $\NEL$}
    \label{fig:NEL}
  \end{center}
\end{figure}

\begin{definition}{} 
  \dfn{System $\NEL$} is shown in Fig.~\ref{fig:NEL}, where the rule
  $\unr$ is called \dfn{unit}.
\end{definition}

As an immediate consequence of Propositions~\ref{prop:atomic_int} and 
\ref{prop:updown} we get:

\begin{proposition}\label{prop:str-equ}
  $\NEL\cup\{\intru\}$ and\/ $\SNEL\cup\{\unr\}$ are strongly
  equivalent.
\end{proposition}

\begin{definition}\label{def:proof} 
    A derivation with no premise is called a \dfn{proof}, denoted by $\Pi$.  A
    system $\sysS$ \dfn{proves} $R$ if there is in $\sysS$ a proof
    $\Pi$ whose conclusion is $R$, written as 
    $$\mysmall\vcstrpr{\sysS}{\Pi}{R}\quadfs$$
    We say
    that a rule $\rho$ is \dfn{admissible} for the system $\sysS$ if
    $\rho\notin\sysS$ and for every proof
    ${\mysmall\vcstrpr{\sysS\cup\{\rho\}}{\Pi}{R}}$
    there is a proof
    ${\mysmall\vcstrpr{\sysS}{\Pi'}{R}}$.
    Two systems are \dfn{equivalent} if they
    prove the same structures.
\end{definition}

Except for $\atiru$ and $\weakru$, all rules in the systems $\SNEL$
and $\NEL$ enjoy a kind of subformula property (which we treat as an
asymmetric property, by going from conclusion to premise): premises
are made of substructures of the conclusions.
To get cut elimination, so as to have a system whose rules all enjoy
the subformula property, we could just get rid of $\atiru$ and
$\weakru$, by proving their admissibility for the other rules.  But we
can do more than that: the whole up fragment of $\SNEL$ (except for
$\swir$ which also belongs to the down fragment) is admissible.  This
entails a {\it modular\/} scheme for proving cut elimination.  We
state here the cut elimination theorem, but its complete proof is
shown in the accompanying paper~\cite{SIS-splitting}.

\begin{theorem}\label{thm:cutelim}
    System\/ $\NEL$ is equivalent to\/ $\SNEL\cup\set{\unr}$.
\end{theorem}

\begin{corollary}\label{cor:cutelim}
    The rule\/ $\intru$ is admissible for system\/ $\NEL$.
\end{corollary}

Any linear implication $T\limp R$, i.e., $\pars{\neg T\bc R}$, is
related to derivability by:

\begin{corollary}\label{cor:nel_snel}
    For any two structures\/ $T$ and\/ $R$, we have
  $$\mysmall
  \simplederi{T}{\strut\SNEL}{}{R}
  \qquad\mbox{\normalsize if and only if}\qquad
  \vcstrpr{\strut\NEL}{}{\pars{\neg T\bc R}}
  \quadfs
  $$
\end{corollary}

\begin{proof}
    For the first direction, perform the following transformations:
    \vskip-2.5ex
    $$\mysmall
    \simplederi{T}{\strut\SNEL}{\Delta}{R}
    \quad
    \stackrel{1}{\leadsto}
    \quad
    \simplederi{\pars{\neg T\bc T}}{\SNEL}{\Delta'}{\pars{\neg T\bc R}} 
    \quad
    \stackrel{2}{\leadsto}
    \quad
    \mbox{\raisebox{3.5ex}{$
    \vcstrder{\SNEL}{\Delta'}{\pars{\neg T\bc R}}{
    \root{\intrd}{\pars{\neg T\bc T}} {
      \root{\unr}{\,\aprun\,}{\leaf{}}}}
    $}}
    \quad
    \stackrel{3}{\leadsto}
    \quad
    \vcstrpr{\strut\NEL}{\Pi}{\pars{\neg T\bc R}} 
    \quadfs
    $$
    In the first step we replace each structure $S$ occurring inside 
    $\Delta$ by $\pars{\neg T\bc S}$, or, in other words, the
    derivation $\Delta'$ is obtained by putting $\Delta$ into the
    context $\pars{\neg T\bc\conhole}$. This
    is then transformed into a proof by adding an instance of $\intrd$ 
    and $\unr$. Then we apply Proposition~\ref{prop:atomic_int} and 
    cut elimination (Theorem~\ref{thm:cutelim})
    to obtain a proof in system $\NEL$.
    For the other direction, we proceed as follows:
    $$\mysmall
    \vcstrpr{\strut\NEL}{\strut\Pi}{\pars{\neg T\bc R}}
    \quadlto
    \simplederi{\aprun}
    {\strut\NEL\setminus\{\unr\}}{\strut\Delta}{\pars{\neg T\bc R}}
    \quadlto
    \vcdernote{\intru}{}{R}{
      \root{\swir}{\nrdx{\pars{\aprs{T\bc  \neg T}\bc R}}}{
	\stem{\strut\NEL\setminus\set{\unr}}{\strut\Delta'}
	     {\aprs{T\bc \pars{\neg T\bc R}}} {
	       \leaf{T}
    }}}
    \quadlto
    \simplederi{T}{\strut\SNEL}{}{R}
    \quadcm
    $$
    where the first two steps are trivial, and the last one is an
    application of Proposition~\ref{prop:atomic_int}.
\end{proof}

It is easy to prove that system $\NEL$ is a conservative extension of
$\BV$ and of $\MELL$ plus $\mix$ and $\mixzero$ (see
\cite{gug:SIS,dissvonlutz}).  The locality properties shown in
\cite{gug:str:01,str:MELL} still hold in this system, of course.  In
particular, the promotion rule is local, as opposed to the same rule
in the sequent calculus.

\section{Decomposition}\label{sec:decomposition}

The new top-down symmetry of derivations in the calculus of structures
allows us to study properties that are not observable in the sequent
calculus.  The most remarkable results so far are decomposition
theorems.  In general, a decomposition theorem says that a given
system $\sysS$ can be divided into $n$ pairwise disjoint subsystems
$\sysS_1$, \dots, $\sysS_n$ such that every derivation $\Delta$ in
system $\sysS$ can be rearranged as composition of $n$ derivations
$\Delta_1$, \dots, $\Delta_n$, where $\Delta_i$ uses only rules of
$\sysS_i$, for every $1\le i\le n$.

System $\SNEL$ can be decomposed into eleven subsystems, and there are
many different possibilities to transform a derivation into eleven
subderivations. We state here only four of them, but, due to the
modular proof, the others are evident.

\begin{theorem}[(Decomposition)]\label{thm:decomposition}
  For every derivation 
  $\mysmall\downsmash{\simplederi{T}{\Delta}{\SNEL}{R}}$
  there are derivations
  $$\mytiny
  \hskip-2em
  \vcstrder{\set{\eru}}{}{R}{
  \stem{\set{\dmrd}}{}{Q_1}{
  \stem{\set{\absrd}}{}{Q_2}{
  \stem{\set{\atiru}}{}{Q_3}{
  \stem{\set{\weakru}}{}{Q_4}{
  \stem{\SNELhc}{}{Q_5}{
  \stem{\set{\weakrd}}{}{P_5}{
  \stem{\set{\atird}}{}{P_4}{
  \stem{\set{\absru}}{}{P_3}{
  \stem{\set{\dmru}}{}{P_2}{
  \stem{\set{\erd}}{}{P_1}{
  \leaf{T}
  }}}}}}}}}}}
  \qquad\qqqqquad
  \vcstrder{\set{\dmrd}}{}{R}{
  \stem{\set{\absrd}}{}{V_1}{
  \stem{\set{\eru}}{}{V_2}{
  \stem{\set{\weakru}}{}{V_3}{
  \stem{\set{\atiru}}{}{V_4}{
  \stem{\SNELhc}{}{V_5}{
  \stem{\set{\atird}}{}{U_5}{
  \stem{\set{\weakrd}}{}{U_4}{
  \stem{\set{\erd}}{}{U_3}{
  \stem{\set{\absru}}{}{U_2}{
  \stem{\set{\dmru}}{}{U_1}{
  \leaf{T}
  }}}}}}}}}}}
  \qquad\qqqqquad
  \vcstrder{\set{\eru}}{}{R}{
  \stem{\set{\dmrd}}{}{Z_1}{
  \stem{\set{\absrd}}{}{Z_2}{
  \stem{\set{\weakrd}}{}{Z_3}{
  \stem{\set{\atiru}}{}{Z_4}{
  \stem{\SNELhc}{}{Z_5}{
  \stem{\set{\atird}}{}{W_5}{
  \stem{\set{\weakru}}{}{W_4}{
  \stem{\set{\absru}}{}{W_3}{
  \stem{\set{\dmru}}{}{W_2}{
  \stem{\set{\erd}}{}{W_1}{
  \leaf{T}
  }}}}}}}}}}}
  \qquad\qqqqquad
  \vcstrder{\set{\dmrd}}{}{R}{
  \stem{\set{\absrd}}{}{R_1}{
  \stem{\set{\weakrd}}{}{R_2}{
  \stem{\set{\eru}}{}{R_3}{
  \stem{\set{\atiru}}{}{R_4}{
  \stem{\SNELhc}{}{R_5}{
  \stem{\set{\atird}}{}{T_5}{
  \stem{\set{\erd}}{}{T_4}{
  \stem{\set{\weakru}}{}{T_3}{
  \stem{\set{\absru}}{}{T_2}{
  \stem{\set{\dmru}}{}{T_1}{
  \leaf{T}
  }}}}}}}}}}}
  $$%
\end{theorem}

For simplicity we will in the following call the four statements
first, second, third, and fourth decomposition (from left to right).
The fourth decomposition is crucial for the cut elimination proof
in~\cite{SIS-splitting}.

\begin{figure}[t]
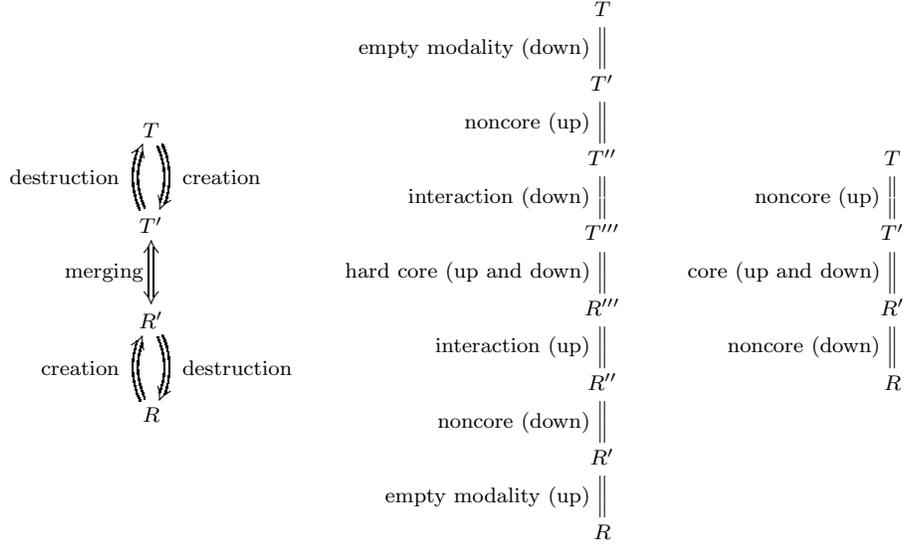

  \begin{center}
    \def\sysbox#1{\hbox{#1}}
    $
    \vcenter{\hbox{\xy\xygraph{[]!{0;<3pc,0pc>:} {T}
	(-@/^.5pc/@{=>}^<>(.5){\sysbox{~creation}}   [d] {T'}
	,-@/_.5pc/@{<=}_<>(.5){\sysbox{destruction }} [d] {\phantom{T'}}
	-@{<=>}_<>(.5){\sysbox{\ merging}}            [d] {R'}
	-@/^.5pc/@{=>}^<>(.5){\sysbox{~destruction}} [d] {R}
	-@/^.5pc/@{=>}^<>(.5){\sysbox{creation }} [u] {\phantom{R'}}
	)
      }\endxy}}    
    \qquad
    \vcstrder{\mbox{empty modality (up)}}{}{R}{
      \stem{\mbox{noncore (down)}}{}{R'}{
	\stem{\mbox{interaction (up)}}{}{R''}{
	  \stem{\mbox{hard core (up and down)}}{}{R'''}{
	    \stem{\mbox{interaction (down)}}{}{T'''}{
	      \stem{\mbox{noncore (up)}}{}{T''}{
		\stem{\mbox{empty modality (down)}}{}{T'}{
		  \leaf{T}}}}}}}}
    \qqquad
    \vcstrder{\mbox{noncore (down)}}{}{R}{
      \stem{\mbox{core (up and down)}}{}{R'}{
	\stem{\mbox{noncore (up)}}{}{T'}{
	  \leaf{T}}}}
    $\\
    \caption{Readings of the decompositions}
    \label{fig:read-decomp}
  \end{center}
\end{figure}

Apart from a decomposition into eleven subsystems, the first and the
second decomposition can also be read as a decomposition into three
subsystems that could be called \dfn{creation}, \dfn{merging}\/ and
\dfn{destruction}.  In the creation subsystem, each rule increases the
size of the structure; in the merging system, each rule does some
rearranging of substructures, without changing the size of the
structures; and in the destruction system, each rule decreases the
size of the structure. Here, the size of the structure incorporates
not only the number of atoms in it, but also the modality-depth for
each atom.  In a decomposed derivation, the merging part is in the
middle of the derivation, and (depending on your preferred reading of
a derivation) the creation and destruction are at the top and at the
bottom, as shown in the left of Figure~\ref{fig:read-decomp}. In
system $\SNEL$ the merging part contains the rules $\swir$, $\seqrd$,
$\seqru$, $\promrd$ and $\promru$, which coincides with the hard core.
In the top-down reading of a derivation, the creation part contains
the rules $\erd$, $\dmru$, $\absru$, $\weakrd$ and $\atird$, and the
destruction part consists of $\eru$, $\dmrd$, $\absrd$, $\weakru$ and
$\atiru$.  In the bottom-up reading, creation and destruction are
exchanged.

Note that this kind of decomposition (creation, merging, destruction)
is quite typical for logical systems presented in the calculus of
structures, and is not restricted to system $\SNEL$.  It holds, for
example, also for systems $\SBV$ and $\SELS$
\cite{gug:str:01,str:MELL}, for classical logic
\cite{brunnler:tiu:01}, and for full propositional linear logic
\cite{str:02}.

The third decomposition allows a separation between hard core and
noncore of the system, such that the up fragment and the down fragment
of the noncore are not merged, as it is the case in the first and
second decomposition.  More precisely, we can separate the seven
subsystems shown in the middle of Figure~\ref{fig:read-decomp}. The
fourth decomposition is even stronger in this respect: it allows a
complete separation between core and noncore, as shown on the right of
Figure~\ref{fig:read-decomp}.  This kind of decomposition is usually
more difficult to achieve than the decomposition into
creation--merging--destruction. In fact, it is not known whether it
holds for full linear logic. Furthermore, the separation between
\emph{non-core up} and \emph{non-core down} has not been achieved
in~\cite{str:MELL} for system~$\SELS$. But it is easy to see how the
proof in this paper can be adapted to the case of system~$\SELS$
presented in~\cite{str:MELL}. For classical logic such a decomposition
can be proved by using the cut-elimination result for the sequent
calculus $\LK$ together with the results in~\cite{brunnler:06:ce}. But
there is no known direct proof in the calculus of structures.

This decomposition into noncore-up, core, and noncore-down also plays
a crucial rule for the cut elimination argument
in~\cite{SIS-splitting}. Recall that cut elimination means to
get rid of the entire up-fragment. Because of the decomposition, the
elimination of the non-core up-fragment is now trivial. Furthermore,
recall that for cut elimination in the sequent calculus, the most
problematic cases are usually the ones where cut interacts with rules
like contraction and weakening, and that in our system these rules
appear as the non-core down rules. In the fourth decomposition these
are \emph{below} the actual cut rules (\ie the core up rules,
cf. Propositions \ref{prop:atomic_int}, \ref{prop:updown},
and~\ref{prop:str-equ}) and can therefore no longer interfere with the
cut elimination. This considerably simplifies our cut elimination
argument in~\cite{SIS-splitting}.

However, it is well-known that there is no free lunch. We cannot
expect that the proof of the decomposition theorem is trivial. At
least, we have to expect problems when the non-core rules (which in
case of $\SNEL$ do all deal with the modalities $\loc$ and~$\lwn$) do
interact with the rules $\promrd$ and $\promru$ (which are the only
core rules that properly deal with $\loc$ and~$\lwn$). The good news is
that these are indeed the only cases where the proof of the decomposition
theorem becomes problematic.

We will now continue with a very brief sketch of the proof and in the
remainder of this paper we will fill in the details.

\begin{figure}[t]
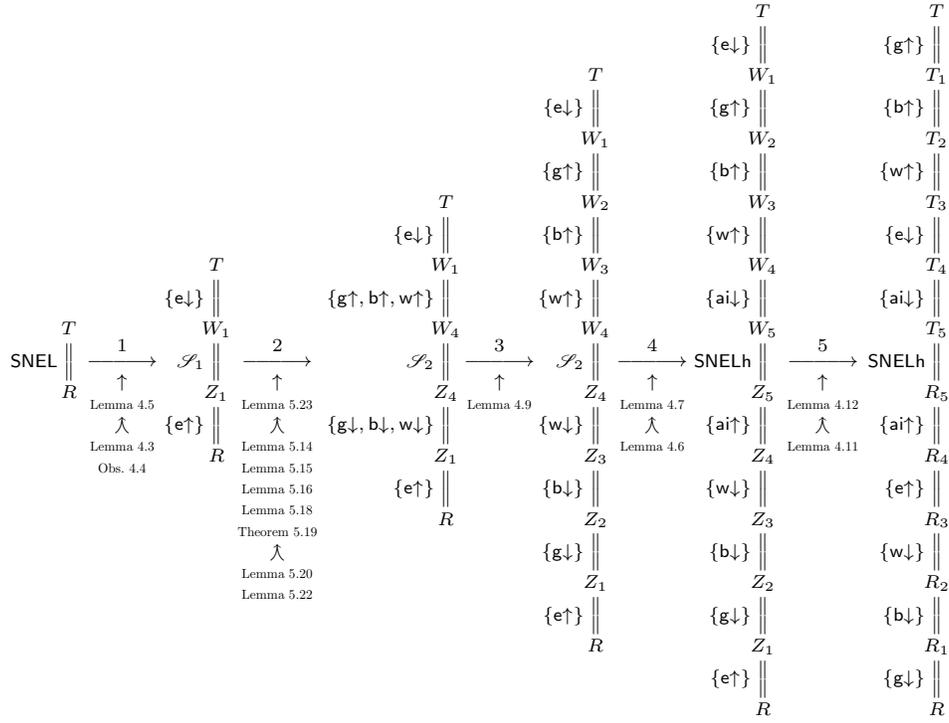

  \begin{center}
    \scriptsize
    \hbox{$
    \vcstrder{\SNEL}{}{R}{
      \leaf{T}
    } 
    \quadsteptol1{%
      \begin{array}{c}
	\lembox{Lemma~\ref{lem:step1}}\\
	\curlywedgeuparrow\\
	\lembox{Lemma~\ref{lem:perm-er}}\\
	\lembox{Obs.~\ref{obs:er}}
      \end{array}}
    \vcstrder{\set{\eru}}{}{R}{
      \stem{\sysS_1}{}{Z_1}{
	\stem{\set{\erd}}{}{W_1}{
	  \leaf{T}
    }}} 
    \!\quadsteptol2{%
      \begin{array}{c}
	\lembox{Lemma~\ref{lem:step2}}\\
	\curlywedgeuparrow\\
	\lembox{Lemma~\ref{lem:singlestep-up}}\\
	\lembox{Lemma~\ref{lem:singlestep-down}}\\
	\lembox{Lemma~\ref{lem:nocycle-term}}\\
	\lembox{Lemma~\ref{lem:nocycle}}\\
	\lembox{Theorem~\ref{thm:no-unforked}}\\
	\curlywedgeuparrow\\
	\lembox{Lemma~\ref{lem:bv_no_cycle}}\\
	\lembox{Lemma~\ref{lem:cycle-cycle}}   
      \end{array}}
    \vcstrder{\set{\eru}}{}{R}{
      \stem{\set{\dmrd,\absrd,\weakrd}}{}{Z_1}{
	\stem{\sysS_2}{}{Z_4}{
	  \stem{\set{\dmru,\absru,\weakru}}{}{W_4}{
	    \stem{\set{\erd}}{}{W_1}{
	      \leaf{T}
    }}}}} 
    \quadsteptol3{\lembox{Lemma~\ref{lem:step3}}}
    \vcstrder{\set{\eru}}{}{R}{
      \stem{\set{\dmrd}}{}{Z_1}{
	\stem{\set{\absrd}}{}{Z_2}{
	  \stem{\set{\weakrd}}{}{Z_3}{
	    \stem{\sysS_2}{}{Z_4}{
	      \stem{\set{\weakru}}{}{W_4}{
		\stem{\set{\absru}}{}{W_3}{
		  \stem{\set{\dmru}}{}{W_2}{
		    \stem{\set{\erd}}{}{W_1}{
		      \leaf{T}
    }}}}}}}}} 
    \quadsteptol4{%
      \begin{array}{c}
	\lembox{Lemma~\ref{lem:step4}}\\
	\curlywedgeuparrow\\
	\lembox{Lemma~\ref{lem:perm-weak-atir}}
      \end{array}}
    \vcstrder{\set{\eru}}{}{R}{
      \stem{\set{\dmrd}}{}{Z_1}{
	\stem{\set{\absrd}}{}{Z_2}{
	  \stem{\set{\weakrd}}{}{Z_3}{
	    \stem{\set{\atiru}}{}{Z_4}{
	      \stem{\SNELhc}{}{Z_5}{
		\stem{\set{\atird}}{}{W_5}{
		  \stem{\set{\weakru}}{}{W_4}{
		    \stem{\set{\absru}}{}{W_3}{
		      \stem{\set{\dmru}}{}{W_2}{
			\stem{\set{\erd}}{}{W_1}{
			  \leaf{T}
    }}}}}}}}}}} 
    \quadsteptol5{%
      \begin{array}{c}
	\lembox{Lemma~\ref{lem:step5}}\\
	\curlywedgeuparrow\\
	\lembox{Lemma~\ref{lem:fbw-e}}
      \end{array}}
    \vcstrder{\set{\dmrd}}{}{R}{
      \stem{\set{\absrd}}{}{R_1}{
	\stem{\set{\weakrd}}{}{R_2}{
	  \stem{\set{\eru}}{}{R_3}{
	    \stem{\set{\atiru}}{}{R_4}{
	      \stem{\SNELhc}{}{R_5}{
		\stem{\set{\atird}}{}{T_5}{
		  \stem{\set{\erd}}{}{T_4}{
		    \stem{\set{\weakru}}{}{T_3}{
		      \stem{\set{\absru}}{}{T_2}{
			\stem{\set{\dmru}}{}{T_1}{
			  \leaf{T}
    }}}}}}}}}}}
    $}
  \end{center}
  \caption{Obtaining the third and fourth decomposition}
  \label{fig:third}
\end{figure}

\begin{figure}[t]
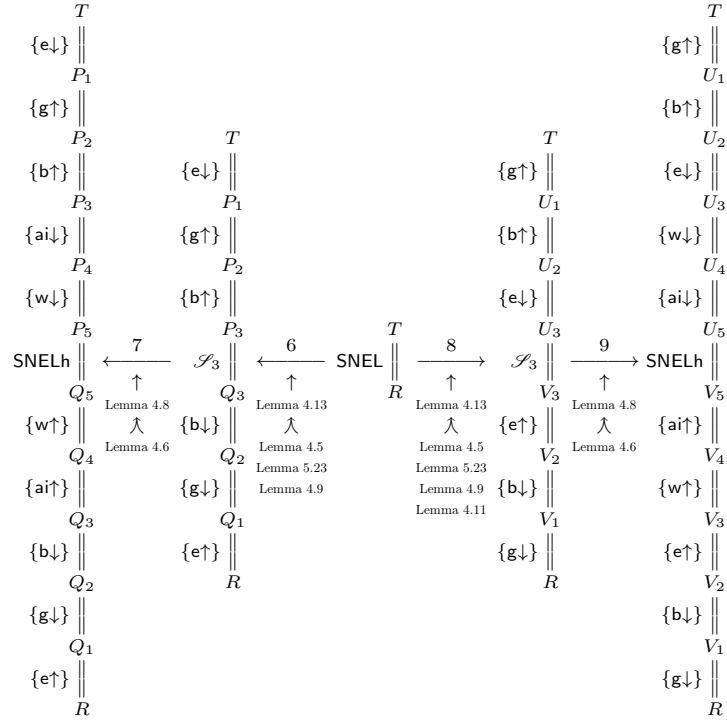

  \begin{center}
    \scriptsize
    $
    \vcstrder{\set{\eru}}{}{R}{
      \stem{\set{\dmrd}}{}{Q_1}{
	\stem{\set{\absrd}}{}{Q_2}{
	  \stem{\set{\atiru}}{}{Q_3}{
	    \stem{\set{\weakru}}{}{Q_4}{
	      \stem{\SNELhc}{}{Q_5}{
		\stem{\set{\weakrd}}{}{P_5}{
		  \stem{\set{\atird}}{}{P_4}{
		    \stem{\set{\absru}}{}{P_3}{
		      \stem{\set{\dmru}}{}{P_2}{
			\stem{\set{\erd}}{}{P_1}{
			  \leaf{T}
    }}}}}}}}}}}
    \quadstepfroml7{%
      \begin{array}{c}
	\lembox{Lemma~\ref{lem:step79}}\\
	\curlywedgeuparrow\\
	\lembox{Lemma~\ref{lem:perm-weak-atir}}
      \end{array}}
    \vcstrder{\set{\eru}}{}{R}{
      \stem{\set{\dmrd}}{}{Q_1}{
	\stem{\set{\absrd}}{}{Q_2}{
	  \stem{\sysS_3}{}{Q_3}{
	    \stem{\set{\absru}}{}{P_3}{
	      \stem{\set{\dmru}}{}{P_2}{
		\stem{\set{\erd}}{}{P_1}{
		  \leaf{T}
    }}}}}}}
    \quadstepfroml6{%
      \begin{array}{c}
	\lembox{Lemma~\ref{lem:step68}}\\
	\curlywedgeuparrow\\
	\lembox{Lemma~\ref{lem:step1}}\\
	\lembox{Lemma~\ref{lem:step2}}\\
	\lembox{Lemma~\ref{lem:step3}}
      \end{array}
    }
    \vcstrder{\SNEL}{}{R}{
      \leaf{T}
    } 
    \quadsteptol8{%
      \begin{array}{c}
	\lembox{Lemma~\ref{lem:step68}}\\
	\curlywedgeuparrow\\
	\lembox{Lemma~\ref{lem:step1}}\\
	\lembox{Lemma~\ref{lem:step2}}\\
	\lembox{Lemma~\ref{lem:step3}}\\
	\lembox{Lemma~\ref{lem:fbw-e}}
      \end{array}}
    \vcstrder{\set{\dmrd}}{}{R}{
      \stem{\set{\absrd}}{}{V_1}{
	\stem{\set{\eru}}{}{V_2}{
	  \stem{\sysS_3}{}{V_3}{
	    \stem{\set{\erd}}{}{U_3}{
	      \stem{\set{\absru}}{}{U_2}{
		\stem{\set{\dmru}}{}{U_1}{
		  \leaf{T}
    }}}}}}}
    \quadsteptol9{%
      \begin{array}{c}
	\lembox{Lemma~\ref{lem:step79}}\\
	\curlywedgeuparrow\\
	\lembox{Lemma~\ref{lem:perm-weak-atir}}
      \end{array}}
    \vcstrder{\set{\dmrd}}{}{R}{
      \stem{\set{\absrd}}{}{V_1}{
	\stem{\set{\eru}}{}{V_2}{
	  \stem{\set{\weakru}}{}{V_3}{
	    \stem{\set{\atiru}}{}{V_4}{
	      \stem{\SNELhc}{}{V_5}{
		\stem{\set{\atird}}{}{U_5}{
		  \stem{\set{\weakrd}}{}{U_4}{
		    \stem{\set{\erd}}{}{U_3}{
		      \stem{\set{\absru}}{}{U_2}{
			\stem{\set{\dmru}}{}{U_1}{
			  \leaf{T}
    }}}}}}}}}}}
    $
  \end{center}
  \caption{Obtaining the first and second decomposition}
  \label{fig:first}
\end{figure}

\begin{proof}[of Theorem~\ref{thm:decomposition} (Sketch)]
  The third and fourth decomposition are obtained via the five steps
  shown in Figure~\ref{fig:third}, where
  $\sysS_1=\SNEL\setminus\set{\erd,\eru}$ and
  $\sysS_2=\set{\atird,\atiru}\cup\SNELhc$. The first and second
  decomposition are reached as shown in Figure~\ref{fig:first}, where
  $\sysS_3=\set{\atird,\atiru,\weakrd,\weakru}\cup\SNELhc$. 
  In these figures we also indicated which lemmas are used for
  achieving which step in the decomposition. Some
  explanation: Step~1 is performed via a rather simple rule
  permutation. The rule $\erd$ is permuted up in the derivation, and
  the rule $\eru$ is permuted down via the dual procedure. The concept
  of permuting rules in the calculus of structures is explained in
  more detail in Section~\ref{sec:permrules}. Step~2 is the most
  critical one. In some sense it can also be considered as a simple
  rule permutation. However, contrary to Step~1, it is not obvious at
  all that Step~2 does terminate: while permuting $\dmru$, $\absru$,
  and $\weakru$ up, new instances of $\dmrd$, $\absrd$, and $\weakrd$
  are introduced, and vice versa. For showing termination, we
  introduce in Section~\ref{sec:flowgraph} the concept of \bqfg. Steps
  3, 4 and~5 are again rather simple rule permutations and are
  detailed out in Section~\ref{sec:permrules} as well. Steps 6 and~8
  are essentially the same as Steps~1--3 and~5 with the only difference
  that the rules $\weakru$ and $\weakrd$ do not need attention. Steps
  7 and~9 are only slight variations of each other and are not more
  complicated than Step~4. They are also done in
  Section~\ref{sec:permrules}. One last remark: Treating the rules
  $\dmru$, $\absru$, $\weakru$ together in Step~2 and separating them
  afterwards in Step~3 has been done on purpose. Treating them
  separately from the very beginning would not give termination in the
  general case. 
\end{proof}

\begin{remark}
  None of the four decompositions relies on the presence of mix nor
  nullary mix. All decompositions presented here work equally well for
  $\MELL$, as presented in~\cite{str:MELL}, where the units $\lone$
  and $\lbot$ are not equivalent to each other. However,
  in~\cite{str:MELL} only our second decomposition is given.
  The proof of the decomposition theorems works equally well for the
  logic which does not have the logical equivalences $\loc\loc
  R\equiv\loc R$ and $\lwn\lwn R\equiv\lwn R$. One just has to remove
  the rules $\dmrd$ and $\dmru$ from the system. 
  As a matter of fact, the structure of the
  modalities (e.g., the fact that there are 7 idempotent modalities in
  $\MELL$ or $\NEL$) does not influence decomposition.
\end{remark}

\section{Permutation of Rules}\label{sec:permrules}

Permutation of rules in the calculus of structures serves the same
purpose as rule permutations on the sequent calculus, with the only
difference that due to the greater flexibility of the formalism,
there are more cases to consider. 

\begin{definition}\label{def:permute}
  A rule $\pi$ \dfn{permutes over}
  a rule $\rho$ (or
  $\rho$ \dfn{permutes under} $\pi$) 
  if 
  $$
  \mbox{for every derivation}
  \quad
  {\mysmall\vcdernote{\pi}{}{P}{\root{\rho}{U}{\leaf{Q}}}}
  \quad
  \mbox{there is a derivation}
  \quad
  {\mysmall\vcdernote{\rho}{}{P}{\root{\pi}{V}{\leaf{Q}}}}
  $$
  for some structure $V$.
\end{definition}

For obtaining our decompositions, this definition is too strict. We
would need, for example, that the rule $\erd$ permutes over all other
rules in the system, which is not the case. We give a weaker concept:

\begin{definition}\label{def:permute-by-system}
  A rule $\pi$ \dfn{permutes over} a rule $\rho$ 
  \dfn{by a system} $\sysS$, if
  $$
  \mbox{for every derivation}
  \quad
  {\mysmall\vcdernote{\pi}{}{P}{\root{\rho}{U}{\leaf{Q}}}}
  \quad
  \mbox{there is a derivation}
  \quad
  {\mysmall\vcstrder{\sysS}{}{P}{\root{\rho}{W}{\root{\pi}{V}{\leaf{Q}}}}}
  $$ for some structures $V$ and $W$. Dually, \dfn{$\rho$ permutes
  under $\pi$ by $\sysS$}, if 
  $$
  \mbox{for every derivation}
  \quad
  {\mysmall\vcdernote{\pi}{}{P}{\root{\rho}{U}{\leaf{Q}}}}
  \quad
  \mbox{there is a derivation}
  \quad
  {\mysmall\vcdernote{\rho}{}{P}{\root{\pi}{V}{\stem{\sysS}{}{W}{\leaf{Q}}}}}
  $$ for some structures $V$ and $W$. 
\end{definition}

Additionally, we will use the following terminology borrowed from term
rewriting. In a rule instance
${\mysmall\vcinf{\rho}{S\cons{\rdx Z}}{S\cons{\cnt W}}}$ we call $Z$ the
\dfn{redex} and $W$ the \dfn{contractum} of the rule's
instance.
If we have $Z=W$, then the rule instance is
called \dfn{trivial}. (This can happen because of the equational
theory and the involvement of the unit $\un$.) In the following we
will assume, \wolg, that the trivial rule instances are removed from
all derivations.

\begin{figure}[t]
  \def\vcma#1{$\vcenter{\xymatrix@C=2em@R=3ex{#1}}$}
  \def\vcmb#1{$\vcenter{\xymatrix@C=2em@R=1ex{#1}}$}
  \def\p{\hbox{$\pi$}}
  \def\r{\hbox{$\rho$}}
  \def\s{\hbox{$\sysS$}}
  \newbox\bulbox
  \setbox\bulbox=\hbox{\phantom{$\bullet$}}
  \def\pbull{\copy\bulbox}
  \begin{center}
    \begin{tabular}{c@{\qquad}c@{\qqqquad}c}
      &local confluence&permutability of rules\\ \\
      have:
      &
      \vcma{
	&\bullet\ar[dl]_\r\ar[dr]^\p\\
	\bullet&&\bullet\\
	&\pbull}
      &
      \vcma{
	&\bullet\ar[dl]_\r\\
	\bullet\ar[dr]_\p&&\pbull\\
	&\bullet}
      \\ \\
      want:
      &
      \vcma{
	&\bullet\ar[dl]_\r\ar[dr]^\p\\
	\bullet\ar@{.>}[dr]_{\s}&&\bullet\ar@{.>}[dl]^{\s}\\
	&\bullet}
      &
      \vcmb{
	&\bullet\ar[ddl]_\r\ar@{->}[dr]^\p\\
	&&\bullet\ar@{->}[dd]^\r\\
	\bullet\ar[ddr]_\p&&\pbull\\
	&&\bullet\ar@{.>}[dl]^{\s}\\
	&\bullet}
      \quad or\quad
      \vcmb{
	&\bullet\ar[ddl]_\r\ar@{.>}[dr]^\s\\
	&&\bullet\ar@{->}[dd]^\p\\
	\bullet\ar[ddr]_\p&&\pbull\\
	&&\bullet\ar@{->}[dl]^\r\\
	&\bullet}
      \\ \\
      &
      $\sysS$ is the full system
      &
      $\sysS$ is as small and restricted as possible
    \end{tabular}
    \caption{The analysis of critical pairs for local
      confluence and
      the permutability of rules}
    \label{fig:critpair}
  \end{center}
\end{figure}
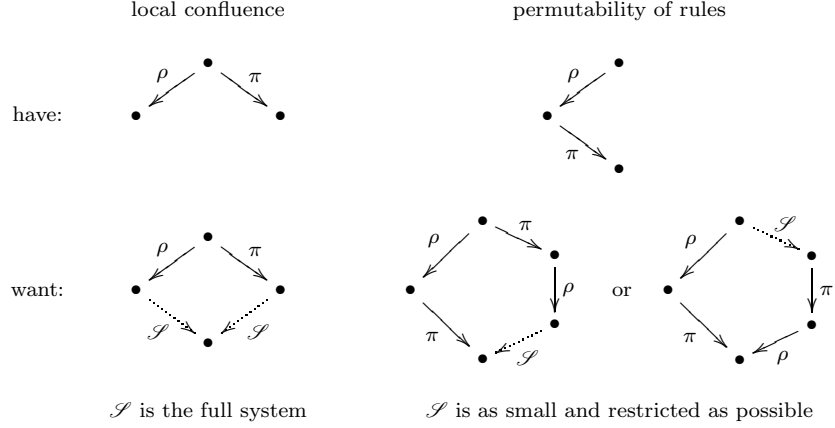

When reading this section, the reader might notice some similarity to the
analysis of critical pairs for local confluence in term rewriting. In fact,
the basic idea is the same but the conceptual goal is different, as it is
shown in Figure~\ref{fig:critpair}.

We now begin by showing that $\erd$ can be permuted over all rules
except $\weakrd$, $\absrd$, $\absru$, and $\dmru$ which will be
discussed later. In each of the following lemmas we also indicate for
which step in the proof of Theorem~\ref{thm:decomposition} they are
needed.

\begin{lemma}[(Step 1 in Fig.~\ref{fig:third})]\label{lem:perm-er}
  The rule $\erd$ permutes over the rules $\eru$, $\atird$, $\atiru$,
  $\swir$, $\seqrd$, $\seqru$, $\promrd$, $\promru$, $\weakru$, and
  $\dmrd$ by the system $\set{\swir,\seqrd,\seqru}$.
\end{lemma}

The proofs of this and the following lemmas about rule permutations
are done by routine case analysis\footnote{This is similar to the
  permutation lemmas in~\cite{str:MELL}, but in here the situation is
  a bit more complicated than in $\MELL$ because of the collapse of
  the units $\lone=\un=\lbot$.} and the reader is invited to skip
these proofs in the first reading.
In such a case analysis most cases are trivial and some
cases are nontrivial. For the sake of completeness, this time we
explain the case analysis in detail, and for similar lemmas that come
later, we show only the nontrivial cases.

\begin{proof}[of Lemma~\ref{lem:perm-er}]
  Consider
  $$\mysmall
  \dernote
      {\erd}{\quadcm}
      {S'\cons{Z'}}
      {\root
	{\rho}
	{S\cons{Z}}
	{\leaf
	  {S\cons{W}}
      }}
  $$ where
  $\rho\in\set{\eru,\atird,\atiru,\swir,\seqrd,\seqru,\promrd,\promru,
  \weakru,\dmrd}=\SNEL\setminus\set{\erd,\weakrd,\absrd,\absru,\dmru}$. We
  have to check all possibilities where the contractum $\un$ of $\erd$
  can appear inside $S\cons{Z}$. We start with the two trivial cases:
  \begin{enumerate}[(i)]
  \item\label{cas:cont} The contractum $\un$ of $\erd$ is inside the
    context $S\conhole$. That means that $Z'=Z$, and we can replace
    $$\mysmall
    \vcenter{\dernote
      {\erd}{}
      {S'\cons{Z}}
      {\root
	{\rho}
	{S\cons{Z}}
	{\leaf
	  {S\cons{W}}}
      }
    }
    \quadto
    \vcenter{\dernote
      {\rho}{}
      {S'\cons{Z}}
      {\root
	{\erd}
	{S'\cons{W}}
	{\leaf
	  {S\cons{W}}}}
    }
    $$
  \item\label{cas:pass} The contractum $\un$ of $\erd$ appears
    inside $Z$, but only inside a substructure of $Z$ that is not
    affected by the rule $\rho$. Instead of getting too formal, we
    show an example:
    $$\mysmall
    \vcenter{\dernote
      {\erd}{}
      {S\pars{\aprs{R\cons{\loc\un}\bc T}\bc U}}
      {\root
	{\swir}
	{S\pars{\aprs{R\cons{\un}\bc T}\bc U}}
	{\leaf
	  {S\aprs{\pars{R\cons{\un}\bc U}\bc T}}
    }}}
    \quadto
    \vcenter{\dernote
      {\swir}{}
      {S\pars{\aprs{R\cons{\loc\un}\bc T}\bc U}}
      {\root
	{\erd}
	{S\aprs{\pars{R\cons{\loc\un}\bc U}\bc T}}
	{\leaf
	  {S\aprs{\pars{R\cons{\un}\bc U}\bc T}}
    }}}
    $$ The cases where the $\un$ appears inside $U$ or $T$ are
    similar. The same situation can occur with the rules $\seqrd$,
    $\seqru$, $\promrd$, $\promru$, and $\dmrd$.
  \end{enumerate}
  The next case is in fact a subcase of \eqref{cas:cont}, but for
  didactic reasons we list it separately.
  \begin{enumerate}[(i)]
    \setcounter{enumi}{2}
  \item The contractum $\un$ of $\erd$ is the redex of $\rho$ (which
    is one of $\eru$, $\atiru$, $\weakru$). Then we have
    $$\mysmall
    \vcenter{\dernote
      {\erd}{}
      {S\cons{\loc\un}}
      {\root
	{\atiru}
	{S\cons{\un}}
	{\leaf
	  {S\aprs{a\bc \neg a}}}
    }}
    \quadto
    \vcenter{\dernote
      {=}{}
      {S\cons{\loc\un}}
      {\root
	{\atiru}
	{S\pars{\un\bc \loc\un}}
	{\root
	  {\erd}
	  {S\pars{\aprs{a\bc \neg a}\bc \loc\un}}
	  {\root
	    {=}
	    {S\pars{\aprs{a\bc \neg a}\bc \un}}
	    {\leaf
	      {S\aprs{a\bc \neg a}}}
    }}}}
    $$
  \end{enumerate}
  Finally, we come to the case which is nontrivial. It is the one
  where we need the system $\set{\swir,\seqrd,\seqru}$.
  \begin{enumerate}[(i)]
    \setcounter{enumi}{3}
  \item\label{cas:nontriv} The contractum $\un$ of $\erd$ actively
    interferes with the rule $\rho$. This can happen because of the
    equational theory for $\un$. First, let $\rho=\atird$ and consider
    the two derivations:
      $$\mysmall
      \vcenter{\dernote
	{\erd}{}
	{S\pars{\aprs{a\bc \loc\un}\bc \neg a}}
	{\root
	  {=}
	  {S\pars{\aprs{a\bc \un}\bc \neg a}}
	  {\root
	    {\atird}
	    {S\pars{a\bc \neg a}}
	    {\leaf
	      {S\cons{\un}}
      }}}}
      \qquand
      \vcenter{\dernote
	{\erd}{}
	{S\pars{\seqs{a\bc \loc\un}\bc \neg a}}
	{\root
	  {=}
	  {S\pars{\seqs{a\bc \un}\bc \neg a}}
	  {\root
	    {\atird}
	    {S\pars{a\bc \neg a}}
	    {\leaf
	      {S\cons{\un}}
      }}}}
      $$
      They can be replaced by
      $$\mysmall
      \vcenter{\dernote
	{\swir}{}
	{S\pars{\aprs{a\bc \loc\un}\bc \neg a}}
	{\root
	  {\atird}
	  {S\aprs{\pars{a\bc \neg a}\bc \loc\un}}
	  {\root
	    {\erd}
	    {S\aprs{\un\bc \loc\un}}
	    {\root
	      {=}
	      {S\aprs{\un\bc \un}}
	      {\leaf
		{S\cons{\un}}
      }}}}}
      \qquand
      \vcenter{\dernote
	{\seqrd}{}
	{S\pars{\seqs{a\bc \loc\un}\bc \neg a}}
	{\root
	  {\atird}
	  {S\seqs{\pars{a\bc \neg a}\bc \loc\un}}
	  {\root
	    {\erd}
	    {S\seqs{\un\bc \loc\un}}
	    {\root
	      {=}
	      {S\seqs{\un\bc \un}}
	      {\leaf
		{S\cons{\un}}
      }}}}}
      $$ respectively. Here we used the rules $\swir$ and $\seqrd$ to
      move the redex $\loc\un$ of $\erd$ out of the way of the rule
      $\atird$ such that the situation could be handled similarly to
      case \eqref{cas:cont}. A similar situation can occur with the
      rules $\swir$, $\promrd$, and~$\seqrd$. We will not show all
      possibilities here, but it should be clear that they all work
      because of the same principle. We content ourselves of
      presenting only the most complicated case (where $\rho=\seqrd$):
      $$\myvsmall
      \qlapm{\vcenter{\dernote
	{\erd}{} 
	{S\pars{\seqs{R\bc \aprs{\seqs{R'\bc T}\bc \rdx{\loc\un}}\bc T'}\bc \seqs{U\bc V}}} 
	{\root
	  {=}
	  {S\pars{\seqs{R\bc \aprs{\seqs{R'\bc T}\bc \un}\bc T'}\bc \seqs{U\bc V}}} 
	  {\root
	    {\seqrd}
	    {S\rdx{\pars{\seqs{R\bc R'\bc T\bc T'}\bc \seqs{U\bc V}}}} 
	    {\leaf
	      {S\seqs{\pars{\seqs{R\bc R'}\bc U}\bc \pars{\seqs{T\bc T'}\bc V}}}
      }}}}
      \to
      \vcenter{\dernote
	{\seqru}{} 
	{S\pars{\seqs{R\bc \rdx{\aprs{\seqs{R'\bc T}\bc \loc\un}\bc T'}}\bc \seqs{U\bc V}}} 
	{\root
	  {\seqru}
          {S\pars{\rdx{\seqs{R\bc \aprs{\seqs{R'\bc T\bc T'}\bc \loc\un}}}\bc \seqs{U\bc V}}} 
	  {\root
	    {\swir}
            {S\rdx{\pars{\aprs{\seqs{R\bc R'\bc T\bc T'}\bc \loc\un}\bc \seqs{U\bc V}}}} 
	    {\root
	      {\seqrd}
              {S\aprs{\rdx{\pars{\seqs{R\bc R'\bc T\bc T'}\bc \seqs{U\bc V}}}\bc \loc\un}} 
	      {\root
		{\erd}
		{S\aprs{\seqs{\pars{\seqs{R\bc R'}\bc U}\bc \pars{\seqs{T\bc T'}\bc V}}\bc 
		    \rdx{\loc\un}}} 
		{\root
		  {=}
		  {S\aprs{\seqs{\pars{\seqs{R\bc R'}\bc U}\bc \pars{\seqs{T\bc T'}\bc V}}\bc 
		      \un}} 
		  {\leaf
		    {S\seqs{\pars{\seqs{R\bc R'}\bc U}\bc \pars{\seqs{T\bc T'}\bc V}}}
      }}}}}}}}
      \quad
      $$ Here, two instances of $\seqru$ and one instance of $\swir$
      are needed to move the $\loc\un$ out of the way of $\seqrd$. A
      different situation can occur when $\rho=\promru$. Consider the
      two derivations
      $$\mysmall
      \vcenter{\dernote
	{\erd}{} 
	{S\cons{\lwn\aprs{R\bc \pars{\aprs{R'\bc T}\bc \rdx{\loc\un}}\bc T'}}} 
	{\root
	  {=}
	  {S\cons{\lwn\aprs{R\bc \pars{\aprs{R'\bc T}\bc \un}\bc T'}}} 
	  {\root
	    {\promru}
	    {S\cons{\rdx{\lwn\aprs{R\bc R'\bc T\bc T'}}}} 
	    {\leaf
	      {S\aprs{\lwn \aprs{R\bc R'}\bc \loc\aprs{T\bc T'}}}
      }}}}
      \qquand
      \vcenter{\dernote
	{\erd}{} 
	{S\cons{\lwn\aprs{R\bc \seqs{\aprs{R'\bc T}\bc \rdx{\loc\un}}\bc T'}}} 
	{\root
	  {=}
	  {S\cons{\lwn\aprs{R\bc \seqs{\aprs{R'\bc T}\bc \un}\bc T'}}} 
	  {\root
	    {\promru}
	    {S\cons{\rdx{\lwn\aprs{R\bc R'\bc T\bc T'}}}} 
	    {\leaf
	      {S\aprs{\lwn \aprs{R\bc R'}\bc \loc\aprs{T\bc T'}}}
      }}}}
      $$
      which can be replaced by:
      $$\mysmall
      \vcenter{\dernote
	{\swir}{} 
	{S\cons{\lwn\aprs{R\bc \rdx{\pars{\aprs{R'\bc T}\bc \loc\un}}\bc T'}}} 
	{\root
	  {\promru}
	  {S\cons{\rdx{\lwn\aprs{R\bc \pars{R'\bc \loc\un}\bc T\bc T'}}}} 
	  {\root
	    {\erd}
	    {S\aprs{\lwn\aprs{R\bc \pars{R'\bc \rdx{\loc\un}}}\bc \loc\aprs{T\bc T'}}} 
	    {\root
	      {=}
	      {S\aprs{\lwn\aprs{R\bc \pars{R'\bc \un}}\bc \loc\aprs{T\bc T'}}} 
	      {\leaf
		{S\aprs{\lwn \aprs{R\bc R'}\bc \loc\aprs{T\bc T'}}}
      }}}}}
      \qquand
      \vcenter{\dernote
	{\seqru}{} 
	{S\cons{\lwn\aprs{R\bc \rdx{\seqs{\aprs{R'\bc T}\bc \loc\un}}\bc T'}}} 
	{\root
	  {\promru}
	  {S\cons{\rdx{\lwn\aprs{R\bc \seqs{R'\bc \loc\un}\bc T\bc T'}}}} 
	  {\root
	    {\erd}
	    {S\aprs{\lwn\aprs{R\bc \seqs{R'\bc \rdx{\loc\un}}}\bc \loc\aprs{T\bc T'}}} 
	    {\root
	      {=}
	      {S\aprs{\lwn\aprs{R\bc \seqs{R'\bc \un}}\bc \loc\aprs{T\bc T'}}} 
	      {\leaf
		{S\aprs{\lwn \aprs{R\bc R'}\bc \loc\aprs{T\bc T'}}}
      }}}}}
      $$ Here the $\loc\un$ has not been moved to the outside but to
      the inside, such that the permutation could be handled as in
      case~\eqref{cas:pass} above. A similar situation can occur with
      the rules $\rho=\swir,\seqrd,\seqru$. Again, we do not show all
      possibilities. But the reader should be able to convince himself
      that it is always possible to move the $\loc\un$ out of the way
      of $\rho$.\footnote{A complete list of all possible cases can be
      found in \cite{dissvonlutz}.} \qed
  \end{enumerate}
\end{proof}

\begin{observation}[(Step 1 in Fig.~\ref{fig:third})]\label{obs:er}
  Let us now see how to permute $\erd$ over the rules $\weakrd$,
  $\absrd$, $\absru$, and $\dmru$, which have been left out in
  Lemma~\ref{lem:perm-er}. The nontrivial cases are as follows:
  \begin{itemize}[---]
  \item for $\weakrd$:\proofadjust
    \begin{equation}
      \label{eq:e-weakrd}
      \mysmall
      \vcdernote
          {\erd}{} 
          {S\cons{\lwn R\cons{\rdx{\loc\un}}}} 
          {\root
	    {\weakrd}
	    {S\cons{\rdx{\lwn R\cons{\un}}}} 
	    {\leaf
	      {S\cons{\un}}
          }}
          \qquadto
          \vcdernote
              {\weakrd}{} 
              {S\cons{\rdx{\lwn R\cons{\loc\un}}}} 
              {\leaf
	        {S\cons{\un}}
              }
    \end{equation}
  \item for $\absrd$:\proofadjust
    \begin{equation}
      \label{eq:e-absrd}
      \mysmall
      \vcdernote
          {\erd}{} 
          {S\cons{\lwn R\cons{\rdx{\loc\un}}}} 
          {\root
	    {\absrd}
	    {S\cons{\rdx{\lwn R\cons{\un}}}} 
	    {\leaf
	      {S\pars{\lwn R\cons{\un}\bc R\cons{\un}}}
          }}
          \qquadto
          \vcdernote
              {\absrd}{} 
              {S\cons{\rdx{\lwn R\cons{\loc\un}}}} 
              {\root
	        {\erd\bc \erd}
	        {S\pars{\lwn R\cons{\loc\un}\bc R\cons{\loc\un}}}
	        {\leaf
	          {S\pars{\lwn R\cons{\un}\bc R\cons{\un}}}
              }}
    \end{equation}
  \item for $\absru$:\proofadjust
    \begin{equation}
      \label{eq:e-absru}
      \mysmall
      \vcdernote
          {\erd}{} 
          {S\aprs{\loc R\cons{\rdx{\loc\un}}\bc R\cons{\un}}} 
          {\root
	    {\absru}
	    {S\aprs{\loc R\cons{\rdx{\un}}\bc R\cons{\un}}} 
	    {\leaf
	      {S\cons{\loc R\cons{\un}}} 
          }}
          \qquadto
          \vcdernote
              {\weakru}{} 
              {S\aprs{\loc R\cons{\loc\un}\bc R\cons{\rdx{\un}}}} 
              {\root
	        {\absru}
	        {S\aprs{\loc R\cons{\loc\un}\bc R\cons{\loc\un}}} 
	        {\root
	          {\erd}
	          {S\cons{\loc R\cons{\loc\un}}} 
	          {\leaf
	            {S\cons{\loc R\cons{\un}}} 
              }}}
    \end{equation}
  \item for $\dmru$:\proofadjust
    \begin{equation}
    \label{eq:e-dmru}
      \mysmall
    \vcdernote{\erd}{}{S\cons{\loc\aprs{\loc\un\bc\loc R}}}{
      \root{=}{S\cons{\loc\aprs{\un\bc\loc R}}}{
	\root{\dmru}{S\cons{\loc\loc R}}{
	  \leaf{S\cons{\loc R}}}}}
	\qquadto
	\vcdernote{\weakru}{}{S\cons{\loc\aprs{\loc\un\bc\loc R}}}{
	  \root{\absru}{S\cons{\loc\aprs{\loc\loc R\bc\loc R}}}{
	    \root{\dmru}{S\cons{\loc\loc\loc R}}{
	      \root{\dmru}{S\cons{\loc\loc R}}{
		\leaf{S\cons{\loc R}}}}}}
    \end{equation}
  \end{itemize}
  Note that these cases do not follow the statement of Definitions
  \ref{def:permute} or~\ref{def:permute-by-system}, which is the reason
  why they have been left out in Lemma~\ref{lem:perm-er}. 
\end{observation}

\begin{lemma}[(Step 1 in Fig.~\ref{fig:third})]\label{lem:step1}
  Every $\SNEL$ derivation $\Delta$ can be decomposed as indicated
  in Step~1 in Figure~\ref{fig:third}.
\end{lemma}

\begin{proof}
  We show by induction that all instances of $\erd$ in $\Delta$ can be
  permuted to the top. For each instance $\pi$ of $\erd$ in $\Delta$
  we define $\be{\pi}$ to be the number of instances of $\absrd$ that
  appear above $\pi$ in $\Delta$. We say that $\pi$ is at the top of
  $\Delta$, if the only rules that appear above $\pi$ in $\Delta$ are
  instances of $\erd$. We let the $\erd$-rank of $\Delta$ be the
  multiset $\erank{\Delta}=\mset{\;\be{\pi}\mid\text{$\pi$ is an
      instance $\erd$ in $\Delta$ and $\pi$ is not at the top}\;}$.
  By $\delta_{\erd}(\Delta)$ we denote the height of the derivation
  above the topmost $\erd$ that is not at the top of $\Delta$.  As
  induction measure we use the pair
  $\tuple{\erank{\Delta},\delta_{\erd}(\Delta)}$ with the
  lexicographic ordering, and the multiset ordering for the first
  component. This measure always goes down if we permute up the
  topmost instance of $\erd$ that is not at the top (by using
  Lemma~\ref{lem:perm-er} and Observation~\ref{obs:er}).  Dually, all
  instances of $\eru$ can be permuted to the bottom.
\end{proof}

We now continue with Step~4. For this, recall that $\SNELhc=
  \set{\swir,\seqrd,\seqru,\promrd,\promru}$.
 
\begin{lemma}[(Steps 4,7,9 in Fig.~\ref{fig:third},\ref{fig:first})]
  \label{lem:perm-weak-atir}
  The rules $\weakrd$ and $\atird$ permute over the rules $\eru$,
  $\atird$, $\atiru$, $\swir$, $\seqrd$, $\seqru$, $\promrd$,
  $\promru$, $\weakru$, and $\dmrd$ by the system
  $\set{\swir,\seqrd,\seqru}$.
\end{lemma}

\begin{proof}
  The contractum of $\weakrd$ and $\atird$ is the same as of $\erd$,
  namely $\un$. Hence, this proof is the same as the one for
  Lemma~\ref{lem:perm-er}. 
\end{proof}

\begin{lemma}[(Step 4 in Fig.~\ref{fig:third})]\label{lem:step4}
  Step~4 in the proof of Theorem~\ref{thm:decomposition} can be
  performed as indicated in Figure~\ref{fig:third}.
\end{lemma}

\begin{proof}
  We show by induction that in a derivation $\Delta$ in the system
  $\set{\atird,\atiru}\cup\SNELhc=
  \set{\atird,\atiru,\swir,\seqrd,\seqru,\promrd,\promru}$ all
  instances of $\atird$ can be permuted to the top. As induction
  measure we use the pair
  $\tuple{n_{\atird}(\Delta),\delta_{\atird}(\Delta)}$ with the
  lexicographic ordering, where $n_{\atird}(\Delta)$ is the number of
  $\atird$ instances in $\Delta$ that are not yet at the top of the
  derivation, and $\delta_{\atird}(\Delta)$ is the height of the
  derivation above the topmost such $\atird$ in $\Delta$.  This measure
  always goes down if we permute the topmost instance of $\atird$ up,
  by using Lemma~\ref{lem:perm-weak-atir}. This measure is simpler
  than the one in the proof of Lemma~\ref{lem:step1} because we do not
  have to deal with~$\absrd$.  Dually, all $\atiru$ can be permuted to
  the bottom.
\end{proof}

\begin{lemma}[(Steps 7,9 in Fig.~\ref{fig:first})]\label{lem:step79}
  Steps 7 and~9 in the proof of Theorem~\ref{thm:decomposition} can be
  performed as indicated in Figure~\ref{fig:first}.
\end{lemma}

\begin{proof}
  This proof is similar to the proof of Lemma~\ref{lem:step4} and uses
  Lemma~\ref{lem:perm-weak-atir}. Note that for Step~7, we
  additionally need to permute $\atird$ over $\weakrd$, for which the
  only nontrivial case is similar to~\eqref{eq:e-weakrd}.
\end{proof}

We will now continue with Step~3, for which the following lemma (and
its dual) is sufficient.

\begin{lemma}[(Step 3 in Fig.~\ref{fig:third})]
  \label{lem:fbw-sep}\label{lem:step3}
  All derivations 
  $\;\smash{\mysmall\vcstrder{\set{\dmru,\absru,\weakru}}{}{W_4}{\leaf{W_1}}}$ 
  can be decomposed as indicated in Step~3 in Figure~\ref{fig:third}.
\end{lemma}

\begin{proof}
  This is again a simple rule permutation. First, all instances of
  $\dmru$ are permuted up to the top. The trivial cases are as in
  Lemma~\ref{lem:perm-er}. The only nontrivial cases are the following:
  \begin{equation}\label{eq:fb1}
    \mysmall
    \vcdernote{\dmru}{}{S\aprs{\loc\loc R\bc R}}{
      \root{\absru}{S\aprs{\loc R\bc R}}{
	\leaf{S\cons{\loc R}}
    }}
    \quadto
    \vcdernote{=}{}{S\aprs{\loc\loc R\bc R}}{
      \root{\weakru}{S\aprs{\loc\loc R\bc \un\bc R}}{
	\root{\absru}{S\aprs{\loc\loc R\bc \loc R\bc R}}{
	  \root{\absru}{S\aprs{\loc\loc R\bc \loc R}}{
	    \root{\dmru}{S\cons{\loc\loc R}}{
	      \leaf{S\cons{\loc R}}
    }}}}}
  \end{equation}
  \begin{equation}\label{eq:fb2}
    \mysmall
    \vcdernote{\dmru}{}{S\aprs{\loc R\cons{\loc\loc T}\bc R\cons{\loc T}}}{
      \root{\absru}{S\aprs{\loc R\cons{\loc T}\bc R\cons{\loc T}}}{
	\leaf{S\cons{\loc R\cons{\loc T}}}
    }}
    \quadto
    \vcdernote{=}{}{S\aprs{\loc R\cons{\loc\loc T}\bc R\cons{\loc T}}}{
      \root{\weakru}{S\aprs{\loc R\cons{\loc\loc T}\bc R\aprs{\un\bc \loc T}}}{
	\root{\absru}{S\aprs{\loc R\cons{\loc\loc T}\bc 
	    R\aprs{\loc\loc T\bc \loc T}}}{
	  \root{\absru}{S\aprs{\loc R\cons{\loc\loc T}\bc R\aprs{\loc\loc T}}}{
	    \root{\dmru}{S\cons{\loc R\cons{\loc\loc T}}}{
	      \leaf{S\cons{\loc R\cons{\loc T}}}
    }}}}}
  \end{equation}
  Finally, all $\weakru$ are permuted under the $\absru$, where
  \begin{equation}
    \mysmall
    \label{eq:bwu}
    \vcdernote{\absru}{}{S\aprs{\loc R\cons{\un}\bc R\cons{\un}}}{
      \root{\weakru}{S\cons{\loc R\cons{\un}}}{
	\leaf{S\cons{\loc R\cons{\loc T}}}}}
    \quadto
    \vcdernote{\weakru,\weakru}{}{S\aprs{\loc R\cons{\un}\bc R\cons{\un}}}{
      \root{\absru}{S\aprs{\loc R\cons{\loc T}\bc R\cons{\loc T}}}{
	\leaf{S\cons{\loc R\cons{\loc T}}}}}
  \end{equation}
  is the only 
  nontrivial case. 
\end{proof}

\begin{remark}
  Note that the decomposition of Lemma~\ref{lem:fbw-sep} does not
  allow much variation. We can neither permute $\absru$ over $\dmru$,
  nor can we permute $\weakru$ over $\absru$, as the following
  examples show:
  $$
    \mysmall
  \vcdernote{\absru}{}{\aprs{\loc\loc a\bc\loc a}}{
    \root{\dmru}{\loc\loc a}{\leaf{\loc a}}}
  \qqquad\mbox{\normalsize and}\qqquad
  \vcdernote{\weakru}{}{a}{
    \root{\absru}{\aprs{\loc a\bc a}}{\leaf{\loc a}}}
  $$
\end{remark}

\begin{lemma}[(Step 5 in Fig.~\ref{fig:third})]\label{lem:fbw-e}
  The rules $\dmru$, $\absru$, and $\weakru$ permute over~$\erd$.
\end{lemma}

\begin{proof}
  The only nontrivial cases are the following.
  $$
  \mysmall
   \vcdernote{\dmru}{}{S\cons{\loc\loc\un}}{
      \root{\erd}{S\cons{\loc\un}}{
	\leaf{S\cons{\un}}}}
    \to
    \vcdernote{\erd}{}{S\cons{\loc\loc\un}}{
      \root{\erd}{S\cons{\loc\un}}{
	\leaf{S\cons{\un}}}}
    \qqquad
    \vcdernote{\absru}{}{S\aprs{\loc\un\bc \un}}{
      \root{\erd}{S\cons{\loc\un}}{
	\leaf{S\cons{\un}}}}
    \to
    \vcdernote{=}{}{S\aprs{\loc\un\bc \un}}{
      \root{\erd}{S\cons{\loc\un}}{
	\leaf{S\cons{\un}}}}
    \qqquad
    \vcdernote{\weakru}{}{S\cons{\un}}{
      \root{\erd}{S\cons{\loc\un}}{
	\leaf{S\cons{\un}}}}
    \to
    \vcdernote{=}{}{S\cons{\un}}{
      \leaf{S\cons{\un}}}
    $$
  In all of them the instance of $\dmru$, $\absru$, and
  $\weakru$, which is permuted up disappears. The trivial cases are
  as in case~(i) of Lemma~\ref{lem:perm-er}. Case~(ii) in the proof
  of that lemma cannot occur here. 
\end{proof}

\begin{lemma}[(Step 5 in Fig.~\ref{fig:third})]\label{lem:step5}
  Step~5 in the proof of Theorem~\ref{thm:decomposition} can be
  performed as indicated in Figure~\ref{fig:third}.
\end{lemma}

\begin{proof}
  By Lemma~\ref{lem:fbw-e} and its dual.
\end{proof}

\begin{lemma}[(Steps 6,8 in Fig.~\ref{fig:first})]\label{lem:step68}
  Steps 6 and~8 in the proof of Theorem~\ref{thm:decomposition} can be
  performed as indicated in Figure~\ref{fig:first}.
\end{lemma}

\begin{proof}
   Steps 6 and~8, are almost identical to Steps 1 to~3 and~5, with the
   only difference that the rules $\weakru$ and $\weakrd$ are omitted.
\end{proof}

After this tour de force of simple rule permutations, the proof of
Theorem~\ref{thm:decomposition} is completed, except for
Step~2. At first sight one might expect that this can also be done by
simple rule permutations. So, let us
attempt to permute all $\dmru$, $\absru$, and $\weakru$ up to the top
of a derivation.

\begin{casana}[(for permuting $\dmru$, $\absru$, $\weakru$ up)]
    \label{para:perm-fbw}
  Consider a derivation
  $$\mysmall
  \vcdernote
      {\pi}{}
      {P}
      {\root
	{\rho}
	{S\cons{Z}}
	{\leaf
	  {S\cons{W}}
      }}
      \quadcm
  $$ where $\rho\in\SNEL\setminus\set{\dmru,\absru,\weakru,\erd,\eru}$ and
  $\pi\in\set{\dmru,\absru,\weakru}$. The trivial cases \eqref{cas:cont}
  and~\eqref{cas:pass} are as in the proof of Lemma~\ref{lem:perm-er}. Then
  there is another (almost) trivial case which does not correspond to a case
  in the proof of Lemma~\ref{lem:perm-er}.
  \begin{enumerate}[(i)]\setcounter{enumi}{2}
  \item\label{cas:fbw-up-triv} The redex $Z$ of $\rho$ is inside the
    contractum of $\pi$, \ie we have one of the following three
    situations
    $$\mysmall
    \vcdernote{\dmru}{}{S\cons{\loc\loc R\cons{Z}}}{
      \root{\rho}{S\cons{\loc R\cons{Z}}}{
	\leaf{S\cons{\loc R\cons{W}}}
    }}
    \qqquad
    \vcdernote{\absru}{}{S\aprs{\loc R\cons{Z}\bc R\cons{Z}}}{
      \root{\rho}{S\cons{\loc R\cons{Z}}}{
	\leaf{S\cons{\loc R\cons{W}}}
    }}
    \qqquad
    \vcdernote{\weakru}{}{S\cons{\un}}{
      \root{\rho}{S\cons{\loc R\cons{Z}}}{
	\leaf{S\cons{\loc R\cons{W}}}
    }}
    $$
    which can be replaced by
    \begin{equation}\label{eq:rho-up}\mysmall
    \vcdernote{\rho}{}{S\cons{\loc\loc R\cons{Z}}}{
      \root{\dmru}{S\cons{\loc\loc R\cons{W}}}{
	\leaf{S\cons{\loc R\cons{W}}}
    }}
    \qqquad
    \vcdernote{\rho\bc \rho}{}{S\aprs{\loc R\cons{Z}\bc R\cons{Z}}}{
      \root{\absru}{S\aprs{\loc R\cons{W}\bc R\cons{W}}}{
	\leaf{S\cons{\loc R\cons{W}}}
    }}
    \qqquad
    \vcdernote{\weakru}{}{S\cons{\un}}{
	\leaf{S\cons{\loc R\cons{W}}}
    }
    \end{equation}
    respectively.
  \end{enumerate}
  The next case corresponds to case \eqref{cas:nontriv} in the proof
  of Lemma~\ref{lem:perm-er}.
  \begin{enumerate}[(i)]\setcounter{enumi}{3}
  \item The contractum $\loc R$ of $\pi$ actively interferes with the
    redex $Z$ of $\rho$. This can only happen with
    $\rho\in\set{\weakrd,\absrd,\promrd}$. If $\rho$ is $\weakrd$ or
    $\absrd$, then the situation is similar to \eqref{eq:e-weakrd}
    and~\eqref{eq:e-absrd} above. If $\rho=\promrd$, then we have one of
    $$\mysmall
    \vcdernote{\dmru}{}{S\pars{\loc\loc R\bc\lwn T}}{
      \root{\promrd}{S\pars{\loc R\bc \lwn T}}{
	\leaf{S\cons{\loc\pars{R\bc T}}}
    }}
    \qqqquad
    \vcdernote{\absru}{}{S\pars{\aprs{\loc R\bc R}\bc \lwn T}}{
      \root{\promrd}{S\pars{\loc R\bc \lwn T}}{
	\leaf{S\cons{\loc\pars{R\bc T}}}
    }}
    \qqqquad
    \vcdernote{\weakru}{}{S\pars{\un\bc \lwn T}}{
      \root{\promrd}{S\pars{\loc R\bc \lwn T}}{
	\leaf{S\cons{\loc\pars{R\bc T}}}
    }}
    $$
    which can be replaced by, respectively (for the second one, see Remark~\ref{rem:equation}):
    $$\mysmall
    \vcdernote{\dmrd}{}{S\pars{\loc\loc R\bc \lwn T}}{
      \root{\promrd}{S\pars{\loc\loc R\bc \lwn\lwn T}}{
	\root{\promrd}{S\cons{\loc\pars{\loc R\bc \lwn T}}}{
	  \root{\dmru}{S\cons{\loc\loc\pars{R\bc T}}}{
	    \leaf{S\cons{\loc\pars{R\bc T}}}
    }}}}
    \qqquad
    \vcdernote{\absrd}{}{S\pars{\aprs{\loc R\bc R}\bc \lwn T}}{
      \root{\swir}{S\pars{\aprs{\loc R\bc R}\bc \lwn T\bc T}}{
	\root{\swir}{S\pars{\aprs{\pars{\loc R\bc \lwn T}\bc R}\bc T}}{
	  \root{\promrd}{S\aprs{\pars{\loc R\bc \lwn T}\bc \pars{R\bc T}}}{
	    \root{\absru}{S\aprs{\loc\pars{R\bc T}\bc \pars{R\bc T}}}{
	      \leaf{S\cons{\loc\pars{R\bc T}}}
    }}}}}
    \qqquad
    \vcdernote{\weakrd}{}{S\pars{\un\bc \lwn T}}{
      \root{=}{S\pars{\un\bc \un}}{
	\root{\weakru}{S\cons{\un}}{
	  \leaf{S\cons{\loc\pars{R\bc T}}}
    }}}
    $$
  \end{enumerate}
\end{casana}

This means that there is indeed no objection against permuting all
instances of $\dmru$, $\absru$, and $\weakru$ up to the top of a
derivation, and then (by duality) permute all $\dmrd$, $\absrd$, and
$\weakrd$ down to the bottom. However, the problem is that while
permuting $\dmru$, $\absru$, $\weakru$ up, we introduce, in
case~\eqref{cas:nontriv}, new instances of $\dmrd$, $\absrd$,
$\weakrd$, and dually, while permuting $\dmrd$, $\absrd$, $\weakrd$
down, we introduce new instances of $\dmru$, $\absru$, $\weakru$. This
means that this permuting up and down could run forever. At least, it
is not obvious that it terminates eventually, as it is the case with
Steps 1, 4, 7 and~9 in the proof of Theorem~\ref{thm:decomposition}.

Please note that there is no obvious induction measure related to the
size of the derivation that could be used for showing termination. The
up and down permutation of $\weakru$ and $\weakrd$ alone is
unproblematic because at each critical case the disturbing instance of
$\promrd$ or $\promru$ is destroyed (but for convenience we will deal
with all six rules $\dmru$, $\absru$, $\weakru$ and $\dmrd$, $\absrd$,
$\weakrd$ together). The up and down permutation of $\dmru$, $\absru$
and $\dmrd$, $\absrd$ is very problematic, however. The rules $\dmru$
and $\dmrd$ cause a duplication of the disturbing instance of
promotion, and the permutation of $\absru$ and $\absrd$ causes an even
worse increase in the size of the derivation. In fact, the $\rho$ in
the middle derivation in \eqref{eq:rho-up} could be an instance of a
promotion that is disturbing for another $\dmru$ or $\absru$.

Clearly, a different technology is needed here, and this is the
purpose of the next section.

\let\rdx\yrdx

\section{Paths and Cycles in !-?-Flow-Graphs}\label{sec:step2}

For showing the termination of Step 2 in the proof of
Theorem~\ref{thm:decomposition}, we first introduce the notion of
\bqfg\ (in Section~\ref{sec:flowgraph}), which is very similar to
Buss' logical flow graphs~\cite{buss:91}, but instead of considering
all subformulas occurring in the derivation, we consider only the !-
and ?-subformulas. Based on these \bqfgs\ we define for each instance
of $\dmru$, $\dmrd$, $\absru$, $\absrd$, $\weakru$, $\weakrd$ its rank
(in Section~\ref{sec:rank}). The purpose of this rank is, very roughly
speaking, to describe the amount of work that still has to be done to
move this rule instance to its destination at the top or the bottom of
the derivation. The up-rank of a derivation $\Delta$, denoted by
$\ranku{\Delta}$ is then the multiset of the ranks of all instances of
$\dmru$, $\absru$, $\weakru$ occurring in $\Delta$, and the down-rank,
denoted by $\rankd{\Delta}$ is the multiset of the ranks of all
instances of $\dmrd$, $\absrd$, $\weakrd$ in $\Delta$. With this, we
are able to show that the process of permuting up all $\dmru$,
$\absru$, $\weakru$ does terminate (in Section~\ref{sec:perm-again}),
by using as induction measure the pair
$\tuple{\ranku{\Delta},\delta}$, where $\delta$ is the height of
$\Delta$ above the topmost $\dmru$, $\absru$, $\weakru$. This is
similar to the proof of Lemma~\ref{lem:step1}, since $\ranku{\Delta}$
does not always go down in the permutation process. Then we define the
rank of a derivation $\Delta$, denoted by $\rank{\Delta}$, to be the
multiset union of $\ranku{\Delta}$ and $\rankd{\Delta}$. For showing
termination of the whole of Step~2, we need to show that eventually
$\rank{\Delta}$ decreases. We will see that this is indeed the case,
provided that the !-?-flow-graph of $\Delta$ is acyclic (in
Section~\ref{sec:perm-again}).  There are, in principle, two kinds of
cycle in a\bqfg: forked cycles and unforked cycles. First, we show
that the forked cycles disappear during the permutation process (in
Section~\ref{sec:cycles}). Finally (in Section~\ref{sec:nocycles}) we
show that unforked cycles cannot exist inside a !-?-flow-graph, by
using an acyclicity property that has independently been discovered in
the theory of proof nets~\cite{retore:97}.  This then completes the
proof of Theorem~\ref{thm:decomposition}.

\subsection{!-?-Flow-Graphs}\label{sec:flowgraph}

\begin{definition}
  For instances of the rules $\dmrd$, $\dmru$, $\absrd$, $\absru$,
  $\weakrd$, $\weakru$, and $\promrd$, $\promru$ we define their
  \dfn{principal structure} as indicated below with a gray background:
  $$\myssmall
  \begin{array}{c@{\qquad}c@{\qquad}c@{\qqqquad}c}
    \vcinf{\dmrd}{S\cons{\rdx{\lwn T}}}{S\cons{\lwn\lwn T}}
    &
    \vcinf{\absrd}{S\cons{\rdx{\lwn T}}}{S\pars{\lwn T\bc T}}
    &
    \vcinf{\weakrd}{S\cons{\rdx{\lwn T}}}{S\cons{\un}}
    &
    \vcinf{\promrd}{S\pars{\rdx{\loc R}\bc  \lwn T}}{S\cons{\loc\pars{R\bc T}}}
    \\[5ex]
    \vcinf{\dmru}{S\cons{\loc\loc R}}{S\cons{\rdx{\loc R}}}
    &
    \vcinf{\absru}{S\aprs{\loc R\bc R}}{S\cons{\rdx{\loc R}}}
    &
    \vcinf{\weakru}{S\cons{\un}}{S\cons{\rdx{\loc R}}}
    &
    \vcinf{\promru}{S\cons{\lwn\aprs{T\bc R}}}{S\aprs{\rdx{\lwn T}\bc \loc R}}
  \end{array}
  $$ If $\rho\in\set{\dmrd,\absrd,\weakrd}$, then its principal
  structure is the redex $\lwn T$ of $\rho$.  If
  $\rho\in\set{\dmru,\absru,\weakru}$, then its principal structure is
  the contractum $\loc R$ of $\rho$. If $\rho=\promrd$, then its
  principal structure is the $\loc$-substructure of its redex, and if
  $\rho=\promru$, then its principal structure is the
  $\lwn$-substructure of its contractum.
\end{definition}

The basic idea of the \bqfg\ of a derivation is to mark the ``path''
that is taken by the principal structures of instances of $\dmru$,
$\dmrd$, $\absru$, $\absrd$, $\weakru$, $\weakrd$ while they are
traveling up and down in the derivation. Formally, it is
defined as follows:

\begin{definition}\label{def:bqfg}
  Let $\Delta$ be a derivation in $\SNEL$. The \dfn{\bqfg} of $\Delta$
  is a directed graph, denoted by $\fg{\Delta}$, whose vertices are
  the occurrences of $\loc$- and $\lwn$-substructures appearing
  in~$\Delta$. Two such substructures are connected via an edge in
  $\fg{\Delta}$ if they appear inside the premise and the conclusion
  of an inference rule according to the prescriptions in
  Figure~\ref{fig:bqfg}.
\end{definition}

\begin{figure}[!t]
  \begin{center}
    \begin{tabular}{c@{\qquad}c@{\qqqquad\qquad}c}
      &&\\
      (i)
      &
      $\vctpathderivation{
	\inf{\rho}{S'\cons{\noc1 R}}{S\cons{\noc2 R}}}{%
	\slimvec \uvec{oc1}{oc2}}$
      &
      $\vctpathderivation{
	\inf{\rho}{S'\cons{\nwn1 T}}{S\cons{\nwn2 T}}}{%
	\slimvec \dvec{wn2}{wn1}}$
      \\
      &&\\
      (ii)      
      &
      $\vctpathderivation{
	\inf{\rho}{S\cons{\noc1 R\cons{Z}}}{S\cons{\noc2 R\cons{W}}}}{%
	\slimvec \uvec{oc1}{oc2}}$
      &
      $\vctpathderivation{
	\inf{\rho}{S\cons{\nwn1 T\cons{Z}}}{S\cons{\nwn2 T\cons{W}}}}{%
	\slimvec \dvec{wn2}{wn1}}$
      \\
      &&\\
      (iii)
      &
      $\vctpathderivation{
	\inf{\absru}{S\aprs{\noc1 R\bc R}}{S\cons{\noc2 R}}}{%
	\slimvec \uvec{oc1}{oc2}}$
      &
      $\vctpathderivation{
	\inf{\absrd}{S\cons{\nwn1 T}}{S\pars{\nwn2 T\bc T}}}{%
	\slimvec \dvec{wn2}{wn1}}$
      \\
      &&\\
      (iv)
      &
      $\vctpathderivation{
	\inf{\dmru}{S\cons{\noc1\noc3 R}}{S\cons{\noc2 R}}}{%
	\slimvec \uvec{oc1}{oc2} \uvec{oc3}{oc2}}$
      &
      $\vctpathderivation{
	\inf{\dmrd}{S\cons{\nwn1 T}}{S\cons{\nwn2\nwn3 T}}}{%
	\slimvec \dvec{wn2}{wn1} \dvec{wn3}{wn1}}$
      \\
      &&\\
      (v)
      &
      $\vctpathderivation{
	\inf{\absru}{S\aprs{\noc1 V\cons{\noc3 R}\bc V\cons{\noc4 R}}}
	    {S\cons{\noc2 V\cons{\noc5 R}}}}{%
	\slimvec \uvec{oc1}{oc2} \uvec{oc3}{oc5} \uvec{oc4}{oc5} }$
      &
      $\vctpathderivation{
	\inf{\absrd}{S\cons{\nwn1 U\cons{\nwn3 T}}}
	    {S\pars{\nwn2 U\cons{\nwn4 T}\bc U\cons{\nwn5 T}}}}{%
	\slimvec \dvec{wn2}{wn1} \dvec{wn4}{wn3} \dvec{wn5}{wn3} }$
      \\
      &&\\
      (vi)
      &
      $\vctpathderivation{
	\inf{\absrd}{S\cons{\nwn1 U\cons{\noc3 R}}}
	    {S\pars{\nwn2 U\cons{\noc4 R}\bc U\cons{\noc5 R}}}}{%
	\slimvec \dvec{wn2}{wn1} \uvec{oc3}{oc4} \uvec{oc3}{oc5} }$
      &
      $\vctpathderivation{
	\inf{\absru}{S\aprs{\noc1 V\cons{\nwn3 T}\bc V\cons{\nwn4 T}}}
	    {S\cons{\noc2 V\cons{\nwn5 T}}}}{%
	\slimvec \uvec{oc1}{oc2} \dvec{wn5}{wn3} \dvec{wn5}{wn4} }$
      \\
      &&\\
      (vii)
      &
      $\vctpathderivation{
	\inf{\promrd}{S\pars{\noc1 R\bc  \nwn1 T}}{
	  S\cons{\noc2\pars{R\bc T}}}}{%
	\slimvec \uvec{oc1}{oc2} \urdvec{oc1}{wn1} }$
      &
      $\vctpathderivation{
	\inf{\promru}{S\cons{\nwn1\aprs{T\bc R}}}{
	  S\aprs{\nwn2 T\bc \noc2 R}}}{%
	\slimvec \dvec{wn2}{wn1} \druvec{wn2}{oc2} }$
      \\
      &&\\
    \end{tabular}
    \caption{Edges in the \bqfg}
    \label{fig:bqfg}
  \end{center}
\end{figure}

When visualizing the \bqfg\ of a derivation, we draw it inside the
derivation, as indicated in Figure~\ref{fig:bqfg}, and as shown in the
example below (cf.~Remark~\ref{rem:equation}):
\begin{equation}
  \label{eq:short}\myssmall
  \vctpathderivation{
    \dernote{\promru}{}{\nwn2\nwn{12}\aprs{\noc2a\bc a}}{
      \root{\promru}{\nwn3\aprs{\nwn{13}\noc3a\bc\noc{13}a}}{
	\root{\absru}{\aprs{\noc{34}\nwn{14}\noc4a\bc\nwn4\noc{14}a}}{
	  \root{\absrd}{\noc{35}\nwn{15}\noc5a}{
	    \root{\promrd}{\noc{36}\pars{\nwn{16}\noc6a\bc\noc{46}a}}{
	      \leaf{\noc{37}\noc{47}\pars{\noc7a\bc a}}}}}}}}{%
    \slimvec
    \longuvec{oc}{2}{7}
    \longuvec{oc}{13}{14}\uvec{oc14}{oc5}\uvec{oc5}{oc46}\uvec{oc46}{oc47}
    \uldvec{oc46}{wn16}\longdvec{wn}{16}{12}%
    \druvec{wn13}{oc13}
    \dvec{wn15}{wn4}\longdvec{wn}{4}{2}%
    \dluvec{wn4}{oc34}\longuvec{oc}{34}{37}%
  }    
\end{equation}

The first two cases in Figure~\ref{fig:bqfg} are straightforward: The
rule $\rho$ either modifies the context of $\loc R$ or $\lwn T$, or
$\rho$ works inside $\loc R$ or $\lwn T$, without touching the
modality. Cases (iii) and (iv) take care of the modalities that are
actively involved in the redex and contractum of the absorption and
digging rules. Cases (v) and~(vi) involve a duplication of a modality
structure due to absorption, which causes a branching in the \bqfg.
The most interesting case is (vii), because the flow changes its
direction. This corresponds to the introduction of a $\absrd$ in
case~\eqref{cas:nontriv} in \ref{para:perm-fbw}.  Note that in
Figure~\ref{fig:bqfg} the cases (vi) and~(vii) are the only ones where
we have a ``forking'' in the graph. In cases (iv) and~(v) the
situation is better described as ``merging'', and in all other cases
the situation is purely ``linear''.

For two vertices $U$ and $V$ of $\fg{\Delta}$, we write
$\edge[\Delta]{U}{V}$ if there is an edge from $U$ to $V$ in
$\fg{\Delta}$. We use $\pathsym[\Delta]$ to denote the transitive
closure of $\edgesym[\Delta]$, and $\spathsym[\Delta]$ to denote the
reflexive transitive closure of $\edgesym[\Delta]$.
We use the standard notions of paths and
cycles in directed graphs:

\begin{definition}
  A \emph{path} in the \bqfg\ of a derivation $\Delta$ is a sequence
  of vertices $V_0,V_1,\ldots,V_n$, such that
  $\edge[\Delta]{V_{i-1}}{V_i}$ for each $i\in\set{1,\ldots,n}$. A
  \emph{cycle} is a path such that the first vertex and the last
  vertex are identical. The \bqfg\ of a derivation is \emph{acyclic},
  if it does not contain any cycle, i.e., there is no vertex $V$
  with~$\path{V}{V}$. A path $p$ is called \emph{cyclic}, if there is
  a vertex which occurs more than once in~$p$.
\end{definition}

Clearly, every cycle is a cyclic path, and the \bqfg\ of a derivation
is acyclic, if and only if it contains no cyclic path.
To come back to our example in~\eqref{eq:short}, consider the
following four excerpts from its \bqfg:
\begin{equation}\myssmall
  \label{eq:exshort}
  \vctpathderivation{%
    \dernote{\promru}{}{\nwn2\nwn{12}\aprs{\noc2a\bc a}}{
      \root{\promru}{\nwn3\aprs{\nwn{13}\noc3a\bc\noc{13}a}}{
	\root{\absru}{\aprs{\rdx{\noc{34}\nwn{14}\noc4a}\bc\nwn4\noc{14}a}}{
	  \root{\absrd}{\noc{35}\nwn{15}\rdx{\noc5a}}{
	    \root{\promrd}{\noc{36}\pars{\nwn{16}\noc6a\bc\noc{46}a}}{
	      \leaf{\noc{37}\noc{47}\pars{\noc7a\bc a}}}}}}}}{%
    \slimvec
    \uvec{oc5}{oc46}\uldvec{oc46}{wn16}\longdvec{wn}{16}{15}%
    \dvec{wn15}{wn4}\dluvec{wn4}{oc34}%
  }    
  \quad
  \vctpathderivation{
    \dernote{\promru}{}{\nwn2\nwn{12}\aprs{\noc2a\bc a}}{
      \root{\promru}{\nwn3\aprs{\nwn{13}\noc3a\bc\noc{13}a}}{
	\root{\absru}{\aprs{\noc{34}\nwn{14}\noc4a\bc\nwn4\noc{14}a}}{
	  \root{\absrd}{\noc{35}\nwn{15}\noc5a}{
	    \root{\promrd}{\noc{36}\pars{\nwn{16}\noc6a\bc\noc{46}a}}{
	      \leaf{\noc{37}\noc{47}\pars{\noc7a\bc a}}}}}}}}{%
    \slimvec
    \longdvec{wn}{15}{12}%
    \dvec{wn15}{wn4}
  }    
  \quad
  \vctpathderivation{
    \dernote{\promru}{}{\nwn2\nwn{12}\aprs{\noc2a\bc a}}{
      \root{\promru}{\nwn3\aprs{\nwn{13}\noc3a\bc\noc{13}a}}{
	\root{\absru}{\aprs{\noc{34}\nwn{14}\noc4a\bc\nwn4\noc{14}a}}{
	  \root{\absrd}{\noc{35}\nwn{15}\noc5a}{
	    \root{\promrd}{\noc{36}\pars{\nwn{16}\noc6a\bc\noc{46}a}}{
	      \leaf{\noc{37}\noc{47}\pars{\noc7a\bc a}}}}}}}}{%
    \slimvec
    \uvec{oc5}{oc46}\uldvec{oc46}{wn16}\longdvec{wn}{16}{13}%
    \druvec{wn13}{oc13}\longuvec{oc}{13}{14}\uvec{oc14}{oc5}
  }    
  \quad
  \vctpathderivation{
    \dernote{\promru}{}{\nwn2\nwn{12}\aprs{\rdx{\noc2a}\bc a}}{
      \root{\promru}{\nwn3\aprs{\nwn{13}\noc3a\bc\noc{13}a}}{
	\root{\absru}{\aprs{\noc{34}\nwn{14}\noc4a\bc\nwn4\noc{14}a}}{
	  \root{\absrd}{\noc{35}\nwn{15}\noc5a}{
	    \root{\promrd}{\noc{36}\pars{\nwn{16}\noc6a\bc\noc{46}a}}{
	      \leaf{\noc{37}\rdx{\noc{47}\pars{\noc7a\bc a}}}}}}}}}{%
    \slimvec
    \uvec{oc5}{oc46}\uldvec{oc46}{wn16}\longdvec{wn}{16}{13}%
    \druvec{wn13}{oc13}\longuvec{oc}{13}{14}\uvec{oc14}{oc5}%
    \longuvec{oc}{2}{5}\uvec{oc46}{oc47}%
  }    
\end{equation}
The first example shows a path, where the first and the last vertex in
the path are marked with a gray background. The subgraph
indicated in the second example
is not a path (direction matters). The third example shows a cycle,
and the last example a cyclic path (again, first and last vertex are
marked). In particular, the \bqfg\ in~\eqref{eq:short} is not acyclic.

\begin{definition}
  A vertex $V$ in $\fg{\Delta}$ is called a \emph{$\loc$-vertex} if it
  is a $\loc$-structure, and \emph{$\lwn$-vertex} if it is a
  $\lwn$-structure. Note that an edge from a $\loc$-vertex to a
  $\loc$-vertex always goes upwards in a derivation. Hence, we call a
  path that contains only $\loc$-vertices an
  \emph{up-path}. Similarly, a path with only $\lwn$-vertices is
  called a \emph{down-path}. Edges from $\loc$-vertices to
  $\lwn$-vertices or from $\lwn$-vertices to $\loc$-vertices are
  called \emph{flipping edges}. The number of flipping edges in a path
  $p$ is called the \emph{flipping number} of $p$, denoted
  by~$\flip{p}$.
\end{definition}

For example, the path indicated in the leftmost derivation
in~\eqref{eq:exshort} has flipping number~2, and the two paths in the
second derivation in~\eqref{eq:exshort} have both flipping number~0.

\begin{definition}
  Let $\Delta$ be a derivation. A vertex $V$ in $\fg{\Delta}$ is
  called a \dfn{\pee-vertex} if it is the principal structure of a
  $\promrd$ or $\promru$. The vertex $V$ is called a \dfn{\bee-vertex}
  if it is the principal structure of a $\absrd$ or $\absru$. 
\end{definition}

\subsection{The Rank of Rules and Derivations}\label{sec:rank}

In this section we define the rank of a rule instance $\rho$ as a
triple in the set $\set{0,1}\times(\omega+1)\times\omega$ equipped
with the lexicographic ordering, where $\omega=\set{0,1,2,\ldots}$ and
$\omega+1=\omega\cup\set{\omega}$ are both equipped with the natural
ordering. Roughly speaking, the first value is the \emph{status of $\rho$},
indicating whether $\rho$ has already reached its destination at the
top or the bottom of the derivation. The second value, called the
\emph{\pee-number of $\rho$}, is the number of instances $\promrd$ and
$\promru$ that $\rho$ might encounter on its journey. And the third
value, called the \emph{onion \bee-number of $\rho$}, is encoding how often
$\rho$ might get duplicated during the permutation process. This means
we have to count the number of $\absrd$ and $\absru$ that might cause
a duplication of $\rho$, as in case (iii) of~\ref{para:perm-fbw}. To
that end, we need the notions of \emph{onion} and \emph{look-back
  tree}, which are defined below. 

\begin{definition}
  The \dfn{\pee-number} of a path $p$ in $\fg{\Delta}$, denoted by
  $\pe{p}$, is the number of \pee-vertices occurring in $p$.  If $p$
  is cyclic, the vertices with multiple occurrences in $p$ are counted
  as many times as they occur in~$p$.
\end{definition}

For example, the path $p$ indicated in the leftmost example in
\eqref{eq:exshort}, we have $\pe{p}=2$.
The rightmost one has $\pe{p}=3$ if the path passes through the cycle
once, and $\pe{p}=5$ if the path passes through the cycle twice, and
so on.  Note that we do not have $\pe{p}=\flip{p}$ in general. But we
have always $\pe{p}\ge\flip{p}$.

\begin{definition}
  Let $\Delta$ be a derivation and let $V$ be a vertex in
  $\fg{\Delta}$. Then the \emph{\pee-number} of $V$ in $\Delta$,
  denoted by $\pe{V}$, is defined as follows:
  \begin{equation}
    \label{eq:pee}
    \pe{V}=\sup\set{\,\pe{p}\mid \mbox{$p$ is a path starting in $V$}\,}
    \quadfs
  \end{equation}
  For a rule instance $\rho$ in $\Delta$ of the kind $\dmrd$,
  $\absrd$, $\weakrd$, or $\dmru$, $\absru$, $\weakru$, we define its
  \emph{\pee-number}, denoted by $\pe[\Delta]{\rho}$ to be the
  \pee-number of its principal structure.
\end{definition}

In other words, for determining $\pe{V}$, we take the maximum of all
$\pe{p}$, where $p$ ranges over all paths that have $V$ as starting
vertex. If one of these paths is cyclic, then $\pe{V}=\omega$. 

For example, consider again the derivation in~\eqref{eq:short}.  Below
we show it again twice where in each derivation one vertex of the
\bqfg\ is marked. Let us denote them by $V_1$ and $V_2$, respectively.
\begin{equation}\myssmall
  \label{eq:exshort-pee}
  \vctpathderivation{
    \dernote{\promru}{}{\nwn2\nwn{12}\aprs{\noc2a\bc a}}{
      \root{\promru}{\nwn3\aprs{\nwn{13}\noc3a\bc\noc{13}a}}{
	\root{\absru}{\aprs{\noc{34}\nwn{14}\noc4a\bc\rdx{\nwn4\noc{14}a}}}{
	  \root{\absrd}{\noc{35}\nwn{15}\noc5a}{
	    \root{\promrd}{\noc{36}\pars{\nwn{16}\noc6a\bc\noc{46}a}}{
	      \leaf{\noc{37}\noc{47}\pars{\noc7a\bc a}}}}}}}}{%
    \slimvec
    \longdvec{wn}{4}{2}%
    \dluvec{wn4}{oc34}\longuvec{oc}{34}{37}%
  }    
  \qqqqquad
  \vctpathderivation{%
    \dernote{\promru}{}{\nwn2\nwn{12}\aprs{\noc2a\bc a}}{
      \root{\promru}{\nwn3\aprs{\nwn{13}\noc3a\bc\noc{13}a}}{
	\root{\absru}{\aprs{\noc{34}\rdx{\nwn{14}\noc4a}\bc\nwn4\noc{14}a}}{
	  \root{\absrd}{\noc{35}\nwn{15}{\noc5a}}{
	    \root{\promrd}{\noc{36}\pars{\nwn{16}\noc6a\bc\noc{46}a}}{
	      \leaf{\noc{37}\noc{47}\pars{\noc7a\bc a}}}}}}}}{%
    \slimvec
    \longuvec{oc}{5}{7}
    \longuvec{oc}{13}{14}\uvec{oc14}{oc5}\uvec{oc5}{oc46}\uvec{oc46}{oc47}
    \uldvec{oc46}{wn16}\longdvec{wn}{16}{12}%
    \druvec{wn13}{oc13}
    \dvec{wn15}{wn4}\longdvec{wn}{4}{2}%
    \dluvec{wn4}{oc34}\longuvec{oc}{34}{37}%
  }    
\end{equation}
On the left, we have shown all paths starting in $V_1$. There are only
two of them, and both have \pee-number 1 (because $V_1$ is their only
\pee-vertex). Hence $\pe{V_1}=1$. On the right we have shown all paths
starting in $V_2$. Because of the cycle, we have $\pe{V_2}=\omega$.

\begin{definition}\label{def:lbt}
  Let $\Delta$ be a derivation. A \emph{look-back tree} $t$ in
  $\Delta$ is a subgraph of $\fg{\Delta}$ such that
  \begin{itemize}[---]
  \item $t$ is a directed tree whose edges are directed towards the root, 
  \item every path from a leaf to the root in $t$ contains at most one
    flipping edge, and
  \item every branching vertex of $t$ (i.e.,
    every vertex with two incoming edges) is the principal structure of
    an instance of $\dmrd$ or $\dmru$. 
  \end{itemize}
  The \dfn{\bee-number} of a
  look-back tree $t$, denoted by $\be{t}$, is the number of
  \bee-vertices occurring in $t$.
\end{definition}

Note that because of the restriction of the flipping number of paths
in $t$ to 1, a look-back tree cannot be cyclic.

Consider for example the following derivations in which we exhibited
subgraphs of the \bqfg.
\begin{equation}\myssmall
  \label{eq:exa-lbt}
  \vctpathderivation{
    \dernote{\absru}{}{\pars{\aprs{\noc1a\bc a}\bc \rdx{\nwn1b}}}{
      \root{\dmrd}{\pars{\rdx{\noc2a}\bc \nwn2b}}{
	\root{\absrd}{\pars{\noc3a\bc \nwn{13}\rdx{\nwn3b}}}{
	  \root{\promrd}{\pars{\noc4a\bc \nwn{14}\pars{\nwn4b\bc b}}}{
	    \leaf{\noc5\pars{a\bc \nwn5b\bc b}}}}}}}{%
    \slimvec
    \longdvec{wn}{5}{1}%
    \longuvec{oc}{1}{4}%
    \uldvec{oc4}{wn14}\dvec{wn14}{wn13}\dvec{wn13}{wn2}
  } 
  \qqquad
  \vctpathderivation{
    \dernote{\promru}{}{\pars{\nwn{11}\aprs{a\bc b}\bc \nwn1c}}{
      \root{\swir}{\pars{\aprs{\nwn{12}a\bc \noc2b}\bc \nwn2c}}{
	\root{\promrd}{\aprs{\nwn{13}a\bc \pars{\noc3b\bc \nwn3c}}}{
	  \leaf{\aprs{\nwn{14}a\bc \noc4\pars{b\bc c}}}}}}}{%
    \slimvec
    \longdvec{wn}{14}{12}%
    \dluvec{wn12}{oc2}%
    \vecangles{oc2}{oc3}{60}{-120}%
    \uldvec{oc3}{wn3}
    \longdvec{wn}{3}{1}%
  }    
  \qqquad
  \vctpathderivation{
    \dernote{\dmrd}{}{\nwn1a}{
      \root{\absrd}{\nwn{12}\rdx{\nwn2a}}{
	\root{\dmrd}{\pars{\nwn{13}\nwn3a\bc\nwn{23}a}}{
	  \root{\dmrd}{\pars{\nwn{14}\nwn4a\bc\nwn{34}\nwn{24}a}}{
  	    \leaf{\pars{\nwn{45}\nwn{15}\nwn5a\bc\nwn{35}\nwn{25}a}}}}}}}{%
    \slimvec
    \longdvec{wn}{5}{1}%
    \longdvec{wn}{15}{12}\dvec{wn12}{wn1}%
    \longdvec{wn}{25}{23}\dvec{wn23}{wn2}%
    \longdvec{wn}{35}{34}\dvec{wn34}{wn23}\dvec{wn45}{wn14}%
  }
\end{equation}
On the left we have a look-back tree, and its \bee-number is two. Its
root and the two \bee-vertices are marked with a gray background. The
second example in~\eqref{eq:exa-lbt} is not a look-back tree because
of the two flippings in the path. The third example is not a look-back
because there is a branching vertex (marked with gray background) that
is not the principal structure of an instance of $\dmrd$ or $\dmru$.

\begin{definition}
  Let $\Delta$ be a derivation and let $V$ be a vertex in
  $\fg{\Delta}$. We define the \dfn{\bee-number} of $V$, denoted by
  $\be{V}$, as follows:
  \begin{equation}
    \label{eq:bee}
    \be{V}=\sup\set{\,\be{t}\mid 
      \mbox{$t$ is a look-back tree with root $V$}\,}
    \quadfs
  \end{equation}
\end{definition}

Note that for the \pee-number of a
vertex, we look forward in the graph, and for the \bee-number we look
backwards. Furthermore, for the \bee-number we consider only paths
with flipping number $0$ or $1$, and we allow branchings as in case
(iv) of Figure~\ref{fig:bqfg}, but never as in cases (v), (vi), and
(vii) of that Figure.

To see some example, consider again the rightmost derivation
in~\eqref{eq:exa-lbt}. Let us denote the $\lwn a$-occurrence in the
conclusion by $V_3$. The first two derivations below
in~\eqref{eq:exshort-bee} show two look-back tree with $V_3$ as
root. The third derivation shows a look-back tree of the
$\loc\loc\pars{\loc a\bc a}$-vertex in the premise of the derivation
in~\eqref{eq:short}. Let us denote that vertex by~$V_4$.
\begin{equation}\myssmall
  \label{eq:exshort-bee}
  \vctpathderivation{
    \dernote{\dmrd}{}{\rdx{\nwn1a}}{
      \root{\absrd}{\nwn{12}\nwn2a}{
	\root{\dmrd}{\pars{\nwn{13}\nwn3a\bc\nwn{23}a}}{
	  \root{\dmrd}{\pars{\nwn{14}\nwn4a\bc\nwn{34}\nwn{24}a}}{
  	    \leaf{\pars{\nwn{45}\nwn{15}\nwn5a\bc\nwn{35}\nwn{25}a}}}}}}}{%
    \slimvec
    \longdvec{wn}{5}{1}%
    \longdvec{wn}{15}{12}\dvec{wn12}{wn1}%
    \dvec{wn45}{wn14}%
  }
  \qquad
  \vctpathderivation{
    \dernote{\dmrd}{}{\rdx{\nwn1a}}{
      \root{\absrd}{\nwn{12}\nwn2a}{
	\root{\dmrd}{\pars{\nwn{13}\nwn3a\bc\nwn{23}a}}{
	  \root{\dmrd}{\pars{\nwn{14}\nwn4a\bc\nwn{34}\nwn{24}a}}{
  	    \leaf{\pars{\nwn{45}\nwn{15}\nwn5a\bc\nwn{35}\nwn{25}a}}}}}}}{%
    \slimvec
    \longdvec{wn}{2}{1}%
    \longdvec{wn}{25}{23}\dvec{wn23}{wn2}%
    \longdvec{wn}{35}{34}\dvec{wn34}{wn23}%
  }
  \qqqqqquad
  \vctpathderivation{
    \dernote{\promru}{}{\nwn2\nwn{12}\aprs{\noc2a\bc a}}{
      \root{\promru}{\nwn3\aprs{\nwn{13}\noc3a\bc\noc{13}a}}{
	\root{\absru}{\aprs{\noc{34}\nwn{14}\noc4a\bc\nwn4\noc{14}a}}{
	  \root{\absrd}{\noc{35}\nwn{15}\noc5a}{
	    \root{\promrd}{\noc{36}\pars{\nwn{16}\noc6a\bc\noc{46}a}}{
	      \leaf{\rdx{\noc{37}\noc{47}\pars{\noc7a\bc a}}}}}}}}}{%
    \slimvec
    \longdvec{wn}{16}{15}%
    \dvec{wn15}{wn4}
    \dluvec{wn4}{oc34}\longuvec{oc}{34}{37}%
  }    
\end{equation}
We have $\be{V_3}=1$. The first look-back tree has \bee-number 1 and
the second one has \bee-number 0. We have $\be{V_4}=2$ because both
instances, $\absru$ and $\absrd$ have their principal structure as
vertex in the indicated look-back tree. 

\begin{definition}
  Let $\Delta$ be a derivation, and $\loc R$ be a $\loc$-vertex in
  $\fg{\Delta}$, i.e., also a substructure of $\Delta$. The
  \emph{onion} of $\loc R$, denoted by $\onion{\loc R}$, is the set of
  all $\lwn$-vertices in $\fg{\Delta}$ that have exactly this $\loc R$
  as substructure. This means, in particular, that they appear in the
  same line of $\Delta$ as $\loc R$. Dually, we define the
  \emph{onion} of a $\lwn$-vertex $\lwn T$, denoted by $\onion{\lwn
    T}$, to be the set of all $\loc$-vertices that have this structure
  $\lwn T$ as substructure. For every rule instance $\rho$ in $\Delta$
  of the kind $\dmrd$, $\absrd$, $\weakrd$, or $\dmru$, $\absru$,
  $\weakru$, we define its \emph{onion} $\onion[\Delta]{\rho}$ in
  $\Delta$ to be the onion of its principal structure. The \emph{onion
    \bee-number} of $\rho$ in $\Delta$, denoted by
  $\onionbe[\Delta]{\rho}$, is the sum of the \bee-numbers of the
  vertices in its onion, i.e.,
  $$
  \onionbe[\Delta]{\rho}=\sum_{V\in\onion[\Delta]{\rho}}\be{V}
  \quadfs
  $$
\end{definition}

For example, consider the bottommost occurrence of $\loc a$ in the
derivation in~\eqref{eq:short}. It is marked in the rightmost
derivation in~\eqref{eq:exshort}. Its onion consists of the two
?-structures in the conclusion of the derivation. Both have
\bee-number 1. Hence, the onion \bee-number of a rule that had this
$\loc a$ as principal structure would be 2.

Finally, we define the \emph{status} of a rule instance to be either 0
or~1, such that it is~1 if the rule is of the kind $\dmrd$, $\absrd$,
$\weakrd$, or $\dmru$, $\absru$, $\weakru$, and not yet at its final
destination at the top or the bottom of the derivation. The status
is~0 if the rule does not play any further role in the
up-down-permutation. The motivation of this is that Step~2 of our
decomposition process (see Figure~\ref{fig:third}) is completed if and
only if all rules instances in the derivation have status~0.
Formally, the status is defined as follows.
 
\begin{definition}
  Let $\SNEL'=\SNEL\setminus\set{\erd,\eru}$, let $\Delta$ be a
  derivation in $\SNEL'$, and let $\rho$ be a rule instance inside
  $\Delta$. Then $\rho$ splits $\Delta$ into two parts:
  $$\mysmall
  \vcstrder{\SNEL'}{\Delta_2}{P}{
    \root{\rho}{S\cons{Z}}{
      \stem{\SNEL'}{\Delta_1}{S\cons{W}}{
	\leaf{Q}}}}
  $$ The \emph{status} of $\rho$ in $\Delta$, denoted by
  $\status[\Delta]{\rho}$ is 1 if we have one of the following two
  cases:
  \begin{itemize}[---]
  \item $\rho\in\set{\dmru,\absru,\weakru}$ and $\Delta_1$ contains
    an instance of a rule in $\SNEL'\setminus\set{\dmru,\absru,\weakru}$, or
  \item $\rho\in\set{\dmrd,\absrd,\weakrd}$ and $\Delta_2$ contains
    an instance of a rule in $\SNEL'\setminus\set{\dmrd,\absrd,\weakrd}$.
  \end{itemize}
  Otherwise $\status[\Delta]{\rho}=0$.
\end{definition}

The reason for using
$\SNEL'$ is that the rules $\erd$ and $\eru$ are not considered in
Step~2 of Figure~\ref{fig:third}. However, all statements in this
section about $\SNEL'$ are also valid for $\SNEL$.

Now we are using the status, the \pee-number, and the
onion \bee-number of a rule instance to define its rank.

\begin{definition}
  For a rule instance $\rho$ of the kind $\dmrd$, $\absrd$, $\weakrd$
  or $\dmru$, $\absru$, $\weakru$ inside a derivation $\Delta$, we
  define its \dfn{rank}
  $\rank[\Delta]{\rho}\in\set{0,1}\times(\omega+1)\times\omega$ to be
  the triple
  $$
  \rank[\Delta]{\rho}=
  \tuple{\status[\Delta]{\rho},\pe[\Delta]{\rho},\onionbe[\Delta]{\rho}}
  \quadfs
  $$
  For the whole of $\Delta$, we define the \dfn{rank}
  $\rank{\Delta}$ to be the multiset of the ranks of its occurrences of
  $\dmrd,\absrd,\weakrd,\dmru,\absru,\weakru$, i.e.,
  $$ \rank{\Delta}=\mset{\;\rank[\Delta]{\rho}\mid\rho\mathrm{~in~}\Delta
    \mathrm{~and~}\rho\mathrm{~is~one~of~}
    \dmrd,\absrd,\weakrd,\dmru,\absru,\weakru\;}
  \quadfs
  $$
  We define the \dfn{down-rank} of $\Delta$, denoted by
  $\rankd{\Delta}$ by considering only
  the down-rules $\dmrd,\absrd,\weakrd$ in the multiset:
  $$
  \rankd{\Delta}=\mset{\;\rank[\Delta]{\rho}\mid
    \rho\mathrm{~in~}\Delta\mathrm{~and~}
    \rho\mathrm{~is~one~of~}\dmrd,\absrd,\weakrd\;}
  \quadfs
  $$
  Similarly, the \dfn{up-rank} $\ranku{\Delta}$ takes only the up-rules
  $\dmru,\absru,\weakru$ into account:
  $$
  \ranku{\Delta}=\mset{\;\rank[\Delta]{\rho}\mid
    \rho\mathrm{~in~}\Delta\mathrm{~and~}
    \rho\mathrm{~is~one~of~}\dmru,\absru,\weakru\;}
  \quadfs
  $$
\end{definition}

It follows immediately from the definition that
$\rank{\Delta}=\rankd{\Delta}\uplus\ranku{\Delta}$. For example,
in~\eqref{eq:short}, we have that the rank of the $\absrd$ instance is
$\tuple{1,\omega,1}$ and the rank of the $\absru$ instance is
$\tuple{1,0,0}$.  Hence, the rank of the whole derivation is the
multiset $\mset{\;\tuple{1,\omega,1}\;,\;\tuple{1,0,0}\;}$.

\subsection{Permutations Again}\label{sec:perm-again}

In this section we will first analyze the impact of the rule
permutations needed for Step~2 to the rank of the derivation, and see
that Step~2 terminates if the \bqfg\ of the derivation is acyclic.
Now, let us consider again the cases in \ref{para:perm-fbw}. We begin with
the trivial cases (cf.\ the proof of Lemma~\ref{lem:perm-er}).

\begin{casana}[(for permuting $\dmru$, $\absru$, $\weakru$
    up)]\label{para:perm-fbw-II} 
  Let a derivation $\Delta$ be given. As in \ref{para:perm-fbw}, 
  Consider a subderivation
  \begin{equation}\mysmall
    \label{eq:delta}
    \vcdernote
	{\pi}{}
	{P}
	{\root
	  {\rho}
	  {S\cons{Z}}
	  {\leaf
	    {S\cons{W}}
	}}
	\quadcm
  \end{equation}
  where $\rho\in\SNEL'\setminus\set{\dmru,\absru,\weakru}$
  and $\pi\in\set{\dmru,\absru,\weakru}$. In the following case
  analysis we replace (as done in Section~\ref{sec:permrules}) in
  $\Delta$ the subderivation in~\eqref{eq:delta} by a new
  subderivation with the same premise and conclusion. We use $\Delta'$
  to denote the result of this replacement.
  \begin{enumerate}[(i)]
  \item The contractum $\loc R$ of $\pi$ is inside the
    context $S\conhole$. Here is an example with $\pi=\dmru$ and
    $\rho=\swir$:
    \begin{equation*}\mysmall
      \vctpathderivation{
	\dernote{\dmru}{}{\nSp1\cons{\noc1\noc{11}\nR1}\cons{
	      \pars{\aprs{\nP1\bc \nT1}\bc \nU1}}}{
	  \root{\swir}{\nSp2\cons{\noc2\nR2}\cons{
		\pars{\aprs{\nP2\bc \nT2}\bc \nU2}}}{
	    \leaf{\nSp3\cons{\noc3\nR3}
	      \cons{\aprs{\pars{\nP3\bc \nU3}\bc \nT3}}}}}}{%
	\boldvec
	\longline{S}{1}{3}%
	\longline{R}{1}{3}%
	\longline{P}{1}{3}%
	\longline{T}{1}{3}%
	\longline{U}{1}{3}%
	\slimvec
	\longuvec{oc}{1}{3}%
	\uvec{oc11}{oc2}
      }
      \quadto
      \vctpathderivation{
	\dernote{\swir}{}{\nSp1\cons{\noc1\noc{11}\nR1}\cons{
	      \pars{\aprs{\nP1\bc \nT1}\bc \nU1}}}{
	  \root{\dmru}{\nSp2\cons{\noc2\noc{12}\nR2}\cons{
		\aprs{\pars{\nP2\bc \nU2}\bc \nT2}}}{
	    \leaf{\nSp3\cons{\noc3\nR3}
	      \cons{\aprs{\pars{\nP3\bc \nU3}\bc \nT3}}}}}}{%
	\boldvec
	\longline{S}{1}{3}%
	\longline{R}{1}{3}%
	\longline{P}{1}{3}%
	\longline{T}{1}{3}%
	\longline{U}{1}{3}%
	\slimvec
	\longuvec{oc}{1}{3}%
	\uvec{oc11}{oc12} \uvec{oc12}{oc3}
      }
    \end{equation*}
    Here, we used $S'\conhole\conhole$ to denote a context with two
    independent holes, and we used bold light lines to indicate
    bunches of parallel paths going through the derivation.  Clearly,
    in this case, neither $\pe[\Delta]{\pi}$ nor
    $\onionbe[\Delta]{\pi}$ change their value (but
    $\status[\Delta]{\pi}$ could go down). The important fact to
    observe is that the rank of all other rules in $\Delta$ remains
    unchanged in $\Delta'$.  Hence, $\ranku{\Delta'}\le\ranku{\Delta}$
    and $\rankd{\Delta'}=\rankd{\Delta}$.
  \item The contractum $\loc R$ of $\pi$ appears inside the redex $Z$
    of $\rho$, but only inside a substructure of $Z$ that is not
    affected by $\rho$. Again, we exhibit an example with $\pi=\dmru$ and
    $\rho=\swir$:
    \begin{equation*}\mysmall\label{eq:triv2}
      \vctpathderivation{
	\dernote{\dmru}{}{
	  \nS1\pars{\aprs{\nP1\cons{\noc1\noc{11}\nR1}\bc \nT1}\bc 
	    \nU1}}{
	  \root{\swir}{
	    \nS2\pars{\aprs{\nP2\cons{\noc2 \nR2}\bc \nT2}\bc \nU2}}{
	    \leaf{\nS3\aprs{\pars{\nP3\cons{\noc3 \nR3}\bc \nU3}\bc \nT3}}}}}{%
	\boldvec
	\longline{S}{1}{3}%
	\longline{R}{1}{3}%
	\longline{P}{1}{3}%
	\longline{T}{1}{3}%
	\longline{U}{1}{3}%
	\slimvec
	\longuvec{oc}{1}{3}%
	\uvec{oc11}{oc2}
      }
      \quadto
      \vctpathderivation{
	\dernote{\swir}{}{
	  \nS1\pars{\aprs{\nP1\cons{\noc1\noc{11}\nR1}\bc \nT1}\bc 
	    \nU1}}{
	  \root{\dmru}{
	    \nS2\aprs{\pars{\nP2\cons{\noc2\noc{12}\nR2}\bc \nU2}\bc \nT2}}{
	    \leaf{\nS3\aprs{\pars{\nP3\cons{\noc3 \nR3}\bc \nU3}\bc \nT3}}}}}{%
	\boldvec
	\longline{S}{1}{3}%
	\longline{R}{1}{3}%
	\longline{P}{1}{3}%
	\longline{T}{1}{3}%
	\longline{U}{1}{3}%
	\slimvec
	\longuvec{oc}{1}{3}%
	\uvec{oc11}{oc12} \uvec{oc12}{oc3}
      }
    \end{equation*}
    As in the previous case, the values of $\pe[\Delta]{\pi}$ and
    $\onionbe[\Delta]{\pi}$ are not affected. This is trivial for
    $\rho\in\set{\swir,\seqrd,\seqru}$, and we leave it as an
    instructive exercise to the reader to verify it also for
    $\rho=\promrd$. For $\rho=\promru$, the value of
    $\pe[\Delta]{\pi}$ remains unchanged, but $\onionbe[\Delta]{\pi}$
    could go down. As in the previous case, $\status[\Delta]{\pi}$
    could go down, and the rank of all other rules in $\Delta$ remains
    unchanged in $\Delta'$.  Hence, $\ranku{\Delta'}\le\ranku{\Delta}$
    and $\rankd{\Delta'}=\rankd{\Delta}$.
  \item The redex $Z$ of $\rho$ is inside the contractum $\loc R$ of
    $\pi$. 
    \begin{enumerate}[(a)]
    \item If $\pi=\weakru$, then 
      $$\mysmall
      \vctpathderivation{
        \dernote{\weakru}{}{\nSp1\cons{\un}}{
          \root{\rho}{\nSp2\cons{\noc2\nR2\cons{\nZ2}}}{
            \leaf{\nSp3\cons{\noc3\nR3\cons{\nW3}}}}}
      }{%
	  \boldvec%
	  \longline{S}{1}{3}%
	  \longline{R}{2}{3}\uline{Z2}{W3}%
	  \slimvec%
	  \longuvec{oc}{2}{3}%
      }
      \qquadto
      \vctpathderivation{
        \dernote{\weakru}{}{\nSp1\cons{\un}}{
          \leaf{\nSp2\cons{\noc2\nR2\cons{\nW2}}}}
      }{%
	  \boldvec%
	  \longline{S}{1}{2}%
      }
      $$
      We have $\rank{\Delta'}\le\rank{\Delta}$ because $\rho$ is
      removed.
    \item If $\pi=\dmru$, then 
      $$\mysmall
      \vctpathderivation{
        \dernote{\dmru}{}{\nSp1\cons{\noc1\noc{11}\nR1\cons{\nZ1}}}{
          \root{\rho}{\nSp2\cons{\noc2\nR2\cons{\nZ2}}}{
            \leaf{\nSp3\cons{\noc3\nR3\cons{\nW3}}}}}
      }{%
	  \boldvec%
	  \longline{S}{1}{3}%
	  \longline{R}{1}{3}\uline{Z1}{Z2}\uline{Z2}{W3}
	  \slimvec 
	  \longuvec{oc}{1}{3}%
          \uvec{oc11}{oc2}
      }
      \qquadto
      \vctpathderivation{
        \dernote{\rho}{}{\nSp1\cons{\noc1\noc{11}\nR1\cons{\nZ1}}}{
          \root{\dmru}{\nSp2\cons{\noc2\noc{12}\nR2\cons{\nW2}}}{
            \leaf{\nSp3\cons{\noc3\nR3\cons{\nW3}}}}}
      }{%
	  \boldvec%
	  \longline{S}{1}{3}%
	  \longline{R}{1}{3}\uline{Z1}{W2}\uline{W2}{W3}
	  \slimvec 
	  \longuvec{oc}{1}{3}%
          \uvec{oc11}{oc12}\uvec{oc12}{oc3}
      }
      $$ Note that the onion of $\rho$ is changed if $\rho$ is an
      instance of $\weakrd$, $\absrd$, or $\dmrd$, because the number
      of $\loc$ in its context increased. But the \bee-number of the
      $\loc$-vertex in the premise of the derivations above is the
      same as the sum of the \bee-numbers of the two $\loc$-vertices
      in the conclusion. Hence, the onion \bee-number of $\rho$ does
      not change. Therefore $\rank{\Delta'}\le\rank{\Delta}$.
    \item If $\pi=\absru$, then the situation is not entirely trivial,
      because $\rho$ gets duplicated:
	$$\mysmall
	\vctpathderivation{
	  \dernote{\absru}{}{
	    \nSp1\aprs{\noc1\nR1\cons{\nZ1}\bc\nR{11}\cons{\nZ{11}}}}{
	    \root{\rho}{
	      \nSp2\cons{\noc2\nR2\cons{\nZ2}}}{
	      \leaf{
		\nSp3\cons{\noc3\nR3\cons{\nW3}}}}}}{%
	  \boldvec
	  \longline{S}{1}{3}%
	  \longline{R}{1}{3} \uline{R11}{R2}
          \uline{Z1}{Z2}\uline{Z11}{Z2}\uline{Z2}{W3}
	  \slimvec 
	  \longuvec{oc}{1}{3}%
	}
	\quadto
	\vctpathderivation{
	  \dernote{\rho}{}{
	    \nSp0\aprs{\noc0\nR0\cons{\nZ0}\bc\nR{10}\cons{\nZ{10}}}}{
	    \root{\rho}{
	      \nSp1\aprs{\noc1\nR1\cons{\nW1}\bc\nR{11}\cons{\nZ{11}}}}{
	      \root{\absru}{
		\nSp2\aprs{\noc2\nR2\cons{\nW2}\bc\nR{12}\cons{\nW{12}}}}{
		\leaf{
		  \nSp3\cons{\noc3\nR3\cons{\nW3}}}}}}}{%
	  \boldvec
	  \longline{S}{0}{3}
	  \longline{R}{0}{3} \longline{R}{10}{12} \uline{R12}{R3}
	  \uline{Z0}{W1}\longline{W}{1}{3}
          \uline{Z10}{Z11}\uline{Z11}{W12}\uline{W12}{W3}
	  \slimvec 
	  \longuvec{oc}{0}{3}%
	}
	$$
        We distinguish the following cases:
      \begin{enumerate}[(1)]
      \item If $\rho$ does not involve any modalities, i.e.,
	$\rho\in\set{\atird,\atiru,\swir,\seqrd,\seqru}$, then
	situation is similar to cases (i) and~(ii) above. No
        rule changes its rank, except that we could have that
	the status of the $\absru$ goes down. Hence, we have
	$\ranku{\Delta'}\le\ranku{\Delta}$ and
	$\rankd{\Delta'}=\rankd{\Delta}$. 
      \item If $\rho=\promrd$, then 
	\begin{equation*}\mysmall\label{eq:bp}
	\vctpathderivation{
	  \dernote{\absru}{}{
	    \nSp1\aprs{\noc1\nR1\pars{\noc{11}\nP1\bc \nwn1\nT1}\bc 
	      \nR{21}\pars{\noc{21}\nP{21}\bc \nwn{21}\nT{21}}}}{
	    \root{\promrd}{
	      \nSp2\cons{\noc2\nR2\pars{\noc{12}\nP2\bc \nwn2\nT2}}}{
	      \leaf{
		\nSp3\cons{\noc3\nR3\cons{\noc{13}\pars{\nP3\bc \nT3}}}}}}}{%
	  \boldvec
	  \longline{S}{1}{3}%
	  \longline{R}{1}{3} \uline{R21}{R2}
	  \longline{T}{1}{3} \uline{T21}{T2}
	  \longline{P}{1}{3} \uline{P21}{P2}
	  \slimvec
	  \longuvec{oc}{1}{3}%
	  \longuvec{oc}{11}{13} \uvec{oc21}{oc12}
	  \longdvec{wn}{2}{1} \dvec{wn2}{wn21}
	  \urdvec{oc12}{wn2}
	}
	\to
	\vctpathderivation{
	  \dernote{\promrd}{}{
	    \nSp0\aprs{\noc0\nR0\pars{\noc{10}\nP0\bc 
		\nwn0\nT0}\bc\nR{20}\pars{\noc{20}\nP{20}\bc\nwn{20}\nT{20}}}}{
	    \root{\promrd}{
	      \nSp1\aprs{\noc1\nR1\cons{\noc{11}\pars{\nP1\bc\nT1}}\bc 
		\nR{21}\pars{\noc{21}\nP{21}\bc\nwn{21}\nT{21}}}}{
	      \root{\absru}{
		\nSp2\aprs{\noc2\nR2\cons{\noc{12}\pars{\nP2\bc \nT2}}\bc 
		  \nR{22}\cons{\noc{22}\pars{\nP{22}\bc \nT{22}}}}}{
		\leaf{
		  \nSp3\cons{\noc3\nR3\cons{\noc{13}\pars{\nP3\bc \nT3}}}}}}}}{%
	  \boldvec
	  \longline{S}{0}{3} 
	  \longline{R}{0}{3} \longline{R}{20}{22} \uline{R22}{R3}
	  \longline{T}{0}{3} \longline{T}{20}{22} \uline{T22}{T3}
	  \longline{P}{0}{3} \longline{P}{20}{22} \uline{P22}{P3}
	  \slimvec
	  \longuvec{oc}{0}{3}
	  \longuvec{oc}{10}{13} \longuvec{oc}{20}{22} \uvec{oc22}{oc13}
	  \urdvec{oc10}{wn0} \urdvec{oc21}{wn21} \dvec{wn21}{wn20}
	}
	\end{equation*} 
	As before, $\pe[\Delta]{\pi}$ and $\onionbe[\Delta]{\pi}$ do
        not change. However, the \pee-number, as well as the onion
        \bee-number of other rules might go down because some paths
        disappear (for example the one from the left $\loc P$ to the
        right $\lwn T$. Hence, $\ranku{\Delta'}\le\ranku{\Delta}$ and
        $\rankd{\Delta'}\le\rankd{\Delta}$.
      \item If  $\rho=\promru$, then 
	$$\myvsmall
	\vctpathderivation{
	  \dernote{\absru}{}{
	    \nSp1\aprs{\noc1\nR1\cons{\nwn1\aprs{\nT1\bc\nP1}}\bc 
	      \nR{21}\cons{\nwn{21}\aprs{\nT{21}\bc\nP{21}}}}}{
	    \root{\promru}{
	      \nSp2\cons{\noc2\nR2\cons{\nwn{2}\aprs{\nT2\bc\nP2}}}}{
	      \leaf{
		\nSp3\cons{\noc3\nR3\aprs{\nwn{3}\nT3\bc\noc{13}\nP3}}}}}}{%
	  \boldvec
	  \longline{S}{1}{3} 
	  \longline{R}{1}{3} \uline{R21}{R2}
	  \longline{T}{1}{3} \uline{T21}{T2}
	  \longline{P}{1}{3} \uline{P21}{P2}
	  \slimvec
	  \longuvec{oc}{1}{3}
	  \longdvec{wn}{3}{1} \dvec{wn2}{wn21}
	  \druvec{wn3}{oc13}
	}
	\to
	\vctpathderivation{
	  \dernote{\promru}{}{
	    \nSp0\aprs{\noc0\nR0\cons{\nwn0\aprs{\nT0\bc\nP0}}\bc 
	      \nR{20}\cons{\nwn{20}\aprs{\nT{20}\bc\nP{20}}}}}{
	    \root{\promru}{
	      \nSp1\aprs{\noc1\nR1\aprs{\nwn1\nT1\bc\noc{11}\nP1}\bc 
		\nR{21}\cons{\nwn{21}\aprs{\nT{21}\bc\nP{21}}}}}{
	      \root{\absru}{
		\nSp2\aprs{\noc2\nR2\aprs{\nwn2\nT2\bc\noc{12}\nP2}\bc 
		  \nR{22}\aprs{\nwn{22}\nT{22}\bc\noc{22}\nP{22}}}}{
		\leaf{
		  \nSp3\cons{\noc3\nR3\aprs{\nwn{3}\nT3\bc\noc{13}\nP3}}}}}}}{%
	  \boldvec
	  \longline{S}{0}{3} 
	  \longline{R}{0}{3} \longline{R}{20}{22} \uline{R22}{R3}
	  \longline{T}{0}{3} \longline{T}{20}{22} \uline{T22}{T3}
	  \longline{P}{0}{3} \longline{P}{20}{22} \uline{P22}{P3}
	  \slimvec
	  \longuvec{oc}{0}{3}
	  \longuvec{oc}{11}{13} \longuvec{oc}{11}{13} \uvec{oc22}{oc13}
	  \longdvec{wn}{3}{0}\dvec{wn3}{wn22}\longdvec{wn}{22}{20}
	  \druvec{wn1}{oc11} \druvec{wn22}{oc22}
	}
	$$ Again, neither $\pe[\Delta]{\pi}$ nor
	$\onionbe[\Delta]{\pi}$ can change (but $\status[\Delta]{\pi}$
	could go down). No other rule in $\Delta$ changes its rank.
	Although the $\promru$-instance is duplicated, no path
	changes its \pee-number or its \bee-number.  Hence,
	$\ranku{\Delta'}\le\ranku{\Delta}$ and
	$\rankd{\Delta'}=\rankd{\Delta}$.
      \item Finally, we have to consider the case where
	$\rho\in\set{\dmrd,\absrd,\weakrd}$. We show only the case for
	$\rho=\dmrd$: 
	$$\mysmall
	\vctpathderivation{
	  \dernote{\absru}{}{
	    \nSp1\aprs{\noc1\nR1\cons{\nwn1\nT1}\bc 
	      \nR{21}\cons{\nwn{21}\nT{21}}}}{
	    \root{\dmrd}{
	      \nSp2\cons{\noc2\nR2\cons{\nwn{2}\nT2}}}{
	      \leaf{
		\nSp3\cons{\noc3\nR3\aprs{\nwn{3}\nwn{13}\nT3}}}}}}{%
	  \boldvec
	  \longline{S}{1}{3} 
	  \longline{R}{1}{3} \uline{R21}{R2}
	  \longline{T}{1}{3} \uline{T21}{T2}
	  \slimvec
	  \longuvec{oc}{1}{3}
	  \longdvec{wn}{3}{1} \dvec{wn2}{wn21}
	  \dvec{wn13}{wn2}
	}
	\quadto
	\vctpathderivation{
	  \dernote{\dmrd}{}{
	    \nSp0\aprs{\noc0\nR0\cons{\nwn0\nT0}\bc 
	      \nR{20}\cons{\nwn{20}\nT{20}}}}{
	    \root{\dmrd}{
	      \nSp1\aprs{\noc1\nR1\cons{\nwn1\nwn{11}\nT1}\bc 
		\nR{21}\cons{\nwn{21}\nT{21}}}}{
	      \root{\absru}{
		\nSp2\aprs{\noc2\nR2\cons{\nwn2\nwn{12}\nT2}\bc 
		  \nR{22}\cons{\nwn{22}\nwn{32}\nT{22}}}}{
		\leaf{
		  \nSp3\cons{\noc3\nR3\aprs{\nwn{3}\nwn{13}\nT3}}}}}}}{%
	  \boldvec
	  \longline{S}{0}{3} 
	  \longline{R}{0}{3} \longline{R}{20}{22} \uline{R22}{R3}
	  \longline{T}{0}{3} \longline{T}{20}{22} \uline{T22}{T3}
	  \slimvec
	  \longuvec{oc}{0}{3}
	  \longdvec{wn}{3}{0} 
	  \longdvec{wn}{13}{11}\dvec{wn11}{wn0}
	  \dvec{wn3}{wn22}\longdvec{wn}{22}{20}
	  \dvec{wn13}{wn32}\dvec{wn32}{wn21}
	  }
	$$ Again, neither $\pe[\Delta]{\pi}$ nor
	$\onionbe[\Delta]{\pi}$ can change (but $\status[\Delta]{\pi}$
	could go down). Hence, $\ranku{\Delta'}\le\ranku{\Delta}$.
	However, the number of $\dmrd$ instances in the derivation is
	increased. But both new instances of $\dmrd$ have
	strictly smaller rank in $\Delta'$ than the original $\dmrd$
	in $\Delta$, because their onion \bee-number is reduced
	by~$1$. Hence, $\rankd{\Delta'}<\rankd{\Delta}$. The same holds
	for $\rho=\absrd$ and $\rho=\weakrd$. Note that for this, it
	is crucial that the look-back tree of a vertex in the onion
        (that is used for computing the onion \bee-number) is acyclic.
      \end{enumerate}
    \end{enumerate}
  \item The crucial case is where the contractum $\loc R$ of $\pi$
    actively interferes with the redex $Z$ of $\rho$. There are four
    subcases:
    \begin{enumerate}[(a)]
    \item For $\rho=\weakrd$, the situation is dual to case (iii.a). We
      show only the case $\pi=\dmru$:
      \begin{equation*}\mysmall
	\vctpathderivation{
	  \dernote{\dmru}{}{\nS1\cons{\nwn1\nZ1\cons{\noc1\noc{11} \nR1}}}{ 
	    \root{\weakrd}{\nS2\cons{\nwn2\nZ2\cons{\noc2\nR2}}}{
	      \leaf{\nS3\cons{\un}}}}}{%
	  \boldvec
	  \longline{S}{1}{3}%
	  \longline{R}{1}{2}%
	  \longline{Z}{1}{2}%
	  \slimvec
	  \longuvec{oc}{1}{2}%
	  \longdvec{wn}{2}{1}%
	  \uvec{oc11}{oc2}
	}
	\quadto\;
	\vctpathderivation{
	  \dernote{\weakrd}{}{
	    \nS1\cons{\nwn1\nZ1\cons{\noc1\noc{11} \nR1}}}{ 
	    \leaf{\nS2\cons{\un}}}}{%
	  \boldvec
	  \longline{S}{1}{2}%
	}
      \end{equation*}
      We have $\ranku{\Delta'}<\ranku{\Delta}$ because $\pi$
      disappears, and $\rankd{\Delta'}\le\rankd{\Delta}$ because the
      status of the $\weakrd$ could go down.
    \item For $\rho=\dmrd$, the situation is dual to case (iii.b). We
      again show only the case $\pi=\dmru$:
      \begin{equation*}\mysmall
	\vctpathderivation{
	  \dernote{\dmru}{}{\nS1\cons{\nwn1\nZ1\cons{\noc1\noc{11} \nR1}}}{ 
	    \root{\dmrd}{\nS2\cons{\nwn2\nZ2\cons{\noc2\nR2}}}{
	      \leaf{\nS3\cons{\nwn3\nwn{23}\nZ3\cons{\noc3\nR3}}}}}}{%
	  \boldvec
	  \longline{S}{1}{3}
	  \longline{R}{1}{3}
	  \longline{Z}{1}{3}%
	  \slimvec
	  \longuvec{oc}{1}{3}
	  \longdvec{wn}{3}{1}
	  \uvec{oc11}{oc2}\dvec{wn23}{wn2}
	}
	\quadto\;
	\vctpathderivation{
	  \dernote{\dmrd}{}{\nS1\cons{\nwn1\nZ1\cons{\noc1\noc{11} \nR1}}}{ 
	    \root{\dmru}{
	      \nS2\cons{\nwn2\nwn{22}\nZ2\cons{\noc2\noc{12}\nR2}}}{
	      \leaf{\nS3\cons{\nwn3\nwn{23}\nZ3\cons{\noc3\nR3}}}}}}{%
	  \boldvec
	  \longline{S}{1}{3}
	  \longline{R}{1}{3}
	  \longline{Z}{1}{3}%
	  \slimvec
	  \longuvec{oc}{1}{3}
	  \longdvec{wn}{3}{1}
	  \uvec{oc11}{oc12}\uvec{oc12}{oc3}
	  \dvec{wn23}{wn22}\dvec{wn22}{wn1}
	}
      \end{equation*}
      We have $\ranku{\Delta'}\le\ranku{\Delta}$ and
      $\rankd{\Delta'}\le\rankd{\Delta}$ because the status of both
      rules could go down, and nothing else changes, for the same
      reason as explained in (iii.b).
    \item For $\rho=\absrd$ the permutations are dual to the ones in
      case (iii.c.4) above.\\ For $\pi=\weakru$, we have
      \begin{equation*}\mysmall
	\vctpathderivation{
	  \dernote{\weakru}{}{\nS1\cons{\nwn1\nZ1\cons{\un}}}{
	    \root{\absrd}{\nS2\cons{\nwn2\nZ2\cons{\noc2\nR2}}}{
	      \leaf{\nS3\pars{\nwn3\nZ3\cons{\noc3\nR3}\bc 
		  \nZ{13}\cons{\noc{13}\nR{13}}}}}}}{%
	  \boldvec
	  \longline{S}{1}{3} 
	  \longline{R}{2}{3} \uline{R2}{R13} 
	  \longline{Z}{1}{3} \uline{Z2}{Z13}%
	  \slimvec
	  \longuvec{oc}{2}{3} \uvec{oc2}{oc13}
	  \longdvec{wn}{3}{1}%
	}
	\quadto
	\vctpathderivation{
	  \dernote{\absrd}{}{\nS1\cons{\nwn1\nZ1\cons\un}}{
	    \root{\weakru}{\nS2\pars{\nwn2\nZ2\cons\un\bc 
		\nZ{22}\cons\un}}{
	      \root{\weakru}{\nS3\pars{\nwn3\nZ3\cons{\noc3\nR3}\bc 
		\nZ{23}\cons\un}}{
	      \leaf{\nS4\pars{\nwn4\nZ4\cons{\noc4\nR4}\bc 
		  \nZ{24}\cons{\noc{24}\nR{24}}}}}}}}{%
	  \boldvec
	  \longline{S}{1}{4} 
	  \longline{R}{3}{4} 
	  \longline{Z}{1}{4} \uline{Z1}{Z22} \uline{Z22}{Z24} 
	  \slimvec
	  \longuvec{oc}{3}{4} 
	  \longdvec{wn}{4}{1}%
	}
      \end{equation*}
      For $\pi=\dmru$, we have
      \begin{equation*}\mysmall
	\vctpathderivation{
	  \dernote{\dmru}{}{\nS1\cons{\nwn1\nZ1\cons{\noc1\noc{11} \nR1}}}{
	    \root{\absrd}{\nS2\cons{\nwn2\nZ2\cons{\noc2\nR2}}}{
	      \leaf{\nS3\pars{\nwn3\nZ3\cons{\noc3\nR3}\bc 
		  \nZ{13}\cons{\noc{13}\nR{13}}}}}}}{%
	  \boldvec
	  \longline{S}{1}{3} 
	  \longline{R}{1}{3} \uline{R2}{R13} 
	  \longline{Z}{1}{3} \uline{Z2}{Z13} 
	  \slimvec
	  \longuvec{oc}{1}{3} \uvec{oc11}{oc2} \uvec{oc2}{oc13}
	  \longdvec{wn}{3}{1}%
	}
	\quadto
	\vctpathderivation{
	  \dernote{\absrd}{}{\nS1\cons{\nwn1\nZ1\cons{\noc1\noc{11} \nR1}}}{
	    \root{\dmru}{\nS2\pars{\nwn2\nZ2\cons{\noc2\noc{12}\nR2}\bc 
		\nZ{22}\cons{\noc{22}\noc{32}\nR{22}}}}{
	      \root{\dmru}{\nS3\pars{\nwn3\nZ3\cons{\noc3\nR3}\bc 
		\nZ{23}\cons{\noc{23}\noc{33}\nR{23}}}}{
	      \leaf{\nS4\pars{\nwn4\nZ4\cons{\noc4\nR4}\bc 
		  \nZ{24}\cons{\noc{24}\nR{24}}}}}}}}{%
	  \boldvec
	  \longline{S}{1}{4} 
	  \longline{R}{1}{4} \uline{R1}{R22} \uline{R22}{R24} 
	  \longline{Z}{1}{4} \uline{Z1}{Z22} \uline{Z22}{Z24} 
	  \slimvec
	  \longuvec{oc}{1}{4} \uvec{oc11}{oc12} \uvec{oc12}{oc3}
	  \uvec{oc1}{oc22} \longuvec{oc}{22}{24} 
	  \uvec{oc11}{oc32} \uvec{oc32}{oc33}\uvec{oc33}{oc24}
	  \longdvec{wn}{4}{1}%
	}
      \end{equation*}
      And for $\pi=\absru$, we have
      \begin{equation*}\mysmall
	\vctpathderivation{
	  \dernote{\absru}{}{\nS1\cons{\nwn1\nZ1\aprs{\noc1\nR1\bc \nR{11}}}}{
	    \root{\absrd}{\nS2\cons{\nwn2\nZ2\cons{\noc2\nR2}}}{
	      \leaf{\nS3\pars{\nwn3\nZ3\cons{\noc3\nR3}\bc 
		  \nZ{13}\cons{\noc{13}\nR{13}}}}}}}{%
	  \boldvec
	  \longline{S}{1}{3} 
	  \longline{R}{1}{3} \uline{R11}{R2} \uline{R2}{R13} 
	  \longline{Z}{1}{3} \uline{Z2}{Z13} 
	  \slimvec
	  \longuvec{oc}{1}{3} \uvec{oc2}{oc13}
	  \longdvec{wn}{3}{1}%
	}
	\quadto  
	\vctpathderivation{
	  \dernote{\absrd}{}{\nS1\cons{\nwn1\nZ1\aprs{\noc1\nR1\bc \nR{11}}}}{
	    \root{\absru}{\nS2\pars{\nwn2\nZ2\aprs{\noc2\nR2\bc \nR{12}}\bc 
		\nZ{22}\aprs{\noc{22}\nR{22}\bc \nR{32}}}}{
	      \root{\absru}{\nS3\pars{\nwn3\nZ3\cons{\noc3\nR3}\bc 
		\nZ{23}\aprs{\noc{23}\nR{23}\bc \nR{33}}}}{
	      \leaf{\nS4\pars{\nwn4\nZ4\cons{\noc4\nR4}\bc 
		  \nZ{24}\cons{\noc{24}\nR{24}}}}}}}}{%
	  \boldvec
	  \longline{S}{1}{4} 
	  \longline{R}{1}{4} \uline{R11}{R12} \uline{R12}{R3} 
	  \uline{R1}{R22} \longline{R}{22}{24} 
	  \uline{R11}{R32} \uline{R32}{R33}\uline{R33}{R24}
	  \longline{Z}{1}{4} \uline{Z1}{Z22} \longline{Z}{22}{24}%
	  \slimvec
	  \longuvec{oc}{1}{4} \uvec{oc1}{oc22} \longuvec{oc}{22}{24}
	  \longdvec{wn}{4}{1}%
	}
      \end{equation*}
      In all three cases, the rule $\pi$ is duplicated. But both
      copies have a smaller onion \bee-number in $\Delta'$. Hence 
      $\ranku{\Delta'}<\ranku{\Delta}$. As in case (iii.c.4) above,
      this crucially relies on the fact that the look-back tree of a
      vertex is acyclic.
      We also have $\rankd{\Delta'}\le\rankd{\Delta}$ because the
      status of the $\absrd$ could go down, and nothing else changes.
    \item The most interesting case is when $\rho=\promrd$. We
      have the following situations:
      \begin{enumerate}[(1)]
      \item  For $\pi=\dmru$:\proofadjust
	\begin{equation*}\mysmall
	\vctpathderivation{%
	  \dernote{\dmru}{}{\nS1\pars{\nordx{\noc1\noc{11} \nR1},\nwn1 \nT1}}{
	    \root{\promrd}{\nS2\nordx{\pars{\rdc{\noc2 \nR2}\bc \nwn2 \nT2}}}{
	      \leaf{\nS3\cons{\cnt{\noc3\pars{\nR3\bc  \nT3}}}}}}}{%
	  \boldvec
	  \longline{S}{1}{3}
	  \longline{R}{1}{3}
	  \longline{T}{1}{3}%
	  \slimvec
	  \longuvec{oc}{1}{3}
	  \longdvec{wn}{2}{1}
	  \urdvec{oc2}{wn2}
	  \uvec{oc11}{oc2}
	}
	\quadto
	\vctpathderivation{%
	  \dernote{\dmrd}{}{\nS1\pars{\noc1\noc{11} \nR1\bc \nwn1 \nT1}}{
	    \root{\promrd}{
	      \nS2\pars{\noc2\noc{12}\nR2\bc \nwn2\nwn{12}\nT2}}{
	      \root{\promrd}{
		\nS3\cons{\noc3\pars{\noc{13}\nR3\bc \nwn{13}\nT3}}}{
		\root{\dmru}{\nS4\cons{\noc4\noc{14}\pars{\nR4\bc \nT4}}}{
		  \leaf{\nS5\cons{\cnt{\noc5\pars{\nR5\bc  \nT5}}}} }}}}} {%
	  \boldvec
	  \longline{S}{1}{5}
	  \longline{R}{1}{5}
	  \longline{T}{1}{5}%
	  \slimvec
	  \longuvec{oc}{1}{5}
	  \longuvec{oc}{11}{14} \uvec{oc14}{oc5}
	  \longdvec{wn}{2}{1}
	  \longdvec{wn}{13}{12} \dvec{wn12}{wn1}
	  \urdvec{oc2}{wn2}
	  \urdvec{oc13}{wn13}
	}
        \end{equation*}
	A single $\dmru$ is replaced by a $\dmru$ and a $\dmrd$. We
	clearly have $\ranku{\Delta'}\le\ranku{\Delta}$ because the
	status of the $\dmru$-instance could go down. If $\fg{\Delta}$
	is acyclic, then also its \pee-number goes down. Note that no
	other up-rule changes its rank.  We cannot make any statements
	about $\rankd{\Delta}$. But, observe that if $\fg{\Delta}$ is
	acyclic, then the \pee-number of the new $\dmrd$ is strictly
	smaller than the \pee-number of the original $\dmru$. Hence,
	if $\fg{\Delta}$ is acyclic, then
	$\rank{\Delta'}<\rank{\Delta}$. Note that even in the case of
	acyclicity of $\fg{\Delta}$, we \emph{do not} have
	$\rankd{\Delta'}\le\rankd{\Delta}$.
      \item For $\pi=\absru$:\proofadjust
	\begin{equation}\label{eq:bu-pd}
          \mysmall
	\vctpathderivation{%
	  \dernote{\absru}{}{
	    \nS1\pars{\nordx{\aprs{\noc1\nR1\bc \nR{11}}}\bc \nwn1\nT1}}{
	    \root{\promrd}{\nS2\nordx{\pars{\rdc{\noc2 \nR2}\bc \nwn2 \nT2}}}{
	      \leaf{\nS3\cons{\cnt{\noc3\pars{\nR3\bc  \nT3}}}} }}} {%
	  \boldvec
	  \longline{S}{1}{3}
	  \longline{R}{1}{3}
	  \uline{R11}{R2}
	  \longline{T}{1}{3}%
	  \slimvec
	  \longuvec{oc}{1}{3}
	  \longdvec{wn}{2}{1}
	  \urdvec{oc2}{wn2}
	}
	\quadto
	\vctpathderivation{%
	  \dernote{\absrd}{}{
	    \nS1\pars{\aprs{\noc1\nR1\bc \nR{11}}\bc \nordx{\nwn1\nT1}}}{
	    \root{\swir}{\nS2\rdc{\pars{\nordx{\aprs{\noc2\nR2\bc \nR{12}}\bc 
		    \nwn2\nT2\bc \nT{12}}}}}{
	      \root{\swir}{
		\nS3\nordx{\pars{\rdc{\aprs{\pars{\noc3\nR3\bc \nwn3\nT3}\bc 
			\nR{13}}\bc \nT{13}}}}}{
		\root{\promrd}{
		  \nS4\cnt{\aprs{\rdc{\pars{\noc4\nR4\bc \nwn4\nT4}}\bc 
		      \pars{\nR{14}\bc \nT{14}}}}}  {
		  \root{\absru}{
		    \nS5\nordx{\aprs{\rdc{\noc5\pars{\nR5\bc \nT5}}\bc 
			\pars{\nR{15}\bc \nT{15}}}}}   {
		    \leaf{\nS6\cons{\cnt{\noc6\pars{\nR6\bc \nT6}}}} }}}}}} {%
	  \boldvec
	  \longline{S}{1}{6}
	  \longline{R}{1}{6}
	  \longline{R}{11}{15}
	  \uline{R15}{R6}
	  \longline{T}{1}{6}
	  \longline{T}{12}{15}
	  \uline{T1}{T12}
	  \uline{T15}{T6}
	  \slimvec
	  \longuvec{oc}{1}{6}
	  \longdvec{wn}{4}{1}
	  \urdvec{oc4}{wn4}
	}
	\end{equation} 
	This case is similar to the one for $\dmru$ above, but
        slightly more complicated. The $\absru$-instance is replaced
        by a $\absru$ and a $\absrd$. By this, it can happen that the
        onion \bee-number of other down rules in $\Delta$ is
        increased. But note that no up-rule can change its onion
        \bee-number. (This is the reason for allowing one flipping edge
        in a path in the look-back tree in Definition~\ref{def:lbt},
        instead of forbidding any flipping edge. Note that cases
        (iii.c.4) and (iv.c) above would also work without the
        flipping edges in the look-back tree.)  Since, as before, the
        status of the $\absru$ could go down, we have
        $\ranku{\Delta'}\le\ranku{\Delta}$. But since the rank of some
        down-rules can become bigger, we cannot compare
        $\rank{\Delta'}$ with $\rank{\Delta}$. Nonetheless, it is
        important to mention that if $\fg{\Delta}$ is acyclic, then
        the \pee-number of the new $\absrd$ is strictly smaller than
        the \pee-number of the original $\absru$.
      \item For $\pi=\weakru$:\proofadjust
	\begin{equation*}\label{eq:weak-prom}\mysmall
	\vctpathderivation{%
	  \dernote{\weakru}{}{\nS1\pars{\nordx{\un\bc \nwn1\nT1}}}{
	    \root{\promrd}{\nS2\nordx{\pars{\rdc{\noc2 \nR2}\bc \nwn2 \nT2}}}{
	      \leaf{\nS3\cons{\cnt{\noc3\pars{\nR3\bc  \nT3}}}} }}} {%
	  \boldvec
	  \longline{S}{1}{3}
	  \longline{R}{2}{3}
	  \longline{T}{1}{3}%
	  \slimvec
	  \longuvec{oc}{2}{3}
	  \longdvec{wn}{2}{1}
	  \urdvec{oc2}{wn2}
	}
	\quadto
	\vctpathderivation{%
	  \dernote{\weakrd}{}{\nS1\pars{\un\bc \lwn T}}{
	    \root{=}{\nS2\pars{\un\bc \un}}{
	      \root{\weakru}{\nS3\cons{\un}}{
		\leaf{\nS4\cons{\loc\pars{R\bc T}}} }}}}{%
	  \boldvec
	  \longline{S}{1}{4}%
	}
	\end{equation*}
        This case is simpler than the other two because the instance
	of $\promrd$ disappears. Hence, we have
	$\ranku{\Delta'}\le\ranku{\Delta}$ and if $\fg{\Delta}$ is
	acyclic also $\rank{\Delta'}<\rank{\Delta}$. But we do
	\emph{not} have $\rankd{\Delta'}\le\rankd{\Delta}$.
      \end{enumerate}
    \end{enumerate}
  \end{enumerate}
\end{casana}

This case analysis is enough to show the following three lemmas.

\begin{lemma}[(Step 2 in Fig.~\ref{fig:third})]
  \label{lem:singlestep-up}
  $$\mysmall
  \mbox{\normalsize Every derivation }
  \quad
  \simplederi{P}{\SNEL'}{\Delta}{Q}
  \quad
  \mbox{\normalsize can be transformed into}
  \quad
  \upsmash{\vcstrder{\SNEL'\setminus\set{\dmru,\absru,\weakru}}{}{Q}{
    \stem{\set{\dmru,\absru,\weakru}}{}{P'}{
      \leaf{P}}}}
  \quad.
  $$ 
\end{lemma}

\begin{proof}
  This transformation is obtained by permuting all instances of
  $\dmru$, $\absru$, $\weakru$ to the top of a derivation as described
  in \ref{para:perm-fbw-II}. In all cases, we have
  $\ranku{\Delta'}\le\ranku{\Delta}$. Termination now follows for the
  same reasons as in the proof of Lemma~\ref{lem:step1}. We use as
  induction measure the pair $\tuple{\ranku{\Delta},\delta}$ in a
  lexicographic ordering, where $\delta$ is the number of rule
  instances in the derivation above the topmost instance of $\dmru$,
  $\absru$, or $\weakru$ with status~$1$. If we always choose this
  topmost instance of $\dmru$, $\absru$, or $\weakru$ with status~$1$
  for performing the next permutation step, then this measure always
  goes down (because whenever we have
  $\ranku{\Delta'}\not<\ranku{\Delta}$ in Case
  Analysis~\ref{para:perm-fbw-II}, then $\delta$ is reduced by~1).
\end{proof}

\begin{lemma}[(Step 2 in Fig.~\ref{fig:third})]
  \label{lem:singlestep-down} 
  $$\mysmall 
  \mbox{\normalsize Every derivation } \quad
  \simplederi{P}{\SNEL'}{\Delta}{Q} \quad 
  \mbox{\normalsize\ can be
    transformed into} \quad
  \upsmash{\vcstrder{\set{\dmrd,\absrd,\weakrd}}{}{Q}{
    \stem{\SNEL'\setminus\set{\dmrd,\absrd,\weakrd}}{}{Q'}{ \leaf{P}}}}
  \quad.
  $$ 
\end{lemma}

\begin{proof}
  Dual to the previous lemma. 
\end{proof}

\begin{lemma}[(Step 2 in Fig.~\ref{fig:third})]\label{lem:nocycle-term}
  If the \bqfg\ of a derivation   
  $\downsmash{\mysmall\simplederi{P}{\SNEL'}{\Delta}{Q}}$ 
  is
  acyclic, then $\Delta$ can be transformed into a derivation $\Delta'$
  \begin{equation}\mysmall
    \label{eq:up-down-done}
    \vcstrder{\set{\dmrd,\absrd,\weakrd}}{}{P}{
      \stem{\set{\atird,\atiru,\swir,\seqrd,\seqru,\promrd,\promru}}{}{P'}{
        \stem{\set{\dmru,\absru,\weakru}}{}{Q'}{
	  \leaf{Q}}}}
    \qquad.\qqqqquad
  \end{equation}
\end{lemma}

\begin{proof}
  The derivation $\Delta'$ is obtained from $\Delta$ by a sequence of
  transformations:
  \begin{equation}
    \label{eq:up-down}
    \Delta=\Delta_0\leadsto\Delta_1\leadsto\Delta_2\leadsto\Delta_3
    \leadsto\ldots\leadsto\Delta'
    \quadcm
  \end{equation}
  where $\Delta_{i+1}$ is obtained from $\Delta_i$ by permuting all
  instances of $\dmru$, $\absru$, $\weakru$ up to the top of the
  derivation if $i$ is even, and by permuting all instances of
  $\dmrd$, $\absrd$, $\weakrd$ down to the bottom of the derivation if
  $i$ is odd. Each of these single steps is well-defined because of
  Lemma~\ref{lem:singlestep-up} and Lemma~\ref{lem:singlestep-down}.
  Now assume $i$ is even and $i\ge2$. Then there are no instances of
  $\dmru$, $\absru$, or $\weakru$ with status~1 in $\Delta_{i+1}$, and
  all instances of $\dmrd$, $\absrd$, $\weakrd$ in $\Delta_{i+1}$ have
  been introduced by case (iv.d) in \ref{para:perm-fbw-II}. Hence, for
  each $\rho'$ of the kind $\dmrd$, $\absrd$, $\weakrd$ in
  $\Delta_{i+1}$, there is a $\rho$ of the kind $\dmru$, $\absru$,
  $\weakru$ in $\Delta_i$, with $\status[\Delta_i]{\rho}=1$ and
  $\pe[\Delta_i]{\rho}>\pe[\Delta_{i+1}]{\rho'}$, and therefore
  $\rank[\Delta_i]{\rho}>\rank[\Delta_{i+1}]{\rho'}$. Hence
  $\rank{\Delta_i}>\rank{\Delta_{i+1}}$. By a similar argument we can
  conclude that $\rank{\Delta_i}>\rank{\Delta_{i+1}}$ for all odd $i$
  with $i>1$. Since the multiset ordering is well-founded, we can
  conclude that the process indicated in \eqref{eq:up-down} terminates
  eventually. The resulting derivation $\Delta'$ is of the desired
  shape~\eqref{eq:up-down-done}.  
\end{proof}

Note that the argument in the previous proof is necessary because of
case (iv.d.2) in \ref{para:perm-fbw-II}. In all other permutation
steps the rank does not increase.\footnote{The paper~\cite{str:MELL}
  contains statements for $\MELL$ that are similar to Lemmas
  \ref{lem:singlestep-up}--\ref{lem:nocycle-term}, but the proofs here
  are simpler, due to the use of the rank in the measure for ensuring
  termination.}

\subsection{Forked and Unforced Cycles}\label{sec:cycles}

As the derivation in \eqref{eq:short} shows, we cannot hope for a
lemma saying that $\fg{\Delta}$ is always acyclic. Nonetheless, the
decomposition terminates for~\eqref{eq:short}, and the result is shown
in Figure~\ref{fig:short}. Since in that figure, the \bqfg\ is
acyclic, the cycle must have been broken eventually. For understanding
how this is happening, we will now continue with an investigation in
the structure of cycles in the \bqfg, and how they are
broken. Before, we exhibit another example of a derivation with a
cycle in its \bqfg:
\begin{equation}\label{eq:cyclederi}\myssmall
  \vctpathderivation{%
    \dernote{\dmru}{}{\aprs{\noc1\pars{\noc{21}\noc{31}a\bc \nwn1\aprs{b\bc c}\bc 
	  \nwn{21}d}\bc \pars{\nwn{31}b\bc \nwn{41}\aprs{a\bc d}\bc \noc{51}c}}}{
      \root{\promru\bc \promru}{\aprs{\noc2\pars{\noc{22}a\bc \nwn2\aprs{b\bc c}\bc 
	    \nwn{22}d}\bc \pars{\nwn{32}b\bc \nwn{42}\aprs{a\bc d}\bc \noc{52}c}}}{
	\root{\swir\bc \swir}{\aprs{\noc3\pars{\noc{23}a\bc 
	      \aprs{\nwn3 b\bc \noc{13}c}\bc \nwn{23}d}\bc \pars{\nwn{33}b\bc 
	      \aprs{\noc{43}a\bc \nwn{43}d}\bc \noc{53}c}}}{
	  \root{\swir\bc \swir}{\aprs{\noc4\pars{\noc{24}a\bc 
		\aprs{\nwn4 b\bc \noc{14}c}\bc \nwn{24}d}\bc 
	      \pars{\nwn{34}b\bc \noc{44}a}\bc \pars{\noc{54}c\bc \nwn{44}d}}}{
	    \root{\absru}{\aprs{\noc5\aprs{\pars{\nwn5 b\bc \noc{25}a}\bc 
		  \pars{\noc{15}c\bc \nwn{25}d}}\bc 
		\pars{\nwn{35}b\bc \noc{45}a}\bc \pars{\noc{55}c\bc \nwn{45}d}}{\phantom{\Big\vert^\vert}}}{
	      \root{\promrd\bc \promrd}{\noc6\aprs{\pars{\nwn6 b\bc \noc{26}a}\bc 
		  \pars{\noc{16}c\bc \nwn{26}d}}{\phantom{\Big\vert_\vert}}}{
		\leaf{\noc7\aprs{\noc{27}\pars{b\bc a}\bc 
		    \noc{17}\pars{c\bc d}}}}}}}}}}{%
    \thickvec
    \druvec{wn3}{oc13}
    \longuvec{oc}{13}{16}
    \urdvec{oc16}{wn26}
    \dvec{wn26}{wn45}
    \longdvec{wn}{45}{43}
    \dluvec{wn43}{oc43}
    \longuvec{oc}{43}{45}
    \uvec{oc45}{oc26}
    \uldvec{oc26}{wn6}
    \longdvec{wn}{6}{3}%
    \slimvec
    \longuvec{oc}{1}{7}
    \longuvec{oc}{21}{27} \uvec{oc31}{oc22}
    \longuvec{oc}{51}{55} \uvec{oc55}{oc16} \uvec{oc16}{oc17}
    \longdvec{wn}{26}{21}
    \dvec{wn6}{wn35} \longdvec{wn}{35}{31}
    \longdvec{wn}{43}{41}
    \longdvec{wn}{3}{1}%
  }
\end{equation}
This derivation can be used to explain why we have Steps 2 and~3 in
the proof of Theorem~\ref{thm:decomposition} (see
Figure~\ref{fig:third}), instead of doing something like
\proofadjust
$$\mytiny
\vcstrder{\SNEL'}{}{Z_1}{
  \leaf{W_1}
} 
\quad\stackrel{2'}{\loongrightarrow}\quad
\vcstrder{\set{\dmrd}}{}{Z_1}{
  \stem{\SNEL'\setminus\set{\dmrd,\dmru}}{}{Z_2}{
    \stem{\set{\dmru}}{}{W_2}{
      \leaf{W_1}
}}} 
\quad\stackrel{2''}{\loongrightarrow}\quad
\vcstrder{\set{\dmrd}}{}{Z_1}{
  \stem{\set{\absrd}}{}{Z_2}{
    \stem{\SNEL'\setminus\set{\dmrd,\dmru,\absrd,\absru}}{}{Z_3}{
      \stem{\set{\absru}}{}{W_3}{
	\stem{\set{\dmru}}{}{W_2}{
	  \leaf{W_1}
}}}}} 
\quad\cdots
$$ Running Step~$2'$ on the derivation in~\eqref{eq:cyclederi} does
indeed fail because of non-termination. If we apply all the
transformations of~\ref{para:perm-fbw-II} together with the ones in
\eqref{eq:fb1} and~\eqref{eq:fb2} (and their duals), then the
instances of $\dmru$ (and $\dmrd$) get caught in the cycle
in~\eqref{eq:cyclederi}, and the process will run forever.  Only if
the $\absru$ is permuted up together with the $\dmru$, the process
does terminate. The reason is that when the instance of $\absru$ is
permuted over the two $\promrd$ on the top of the derivation, the
cycle is broken, because some edges in the \bqfg\ disappear. This shows
the importance of case (iii.c.2) in~\ref{para:perm-fbw-II}, and
motivates the following definition.

\begin{figure}[!t]
  $$
  \hskip-10em
  \vctpathderivation{
    \dernote{\absrd}{}{\nwn1\nwn{101}\aprs{\noc1a\bc a}}{
      \root{\absrd}{
	\pars{\nwn2\nwn{102}\aprs{\noc2a\bc a}\bc
	  \nwn{302}\aprs{\noc{402}a\bc a}}}{
      \root{\absrd}{
	\pars{\nwn3\nwn{103}\aprs{\noc3a\bc a}\bc
	  \nwn{203}\aprs{\noc{203}a\bc a}\bc
	  \nwn{303}\aprs{\noc{403}a\bc a}}}{
      \root{\absrd}{
	\pars{\nwn4\nwn{104}\aprs{\noc4a\bc a}\bc
	  \nwn{204}\aprs{\noc{204}a\bc a}\bc
	  \nwn{304}\aprs{\noc{404}a\bc a}\bc
	  \aprs{\noc{504}a\bc a}}}{
      \root{\absrd}{
	\pars{\nwn5\nwn{105}\aprs{\noc5a\bc a}\bc
	  \nwn{205}\aprs{\noc{205}a\bc a}\bc
	  \aprs{\noc{305}a\bc a}\bc
	  \nwn{305}\aprs{\noc{405}a\bc a}\bc
	  \aprs{\noc{505}a\bc a}}}{ 
      \root{\promru}{
	\pars{\nwn6\pars{\nwn{106}\aprs{\noc6a\bc a}\bc
	    \aprs{\noc{106}a\bc a}}\bc
	  \nwn{206}\aprs{\noc{206}a\bc a}\bc
	  \aprs{\noc{306}a\bc a}\bc
	  \nwn{306}\aprs{\noc{406}a\bc a}\bc
	  \aprs{\noc{506}a\bc a}}}{ 
      \root{\promru}{
	\pars{\nwn7\pars{\nwn{107}\aprs{\noc7a\bc a}\bc
	    \aprs{\noc{107}a\bc a}}\bc
	  \nwn{207}\aprs{\noc{207}a\bc a}\bc
	  \aprs{\noc{307}a\bc a}\bc
	  \aprs{\nwn{307}\noc{407}a\bc\noc{607}a}\bc
	  \aprs{\noc{507}a\bc a}}}{ 
      \root{\promru}{
	\pars{\nwn8\pars{\nwn{108}\aprs{\noc8a\bc a}\bc
	    \aprs{\noc{108}a\bc a}}\bc
	  \aprs{\nwn{208}\noc{208}a\bc\noc{708}a}\bc
	  \aprs{\noc{308}a\bc a}\bc
	  \aprs{\nwn{308}\noc{408}a\bc\noc{608}a}\bc
	  \aprs{\noc{508}a\bc a}}}{ 
      \root{\swir}{
	\pars{\nwn9\pars{\aprs{\nwn{109}\noc9a\bc\noc{809}a}\bc
	    \aprs{\noc{109}a\bc a}}\bc
	  \aprs{\nwn{209}\noc{209}a\bc\noc{709}a}\bc
	  \aprs{\noc{309}a\bc a}\bc
	  \aprs{\nwn{309}\noc{409}a\bc\noc{609}a}\bc
	  \aprs{\noc{509}a\bc a}}}{ 
      \root{\swir}{
	\pars{\nwn{10}\pars{\aprs{\nwn{110}\noc{10}a\bc\noc{810}a}\bc
	    \aprs{\noc{110}a\bc a}}\bc
	  \aprs{\nwn{210}\noc{210}a\bc\noc{710}a}\bc
	  \aprs{\noc{310}a\bc a}\bc
	  \aprs{\pars{\aprs{\nwn{310}\noc{410}a\bc
		\noc{610}a}\bc\noc{510}a}\bc a}}}{ 
      \root{\swir}{
	\pars{\nwn{11}\pars{\aprs{\nwn{111}\noc{11}a\bc\noc{811}a}\bc
	    \aprs{\noc{111}a\bc a}}\bc
	  \aprs{\nwn{211}\noc{211}a\bc\noc{711}a}\bc
	  \aprs{\noc{311}a\bc a}\bc
	  \aprs{\pars{\nwn{311}\noc{411}a\bc\noc{511}a}\bc\noc{611}a\bc a}}}{ 
      \root{\swir}{
	\pars{\nwn{12}\pars{\aprs{\nwn{112}\noc{12}a\bc\noc{812}a}\bc
	    \aprs{\noc{112}a\bc a}}\bc
	  \aprs{\pars{\aprs{\nwn{212}\noc{212}a\bc\noc{712}a}\bc
	      \noc{312}a}\bc a}\bc
	  \aprs{\pars{\nwn{312}\noc{412}a\bc\noc{512}a}\bc\noc{612}a\bc a}}}{ 
      \root{\swir}{
	\pars{\nwn{13}\pars{\aprs{\nwn{113}\noc{13}a\bc\noc{813}a}\bc
	    \aprs{\noc{113}a\bc a}}\bc
	  \aprs{\pars{\nwn{213}\noc{213}a\bc\noc{313}a}\bc\noc{713}a\bc a}\bc
	  \aprs{\pars{\nwn{313}\noc{413}a\bc\noc{513}a}\bc\noc{613}a\bc a}}}{ 
      \root{\swir}{
	\pars{\nwn{14}\pars{
	    \aprs{\pars{\aprs{\nwn{114}\noc{14}a\bc\noc{814}a}\bc
		\noc{114}a}\bc a}}\bc
	  \aprs{\pars{\nwn{214}\noc{214}a\bc\noc{314}a}\bc\noc{714}a\bc a}\bc
	  \aprs{\pars{\nwn{314}\noc{414}a\bc\noc{514}a}\bc\noc{614}a\bc a}}}{ 
      \root{\promru}{
	\pars{\nwn{15}\aprs{\pars{\nwn{115}\noc{15}a\bc\noc{115}a}\bc 
		\noc{815}a\bc a}\bc
	  \aprs{\pars{\nwn{215}\noc{215}a\bc\noc{315}a}\bc\noc{715}a\bc a}\bc
	  \aprs{\pars{\nwn{315}\noc{415}a\bc\noc{515}a}\bc\noc{615}a\bc a}}}{ 
      \root{\swir}{
	\pars{\aprs{\noc{916}\pars{\nwn{116}\noc{16}a\bc\noc{116}a}\bc 
		\nwn{16}\aprs{\noc{816}a\bc a}}\bc
	  \aprs{\pars{\nwn{216}\noc{216}a\bc\noc{316}a}\bc\noc{716}a\bc a}\bc
	  \aprs{\pars{\nwn{316}\noc{416}a\bc\noc{516}a}\bc\noc{616}a\bc a}}}{ 
      \root{\swir}{
	\pars{\aprs{\noc{917}\pars{\nwn{117}\noc{17}a\bc\noc{117}a}\bc 
		\pars{
		  \nwn{17}\aprs{\noc{817}a\bc a}\bc
		  \aprs{\pars{\nwn{217}\noc{217}a\bc\noc{317}a}\bc
		    \noc{717}a\bc a}}}\bc
	  \aprs{\pars{\nwn{317}\noc{417}a\bc\noc{517}a}\bc\noc{617}a\bc a}}}{ 
      \root{\swir}{
	\pars{\aprs{\noc{918}\pars{\nwn{118}\noc{18}a\bc\noc{118}a}\bc 
	    \pars{\nwn{218}\noc{218}a\bc\noc{318}a}\bc
	    \pars{\nwn{18}\aprs{\noc{818}a\bc a}\bc\aprs{\noc{718}a\bc a}}}\bc
	  \aprs{\pars{\nwn{318}\noc{418}a\bc\noc{518}a}\bc\noc{618}a\bc a}}}{ 
      \root{\swir}{
	\aprs{\noc{919}\pars{\nwn{119}\noc{19}a\bc\noc{119}a}\bc 
	  \pars{\nwn{219}\noc{219}a\bc\noc{319}a}\bc
	  \pars{\nwn{19}\aprs{\noc{819}a\bc a}\bc\aprs{\noc{719}a\bc a}\bc
	    \aprs{\pars{\nwn{319}\noc{419}a\bc\noc{519}a}\bc
	      \noc{619}a\bc a}}}}{ 
      \root{\swir}{
	\aprs{\noc{920}\pars{\nwn{120}\noc{20}a\bc\noc{120}a}\bc 
	  \pars{\nwn{220}\noc{220}a\bc\noc{320}a}\bc
	  \pars{\nwn{320}\noc{420}a\bc\noc{520}a}\bc
	  \pars{\nwn{20}\aprs{\noc{820}a\bc a}\bc
	    \aprs{\noc{720}a\bc a}\bc\aprs{\noc{620}a\bc a}}}}{ 
      \root{\swir}{
	\aprs{\noc{921}\pars{\nwn{121}\noc{21}a\bc\noc{121}a}\bc 
	  \pars{\nwn{221}\noc{221}a\bc\noc{321}a}\bc
	  \pars{\nwn{321}\noc{421}a\bc\noc{521}a}\bc
	  \pars{\nwn{21}\aprs{\noc{821}a\bc a}\bc
	    \aprs{\noc{621}a\bc\pars{\aprs{\noc{721}a\bc a}\bc a}}}}}{ 
      \root{\promrd}{
	\aprs{\noc{922}\pars{\nwn{122}\noc{22}a\bc\noc{122}a}\bc 
	  \pars{\nwn{222}\noc{222}a\bc\noc{322}a}\bc
	  \pars{\nwn{322}\noc{422}a\bc\noc{522}a}\bc
	  \pars{\nwn{22}\aprs{\noc{822}a\bc a}\bc\noc{622}a}\bc
	  \pars{\aprs{\noc{722}a\bc a}\bc a}}}{ 
      \root{\promrd}{
	\aprs{\noc{923}\pars{\nwn{123}\noc{23}a\bc\noc{123}a}\bc 
	  \pars{\nwn{223}\noc{223}a\bc\noc{323}a}\bc
	  \pars{\nwn{323}\noc{423}a\bc\noc{523}a}\bc
	  \noc{623}\pars{\aprs{\noc{823}a\bc a}\bc a}\bc
	  \pars{\aprs{\noc{723}a\bc a}\bc a}}}{ 
      \root{\promrd}{
	\aprs{\noc{924}\pars{\nwn{124}\noc{24}a\bc\noc{124}a}\bc 
	  \pars{\nwn{224}\noc{224}a\bc\noc{324}a}\bc
	  \noc{524}\pars{\noc{424}a\bc a}\bc
	  \noc{624}\pars{\aprs{\noc{824}a\bc a}\bc a}\bc
	  \pars{\aprs{\noc{724}a\bc a}\bc a}}}{ 
      \root{\promrd}{
	\aprs{\noc{925}\pars{\nwn{125}\noc{25}a\bc\noc{125}a}\bc 
	  \noc{325}\pars{\noc{225}a\bc a}\bc
	  \noc{525}\pars{\noc{425}a\bc a}\bc
	  \noc{625}\pars{\aprs{\noc{825}a\bc a}\bc a}\bc
	  \pars{\aprs{\noc{725}a\bc a}\bc a}}}{ 
      \root{\absru}{
	\aprs{\noc{926}\noc{126}\pars{\noc{26}a\bc a}\bc 
	  \noc{326}\pars{\noc{226}a\bc a}\bc
	  \noc{526}\pars{\noc{426}a\bc a}\bc
	  \noc{626}\pars{\aprs{\noc{826}a\bc a}\bc a}\bc
	  \pars{\aprs{\noc{726}a\bc a}\bc a}}}{ 
      \root{\absru}{
	\aprs{\noc{927}\noc{127}\pars{\noc{27}a\bc a}\bc 
	  \noc{527}\pars{\noc{427}a\bc a}\bc
	  \noc{627}\pars{\aprs{\noc{827}a\bc a}\bc a}\bc
	  \pars{\aprs{\noc{727}a\bc a}\bc a}}}{ 
      \root{\absru}{
	\aprs{\noc{928}\noc{128}\pars{\noc{28}a\bc a}\bc 
	  \noc{528}\pars{\noc{428}a\bc a}\bc
	  \noc{628}\pars{\aprs{\noc{828}a\bc a}\bc a}}}{ 
      \root{\absru}{
	\aprs{\noc{929}\noc{129}\pars{\noc{29}a\bc a}\bc 
	  \noc{529}\pars{\noc{429}a\bc a}\bc
	  \noc{629}\pars{\noc{829}a\bc a}}}{ 
      \root{\absru}{
	\aprs{\noc{930}\noc{130}\pars{\noc{30}a\bc a}\bc 
	  \noc{630}\pars{\noc{830}a\bc a}}}{ 
      \leaf{\noc{931}\noc{131}\pars{\noc{31}a\bc a}}
      }}}}}}}}}}}}}}}}}}}}}}}}}}}}}}}
    {\slimvec
      \longuvec{oc}{1}{26}\longurvec{oc}{26}{31}%
      \uvec{oc5}{oc106}\longuvec{oc}{106}{126}\longurvec{oc}{126}{131}%
      \uvec{oc2}{oc203}\longuvec{oc}{203}{217}%
      \vecanglespos{oc217}{oc218}{140}{-40}{.8}
      \longuvec{oc}{218}{226}\uvec{oc226}{oc27}%
      \uvec{oc204}{oc305}\longuvec{oc}{305}{317}%
      \vecanglespos{oc317}{oc318}{120}{-30}{.35}
      \longuvec{oc}{318}{326}\uvec{oc326}{oc127}
      \uvec{oc1}{oc402}\longuvec{oc}{402}{419}
      \vecanglespos{oc419}{oc420}{150}{-30}{.5}
      \longuvec{oc}{420}{427}\longurvec{oc}{427}{429}\uvec{oc429}{oc30}%
      \uvec{oc403}{oc504}\longuvec{oc}{504}{519}%
      \vecanglespos{oc519}{oc520}{140}{-25}{.55}
      \longuvec{oc}{520}{527}\longurvec{oc}{527}{529}\uvec{oc529}{oc130}
      \uldvec{oc523}{wn323}\longdvec{wn}{326}{320}%
      \vecanglespos{wn320}{wn319}{-40}{160}{.45}
      \longdvec{wn}{319}{302}\dvec{wn302}{wn101}
      \druvec{wn307}{oc607}\longuvec{oc}{607}{627}
      \longurvec{oc}{627}{629}\ulvec{oc629}{oc630}\ulvec{oc630}{oc131}
      \vecangles{oc622}{wn22}{140}{30}\longdvec{wn}{22}{20}
      \vecanglespos{wn20}{wn19}{-140}{60}{.9}\dlvec{wn19}{wn18}
      \vecanglespos{wn18}{wn17}{-140}{60}{.4}\dvec{wn17}{wn16}
      \vecanglespos{wn16}{wn15}{-120}{40}{.6}\longdvec{wn}{15}{1}%
      \vecanglespos{wn16}{oc916}{-140}{-40}{.6}
      \longuvec{oc}{916}{926}\longurvec{oc}{926}{931}%
      \uldvec{oc324}{wn224}\longdvec{wn}{224}{218}%
      \vecanglespos{wn218}{wn217}{-60}{150}{.8}
      \longdvec{wn}{217}{203}\dvec{wn203}{wn102}
      \druvec{wn208}{oc708}\longuvec{oc}{708}{719}
      \longurvec{oc}{719}{721}
      \longuvec{oc}{721}{727}\uvec{oc727}{oc828}
      \uldvec{oc125}{wn125}\longdvec{wn}{125}{101}%
      \druvec{wn109}{oc809}\longuvec{oc}{809}{817}
      \longurvec{oc}{817}{819}\vecanglespos{oc819}{oc820}{50}{-130}{.16}
      \longuvec{oc}{820}{827}\longurvec{oc}{827}{829}
      \ulvec{oc829}{oc830}\ulvec{oc830}{oc31}
    }
    $$
    \caption{Result of applying the decomposition to the derivation
    in~\eqref{eq:short}}
    \label{fig:short}
\end{figure}

\begin{definition}
  A cycle $c$ in $\fg{\Delta}$ is called \dfn{forked}
  if there is an instance of
  $$\myssmall
  \vcinf{\absru}{S\aprs{\loc R\bc R}}{S\cons{\loc R}}
  \qqquad\mbox{or}\qqquad
  \vcinf{\absrd}{S\cons{\lwn T}}{S\pars{\lwn T\bc T}}
  $$ inside $\Delta$ such that both copies of $R$ of the redex of the
  $\absru$, or both copies of $T$ in the contractum of $\absrd$
  contain vertices of the cycle. We say that such an instance of
  $\absru$ or $\absrd$ \dfn{forks} the cycle $c$. The number of
  $\absru$ and $\absrd$ that fork a cycle $c$ is called the
  \emph{forking number} of $c$, denoted by $\fork{c}$. A cycle $c$
  with $\fork{c}=0$ is called \dfn{unforked}.
\end{definition}

The cycles in \eqref{eq:short} and~\eqref{eq:cyclederi} are both
forked. The one in \eqref{eq:short} has forking number~2 (since both,
the $\absrd$ and the $\absru$ fork the cycle), and the cycle
in~\eqref{eq:cyclederi} has forking number~1.

\begin{lemma}[(Step 2 in Fig.~\ref{fig:third})]\label{lem:nocycle}
  Let $\Delta$ be a derivation in $\SNEL'$, and let $\Delta'$ be the
  result of first permuting all $\dmrd$, $\absrd$, $\weakrd$ down (via
  Lemma~\ref{lem:singlestep-down}), and then permuting all $\dmru$,
  $\absru$, $\weakru$ up (via Lemma~\ref{lem:singlestep-up}). If
  $\fg{\Delta'}$ is cyclic, then $\fg{\Delta'}$ contains an unforked cycle.
\end{lemma}

\begin{figure}[!t]
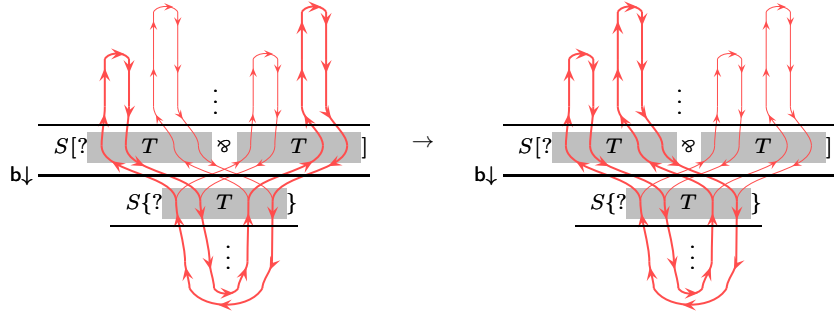

  \def\mdist{}
  \def\normaluparrpos{.8}
  \def\normaldownarrpos{.3}
  \def\thederi{
    \dernote{}{}{\begin{array}{c}
	\phantom{AS:?}
	\mdist\nph2\mdist\nph{12}\,\vdots\,\nph{22}\mdist\nph{32}\mdist\;\\ 
	\phantom{aS:?}
	\mdist\nph1\mdist\nph{11}\mdist\nph{21}\mdist\nph{31}\mdist\;
    \end{array}}{
      \root{\absrd}{\phantom{\Big\vert^{\vert}}
	S\cons{\lwn
	  \rdx{\mdist\nph3\mdist\nph{13}\,T\,\nph{23}\mdist\nph{33}\mdist}}}{
	\root{}{\phantom{\Big\vert_{\vert}}S\pars{\lwn
	    \rdx{\mdist\nph4\mdist\nph{14}\,T\,
	      \nph{24}\mdist\nph{34}\mdist}\bc
	    \rdx{\mdist\nph{44}\mdist\nph{54}\,T\,
	      \nph{64}\mdist\nph{74}\mdist}}}{
	  \leaf{\begin{array}{c}
	      \quad
	      \mdist\nph7\mdist\nph{17}\mdist\nph{27}\mdist\nph{37}\mdist
	      \phantom{\quad\vdots\quad}
	      \mdist\nph{47}\mdist\nph{57}\mdist\nph{67}\mdist\nph{77}\mdist
	      \\ 
	      \quad
	      \mdist\nph6\mdist\nph{16}\mdist\nph{26}\mdist\nph{36}\mdist
	      \phantom{\quad\vdots\quad}
	      \mdist\nph{46}\mdist\nph{56}\mdist\nph{66}\mdist\nph{76}\mdist
	      \\ 
	      \quad
	      \mdist\nph5\mdist\nph{15}\mdist\nph{25}\mdist\nph{35}\mdist
	      \quad\vdots\quad
	      \mdist\nph{45}\mdist\nph{55}\mdist\nph{65}\mdist\nph{75}\mdist
	  \end{array}}
    }}}
  }
  $$
  \vctpathderivation{\thederi}{%
    \thickvec%
    \longuvec{ph}{1}{6}%
    \longdvec{ph}{16}{11}%
    \urdvec{ph6}{ph16}
    \longuvec{ph}{21}{23}%
    \longdvec{ph}{33}{31}%
    \druvec{ph11}{ph21}\dluvec{ph31}{ph1}
    \uvec{ph23}{ph64}\longuvec{ph}{64}{67}%
    \dvec{ph74}{ph33}\longdvec{ph}{77}{74}%
    \urdvec{ph67}{ph77}
    \slimvec%
    \longuvec{ph}{23}{27}%
    \longdvec{ph}{37}{33}%
    \urdvec{ph27}{ph37}
    \uvec{ph3}{ph44}\longuvec{ph}{44}{46}%
    \dvec{ph54}{ph13}\longdvec{ph}{56}{54}%
    \urdvec{ph46}{ph56}
  }
  \quadto\!
  \vctpathderivation{\thederi}{%
    \thickvec%
    \longuvec{ph}{1}{6}%
    \longdvec{ph}{16}{11}%
    \urdvec{ph6}{ph16}
    \longuvec{ph}{21}{27}%
    \longdvec{ph}{37}{31}%
    \urdvec{ph27}{ph37}
    \druvec{ph11}{ph21}\dluvec{ph31}{ph1}
    \slimvec%
    \uvec{ph3}{ph44}\longuvec{ph}{44}{46}%
    \dvec{ph54}{ph13}\longdvec{ph}{56}{54}%
    \urdvec{ph46}{ph56}
    \uvec{ph23}{ph64}\longuvec{ph}{64}{67}%
    \dvec{ph74}{ph33}\longdvec{ph}{77}{74}%
    \urdvec{ph67}{ph77}
  }
  $$
  \caption{The basic idea of the proof of Lemma~\ref{lem:nocycle}}
  \label{fig:forkdown}
\end{figure}

\begin{proof}
  We show the following claim: If $\fg{\Delta'}$ contains a cycle $c$
  with $\fork{c}=n$ for some $n>0$, then it also contains a cycle $c'$
  with $\fork{c'}=n-1$. Clearly, the cycle $c$ must be forked by $n$
  instances of~$\absrd$ that have all been introduced by the
  transformation shown in~\eqref{eq:bu-pd} (because first, all
  $\absrd$ have been permuted down, and then all $\absru$ have been
  permuted up). Now consider the topmost $\absrd$ that forks $c$. The
  introduction of this $\absrd$ causes a duplication of all up-paths
  and down-paths through $T$ (we are still referring
  to~\eqref{eq:bu-pd}). Furthermore, the continued up-permutation of
  the $\absru$ (that caused the introduction of the $\absrd$) causes a
  duplication of all flipping edges connecting up-paths and down-paths
  through $T$ (see cases (iii.c.2) and (iii.c.3)
  in~\ref{para:perm-fbw-II}). This is indicated on the left of
  Figure~\ref{fig:forkdown}, which shows this topmost $\absrd$ and the
  cycle $c$ in bold lines. For every path starting or ending inside
  the right-hand side copy of $T$ in the contractum of the $\absrd$,
  we have a path starting or ending at the same place inside the
  left-hand side copy of $T$. Hence, from $c$, we can construct
  another cycle $c'$, which does not use the right-hand side copy of
  $T$, as it is visualized in bold lines on the right of
  Figure~\ref{fig:forkdown}. Thus, the $\absrd$ does not fork
  $c'$. Hence $\fork{c'}=n-1$.  By induction on $\fork{c}$ it follows
  that $\fg{\Delta'}$ must contain an unforked cycle.
\end{proof}

Let us now state the key
property of \bqfgs, that in the end makes the decomposition possible.

\begin{theorem}[(Step 2 in Fig.~\ref{fig:third})]\label{thm:no-unforked}
  There is no derivation $\Delta$ in $\SNEL$, such that $\fg{\Delta}$
  contains an unforked cycle.
\end{theorem}

The proof of this theorem, which is the final link for completing
Step~2 in the decomposition, is the purpose of the next section.

\subsection{Switch and Seq}\label{sec:nocycles}

The impossibility of unforked cycles in a \bqfg\ is caused by a
fundamental property of derivations in $\MLL$, which remains untouched
by adding \emph{seq}, and which is nothing but the acyclicity
condition for $\MLL$ proof nets. However, under the presence of
\emph{seq}, the formulation of this acyclicity is a bit more
complicated and the proof a bit more involved, since we cannot rely on
the sequent calculus~\cite{tiu:SIS-II}. We state the property we need
in the following lemma (a similar result has already been shown by
Retor\'e~\cite{retore:99}):

\begin{lemma}[(Step 2 in Fig.~\ref{fig:third})]\label{lem:bv_no_cycle}
    Let $n>0$ and let $a_0$ ,$a_1$, \ldots, $a_{n-1}$, $b_0$, $b_1$,
    \ldots, $b_{n-1}$ be $2n$ 
    different atoms. Further, let $W_0,\dots,W_{n-1},Z_0,\dots,Z_{n-1}$ be
    structures, such that 
    \begin{itemize}[---]
    \item $W_i=\pars{a_i\bc b_i}$ or $W_i=\seqs{a_i\bc b_i}$, for every  
      $i=0,\dots,{n-1}$,      
    \item $Z_j=\aprs{b_j\bc a_{j+1}}$ or $Z_j=\seqs{b_j\bc a_{j+1}}$,
      for every $j=0,\dots,{n-1}$ (where the indices are counted
      modulo $n$).
    \end{itemize}
    Then there is no derivation
    \begin{equation}\label{eq:fancy-atomic-cycle}\mysmall
      \simplederi
	  {\aprs{W_0\bc W_1\bc \ldots\bc W_{n-1}}}
	  {\set{\swir,\seqrd,\seqru}}{\Deltat}
	  {\pars{Z_0\bc Z_1\bc \ldots\bc Z_{n-1}}}
    \end{equation}
\end{lemma}

\begin{remark}
  This lemma can be used to prove that every theorem of $\BV$ is also
  a theorem of pomset logic. The other direction is still an open
  problem.
\end{remark}

Before giving the proof of Lemma~\ref{lem:bv_no_cycle}, let us state
and prove the second lemma of this section, which says that an
unforked cycle in the \bqfg\ of a derivation $\Delta$ can be
transformed into a derivation $\Deltat$ as shown
in~\eqref{eq:fancy-atomic-cycle} above.  The basic idea is to remove
from $\Delta$ everything that does not belong to the cycle, and then
construct $\Deltat$ such that the \bqfg\ of $\Delta$ becomes the
\afg\ of $\Deltat$.

To make this technically precise, note that in every cycle $c$ in a \bqfg, 
the following numbers are all equal:
\begin{itemize}[---]
\item the number of maximal $\loc$-up-paths in $c$,
\item the number of maximal $\lwn$-down-paths in $c$,
\item the number of flipping edges in $c$ from a $\loc$-vertex to a
  $\lwn$-vertex, and 
\item the number of flipping edges in $c$ from a $\lwn$-vertex to a
  $\loc$-vertex. 
\end{itemize}
We call this number the \emph{characteristic number} of $c$. For
example, the cycle in the derivation in~\eqref{eq:short} has
characteristic number~1, and the one in~\eqref{eq:cyclederi} has
characteristic number~2. 

\begin{lemma}[(Step 2 in Fig.~\ref{fig:third})]\label{lem:cycle-cycle}
  Let $\Delta$ be a derivation in $\SNEL'$ such that $\fg{\Delta}$
  contains an unforked cycle $c$. Then there is a derivation
  \begin{equation}\label{eq:atomic-cycle}\mysmall
    \vcstrder{\set{\swir,\seqrd,\seqru}}{\Deltat}{
      \pars{\aprs{b_0\bc a_1}\bc \aprs{b_1\bc a_2}\bc \ldots\bc 
	\aprs{b_{n-2}\bc a_{n-1}}\bc\aprs{b_{n-1}\bc a_0}}}{
      \leaf{\aprs{\pars{a_0\bc b_0}\bc \pars{a_1\bc b_1}\bc \ldots\bc 
	  \pars{a_{n-2}\bc b_{n-2}}\bc\pars{a_{n-1}\bc b_{n-1}}}}
    }
  \end{equation}
  for some atoms $a_0,\ldots,a_{n-1},b_0,\ldots,b_{n-1}$, where $n>0$ is the
  characteristic number of $c$.
\end{lemma}

\begin{proof}
  First, we transform $\Delta$ into a derivation
  $\Delta'$ which contains only rules from
  $\SNEL'\setminus\set{\dmrd,\absrd,\weakrd,\weakru}$ and in which the
  cycle is preserved. This is done by moving the rules $\dmrd$,
  $\absrd$, and $\weakrd$ down in the derivation by applying
  Lemma~\ref{lem:singlestep-down}, and by moving all instances of
  $\weakru$ also down in derivation (by applying the dual of
  Lemma~\ref{lem:perm-weak-atir}, together with~\eqref{eq:bwu}):
  \proofadjust
  $$\mysmall
  \vcstrder{\SNEL'}{\Delta}{Q}{
    \leaf{P}}
  \qquadlto
  \vcstrder{\set{\dmrd,\absrd,\weakrd,\weakru}}{\Delta_\downarrow}{Q}{
    \stem{\SNEL'\setminus\set{\dmrd,\absrd,\weakrd,\weakru}}{
      \Delta'}{Q'}{
	\leaf{P}}}
  \quadfs
  $$ 
  Since $c$ is unforked, no transformation step destroys the cycle,
  which is therefore still present in $\fg{\Delta'}$. It cannot be
  inside $\fg{\Delta_\downarrow}$ because there are no flipping edges.
  We continue the proof by marking some structures occurring in
  $\Delta'$:
  \begin{itemize}[---]
  \item We start by marking all $\loc$- and $\lwn$-vertices of
    $c$ by $\bloc$ and $\blwn$, respectively. It cannot happen that a
    $\bloc$- or $\blwn$-structure occurs inside another $\bloc$- or
    $\blwn$-structure, as it is case in the example in~\eqref{eq:short},
    because for having such a situation the cycle must change its
    ``!-?-depth'' twice, once at a $\absrd$, and once at a $\absru$,
    which is not possible since there are no $\absrd$ in $\Delta'$.
  \item Now we replace every $\bloc$ by $\bloc_i$ and
    every $\blwn$ by $\blwn_j$ for some $i,j\in\set{0,\ldots,n-1}$, such
    that
    \begin{itemize}[--]
    \item two $\bloc$-vertices in the same up-path get the same
      index, and two $\blwn$ in the same down-path get the same
      index, and
    \item every flipping edge in $c$ goes from a $\bloc_i$ to a
      $\blwn_i$ vertex, or from a $\blwn_i$ to a $\bloc_{i+1}$ vertex,
      where the addition is modulo $n$.
    \end{itemize}
  \item Note that at every flipping edge from a $\bloc_i$ to a
    $\blwn_i$ vertex there is another edge in $\fg{\Delta}$ also
    starting at $\bloc_i$, which continues the up-path marked by
    $\bloc_i$ up to the top of the derivation. We mark all
    $\loc$-vertices on this path by $\ubloc_i$. Since there are no
    instances of $\absrd$ left in $\Delta'$, the $\ubloc_i$ up-path is
    never forked, and since there are no $\erd$ and no $\weakrd$ in
    $\Delta'$, this path does not end before the top of the derivation. Hence,
    the premise $P$ of $\Delta'$ contains exactly
    $n$ substructures, marked by $\ubloc_0$, $\ubloc_1$, \ldots,
    $\ubloc_{n-1}$. Let us call them $\ubloc_0W_0$, $\ubloc_1W_1$,
    \ldots, $\ubloc_{n-1}W_{n-1}$.
    We also have $n$ instances of $\promrd$ in $\Delta$, marked as follows:
    \begin{equation}\mysmall
      \label{eq:promrd-u}
      \vcinf{\promrd}{
	S\pars{\bloc_i R\bc\blwn_i T}}{
	S\cons{\ubloc_i\pars{R\bc T}}}
    \end{equation}
  \item Now we proceed similarly and mark the
    continuations of the $\blwn_i$-down-paths by $\dblwn_i$, i.e., we
    obtain $n$ instances of $\promru$ marked as
    \begin{equation}\mysmall
      \label{eq:promru-d}
      \vcinf{\promru}{
	S\cons{\dblwn_i\aprs{T\bc R}}}{
	S\aprs{\blwn_i T\bc\bloc_{i+1} R}}
    \end{equation}
    However, note that now it can happen that we meet during the marking
    process a proper forking vertex, due to the presence of $\absru$:
    $$\mysmall
    \vctpathderivation{
      \inf{\absru}{S\aprs{\noc1 V\cons{\nwn3 T}\bc V\cons{\nwn4 T}}}
	  {S\cons{\noc2 V\cons{\nwn5_i^\bdmark T}}}}{%
      \slimvec \uvec{oc1}{oc2} \dvec{wn5}{wn3} \dvec{wn5}{wn4} 
    }
    \quadfs
    $$ then we continue the marking in only one side, namely, into that
    copy of $V\cons{\lwn T}$ in the redex of $\absru$, that contains
    already a $\bloc$-, $\blwn$-, $\ubloc$-, or $\dblwn$-marking. Note
    that it cannot happen that both copies of $V\cons{\lwn T}$ contain
    such a marking because the cycle is unforked. If neither side
    contains a marking, we arbitrarily pick one side. Since there are no
    $\eru$ and $\weakru$ in $\Delta'$, the $\dblwn$-paths cannot end in
    the middle of the derivation. Hence, the conclusion $Q'$ of
    $\Delta'$ contains exactly $n$ different marked $\dblwn$-structures,
    that we denote by $\dblwn_0Z_0$, $\dblwn_1Z_1$, \ldots,
    $\dblwn_{n-1}Z_{n-1}$. 
  \end{itemize}
  Now we remove in $\Delta'$ every modality
  that is not marked, and we replace every atom that is not inside a
  marked structure by the unit~$\un$. The important point is that
  after this rather drastic change we still have a correct
  derivation. Every rule instance in $\Delta'$ remains valid, or
  becomes vacuous, i.e., premise and conclusion are identical. Note
  that here we make crucial use of the fact that the cycle is
  unforked: Doing this deletion to a $\absru$ which forks $c$ would
  yield an incorrect inference step.
  
  Let us call the new derivation
  $\Delta''$. Its premise $P''$ is made from the structures
  $\ubloc_0W_0$, $\ubloc_1W_1$, \ldots, $\ubloc_{n-1}W_{n-1}$ by using
  only the binary connectives $\vlten$, $\vlseq$, and $\vlpar$, and
  its conclusion $Q''$ s made from $\dblwn_0Z_0$, $\dblwn_1Z_1$,
  \ldots, $\dblwn_{n-1}Z_{n-1}$ by using only $\vlten$, $\vlseq$, and
  $\vlpar$.  Now note that for arbitrary structures $A$ and $B$, we
  have the following three derivations:
  $$\mysmall
  \vcdernote{=}{}{\seqs{A\bc B}}{
    \root{\seqru}{\seqs{\aprs{A\bc\un}\bc\aprs{\un\bc B}}}{
      \root{=}{\aprs{\seqs{A\bc\un}\bc\seqs{\un\bc B}}}{
	\leaf{\aprs{A\bc B}}}}}
  \nqquand
  \vcdernote{=}{}{\pars{A\bc B}}{
    \root{\swir}{\pars{\aprs{A\bc\un}\bc B}}{
      \root{=}{\aprs{A\bc\pars{\un\bc B}}}{
	\leaf{\aprs{A\bc B}}}}}
  \nqquand
  \vcdernote{=}{}{\pars{A\bc B}}{
    \root{\seqrd}{\pars{\seqs{A\bc\un}\bc\seqs{\un\bc B}}}{
      \root{=}{\seqs{\pars{A\bc\un}\bc\pars{\un\bc B}}}{
	\leaf{\seqs{A\bc B}}}}}
  $$
  Hence, we can extend $\Delta''$ as follows:
  \begin{equation}\mysmall
    \label{eq:only}
    \vcstrder{\set{\seqrd,\swir}}{}{
      \pars{\dblwn_0Z_0\bc\dblwn_1Z_1\bc\vldots\bc\dblwn_{n-1}Z_{n-1}}}{
      \stem{}{\Delta''}{Q''}{
	\stem{\set{\seqru,\swir}}{}{P''}{
	  \leaf{\aprs{\ubloc_0W_0\bc\ubloc_1W_1\bc\vldots\bc
	      \ubloc_{n-1}W_{n-1}}}}}}
  \end{equation}
  Let us use $\Delta'''$ to denote the derivation
  in~\eqref{eq:only}. We finally obtain $\Deltat$ from $\Delta'''$ by
  replacing every $\bloc_i$-structure by $a_i$, every
  $\blwn_i$-structure by $b_i$, every $\ubloc_i$-structure by
  $\pars{a_i\bc b_i}$, and every
  $\dblwn_i$-structure by $\aprs{b_i,a_{i+1}}$. Clearly, every
  inference rule remains valid, or becomes vacuous, as for example the
  instances of $\promrd$ in~\eqref{eq:promrd-u} and $\promru$
  in~\eqref{eq:promru-d}: 
  $$\mysmall
  \vcinf{\promrd}{
    S\pars{\bloc_i R\bc\blwn_i T}}{
    S\cons{\ubloc_i\pars{R\bc T}}}
  \;\to\;
  \vcinf{=}{S\pars{a_i\bc b_i}}{S\pars{a_i\bc b_i}}
  \nqquand
  \vcinf{\promru}{
    S\cons{\dblwn_i\aprs{T\bc R}}}{
    S\aprs{\blwn_i T\bc\bloc_{i+1} R}}
  \;\to\;
  \vcinf{=}{S\aprs{b_i\bc a_{i+1}}}{S\aprs{b_i\bc a_{i+1}}}
  $$
  If a rule does not become vacuous,
  it must be one of $\swir$, $\seqrd$, and $\seqru$. 
\end{proof}

\begin{proof}[of Lemma~\ref{lem:bv_no_cycle}]
  The proof is carried out by induction on the pair
  $\tuple{n,q}$, where $q$ is the number of seq-structures in the
  conclusion, and we endorse the lexicographic ordering on
  $\Nat\times\Nat$. The base case (i.e., $n=1$) is
  trivial.  For the inductive case we assume \bwoc\ the existence of
  the derivation $\Deltat$ in \eqref{eq:fancy-atomic-cycle} and
  consider the bottommost rule instance $\rho$. There are three cases.
  \begin{enumerate}[(i)]
  \item $\rho=\seqru$. There is only one possibility to apply this rule:
    $$\mysmall
    \vcdernote{\seqru}{}{
      \pars{Z_0\bc \vldots\bc Z_{j-1}\bc 
	\seqs{b_j\bc a_{j+1}}\bc Z_{j+1}\bc \vldots\bc Z_{n-1}}}{
      \stem{\set{\swir,\seqrd,\seqru}}{\Delta'}{
	\pars{Z_0\bc \vldots\bc Z_{j-1}\bc 
	  \aprs{b_j\bc a_{j+1}}\bc Z_{j+1}\bc \vldots\bc Z_{n-1}}}{
	\leaf{\aprs{W_0\bc W_1\bc \ldots\bc W_{n-1}}}
    }}
    $$
    We can apply the induction hypothesis to $\Delta'$ because the
    number $n$ did not change and the number $q$ of seq-structures in
    the conclusion did decrease by 1. Hence we get a contradiction.
  \item $\rho=\seqrd$. There are several possibilities to apply this rule. We
    show here only two representative cases and leave the others to
    the reader because they are very similar. The
    complete case analysis can be found in \cite{dissvonlutz}.
    \begin{enumerate}[(a)]
    \item If we have
      \proofadjust
      $$\mysmall
      \vcdernote{\seqrd}{}{
	\pars{\seqs{b_0\bc a_1}\bc Z_1\bc \vldots\bc Z_{i-1}\bc 
          \seqs{b_i\bc a_{i+1}}\bc Z_{i+1}\bc \vldots\bc Z_{n-1}}}{
        \stem{\set{\swir,\seqrd,\seqru}}{\Delta'}{
          \pars{\seqs{\pars{b_0\bc b_i}\bc \pars{a_1\bc a_{i+1}}}\bc 
            Z_1\bc \vldots\bc Z_{i-1}\bc Z_{i+1}\bc \vldots\bc Z_{n-1}}}{
	\leaf{\aprs{W_0\bc W_1\bc \ldots\bc W_{n-1}}}
      } }
      $$
      then $\Delta'$ remains valid if we replace
      $a_m$ and $b_m$ by $\un$ for every $m>i$ and for $m=0$.
      This gives us the derivation
      $$\mysmall
      \vcstrder{\set{\swir,\seqrd,\seqru}}{\Delta''}{
	\pars{\seqs{b_i\bc a_1}\bc Z_1\bc \vldots\bc Z_{i-1}}}{
	\leaf{\aprs{W_1\bc \vldots\bc W_i}}
      }
      $$ which is a contradiction to the induction hypothesis because
      $i<n$.
    \item Consider
      \proofadjust
      $$\mysmall
      \vcdernote{\seqrd}{}{
	\pars{\seqs{b_0\bc a_1}\bc Z_1\bc \vldots\bc Z_{n-1}}}{
	\stem{\set{\swir,\seqrd,\seqru}}{\Delta'}{
	  \pars{\seqs{b_0\bc \pars{a_1\bc Z_{k_1}\bc \vldots\bc Z_{k_v}}}\bc 
	    Z_{h_1}\bc \vldots\bc Z_{h_s}}}{
	  \leaf{\aprs{W_0\bc W_1\bc \ldots\bc W_{n-1}}}
      }}
      $$
      where $\{1,\dots,{n-1}\}\setminus\{k_1,\dots,k_v\}=\{h_1,\dots,h_s\}$ and
      $s=n-v-1$ and (\wolg) $k_1<k_2<\ldots<k_v$.
      As before, the derivation $\Delta'$ remains valid if we replace
      $a_m$ and $b_m$ by $\un$ for every $m$ with $1\le m\le k_v$.
      Then we get
      $$\mysmall
      \vcstrder{\set{\swir,\seqrd,\seqru}}{\Delta''}{
	\pars{\seqs{b_0\bc a_{k_v+1}}\bc Z_{k_v+1}\bc \vldots\bc Z_{n-1}}}{
	\leaf{\aprs{W_0\bc W_{k_v+1}\bc \vldots\bc W_{n-1}}}}
      $$ which is (as before) a contradiction to the induction
      hypothesis because $v\ge1$.
    \end{enumerate}
  \item $\rho=\swir$. This is similar to the case for $\seqrd$. But note that
    a situation like in (ii.a) cannot happen for $\swir$. \qed
  \end{enumerate}
\end{proof}

\begin{proof}[of Theorem~\ref{thm:no-unforked}]
  The existence of an unforked cycle in $\fg{\Delta}$ implies by
  Lemma~\ref{lem:cycle-cycle} the existence of a derivation as in
  \eqref{eq:atomic-cycle}. By Lemma~\ref{lem:bv_no_cycle}, this
  is impossible.
\end{proof}

Now we can complete the proof of Theorem~\ref{thm:decomposition} by
proving the following lemma.

\begin{lemma}[(Step 2 in Fig.~\ref{fig:third})]\label{lem:step2}
  Step~2 in the proof of Theorem~\ref{thm:decomposition} can be
  performed as indicated in Figure~\ref{fig:third}.
\end{lemma}

\begin{proof}
  We can first permute all $\dmrd$, $\absrd$, $\weakrd$ down, by
  applying Lemma~\ref{lem:singlestep-down}, and then permuting all
  $\dmru$, $\absru$, $\weakru$ up, by applying
  Lemma~\ref{lem:singlestep-up}. The result has an acyclic \bqfg\
  (otherwise, it would have by Lemma~\ref{lem:nocycle} an unforked
  cycle, which is impossible by Theorem~\ref{thm:no-unforked}). Hence,
  we can apply Lemma~\ref{lem:nocycle-term}.
\end{proof}

\section{Perspectives}

We now briefly mention the developments that we expect to be based on 
this work.

Much of the arguments that we use have similarities with the
techniques developed for atomic flows~\cite{gug:gun:flows}, with the
important difference that, here, we are dealing with flows of
modalities rather than atoms. Nonetheless, the similarities suggest
that there might be a common structure that can perhaps be unveiled
and exploited in future research. 

The techniques developed here for decomposition might possibly be
exported to the many modal logics already available in deep inference
(some of which have no known analytic presentation in Gentzen
formalisms).

Both for $\NEL$ and for other modal logics, it should be possible to
use decomposition for investigating interpolation, which is a
classical proof theoretic concern, and that relies on a similar kind
of decomposition (we can consider Herbrand-like theorems as simple
examples of decomposition).

We mentioned the applications of $\BV$ to process algebras and causal
quantum evolution. We expect $\NEL$ to find uses in the same
directions. In the case of process algebras, this is almost obvious,
given that $\NEL$ is Turing-complete and that exponentials have been
justified since their first introduction as ways of controlling
resources (\emph{i.e.}, messages, processes). The logic $\BV$ has also
been used to define $\BV$-categories~\cite{blute:etal:ACS} for
providing an axiomatic description of probabilistic coherence
spaces~\cite{girard:tract}.

What we have in this paper is a basic compositional result, so, we 
expect applications to be very broad in range. That said, we think 
that, perhaps, the most important outcome of this whole research on 
seq and its logical systems is one of extending the limits of proof 
theory and of developing new insight and new techniques.

We witness here an interesting phenomenon: on one hand, 
we have a very simple system in a very simple formalism (\emph{i.e.}, 
$\NEL$ in the calculus of structures); on the other hand, we have a 
very simple property, decomposition. In the middle, connecting the 
two, there's a rich and complex combinatorial phenomenon. Sometimes, 
in similar situations, the mathematics that arises has some lasting 
value, and this is our hope for this paper.

\subsection*{Acknowledgments}

The referees provided unusually deep and outstanding suggestions for
making this paper more accessible and, in general, better. We would
like to thank Luca Roversi for the extremely insightful and technical
comments, and the numerous discussions on this subject, which greatly
helped us understand the matter.


\begin{received}
Received March 2009;
revised February 2010;
accepted June 2010
\end{received}

\end{document}